\documentclass[11pt,reqno]{amsart}

\usepackage{graphicx}
\usepackage{amssymb, amsmath}
\usepackage{amsfonts}
\usepackage{amssymb}
\usepackage{epsfig}
\usepackage{float}
\usepackage{color}
\usepackage{bm}
\usepackage{paralist}

\usepackage{array}

\newtheorem{lem}{Lemma}[section]

\newcommand{\bqn}{\begin{equation}}
\newcommand{\eqn}{\end{equation}}
\newcommand{\beqx}{\begin{equation*}}
\newcommand{\eeqx}{\end{equation*}}
\newcommand{\barr}{\begin{array}}
\newcommand{\earr}{\end{array}}
\newcommand{\beqn}{\begin{eqnarray}}
\newcommand{\eeqn}{\end{eqnarray}}
\newcommand{\beqnx}{\begin{eqnarray*}}
\newcommand{\eeqnx}{\end{eqnarray*}}
\newcommand{\bmt}{\begin{multline}}
\newcommand{\emt}{\end{multline}}

\newcommand{\D}{\partial}

  \newcommand{\phia}{\phi^A}

\newcommand{\ep}{\varepsilon}
\newcommand{\ve}{\varepsilon}

\newcommand{\er}{\eqref}
\newcommand{\lb}{\label}

\def\tilde{\widetilde}

\numberwithin{equation}{section} 

\vspace{1cm}

\begin{document}
\bibliographystyle{siam}

\title[Boundary layers for Vlasov--Poisson system]
{Approximate solutions for the Vlasov--Poisson system with boundary layers}

\author[C.-Y. Jung]{Chang-Yeol Jung}
\address[CJ]{Department of Mathematical Sciences, Ulsan National Institute of Science and Technology, Ulsan 44919, Korea}
\email{cjung@unist.ac.kr}
\author[B. Kwon]{Bongsuk Kwon}
\address[BK]{Department of Mathematical Sciences, Ulsan National Institute of Science and Technology, Ulsan 44919, Korea}
\email{bkwon@unist.ac.kr}
\author[M. Suzuki]{Masahiro Suzuki}
\address[MS]{Department of Computer Science and Engineering, Nagoya Institute of Technology, Nagoya 4668555, Japan}
\email{masahiro@nitech.ac.jp}
\author[M. Takayama]{Masahiro Takayama}
\address[MT]{Department of Mathematics, Keio University, Yokohama 2238522, Japan}
\email{masahiro@math.keio.ac.jp}

\date{\today}



\begin{abstract}
 We construct the approximate solutions to the Vlasov--Poisson system in a half-space, which arises in the study of the quasi-neutral limit problem in the presence of a sharp boundary layer, referred as to the plasma sheath in the context of plasma physics. The quasi-neutrality is an important characteristic of plasmas and its scale is characterized by a small parameter, called the Debye length. 
 We present the approximate equations obtained by a formal expansion in the parameter and study the properties of the approximate solutions. 
 Moreover, we present numerical experiments demonstrating that the approximate solutions converge to those of the Vlasov--Poisson system as the parameter goes to zero.
\end{abstract}
\maketitle   


\section{Introduction}\lb{S1}

We consider the Vlasov--Poisson system  in a one-dimensional half-space:
\begin{subequations}\label{VP1}
\begin{gather}
\D_{t} f + \xi_{1} \partial_{x} f+\partial_{x} \phi \partial_{\xi_{1}} f =0, \quad t>0, \ x>0 , \ \xi \in \mathbb R^{3},
\label{Vlasov1}
\\
\ve \partial_{xx} \phi - \int_{\mathbb R^{3}} f d\xi + e^{-{\phi}}=0, \quad t>0, \ x>0,
\label{Poisson1}
\end{gather}
where $t>0$, $x>0$, and $\xi = (\xi_{1},\xi_{2},\xi_{3})=(\xi_{1},\xi') \in \mathbb R^{3}$ 
are the time variable, space variable, and velocity, respectively.
The motion of positive ions in a plasma, 
 is  often described at the microscopic level by \eqref{VP1}, in which the unknown functions $f = f(t,x,\xi)$ and $-\phi=-\phi(t,x)$ stand 
for the velocity distribution of positive ions
and the electrostatic potential, respectively. 
The small parameter $0<\ve\ll1$ is the square of the rescaled Debye length.  
It is assumed that the number density of electrons obeys the Boltzmann relation $n_{e}=e^{-{\phi}}$.

 When a negatively charged material is immersed in a plasma, it is known by experiments that the ion-rich layer is formed near the surface shielding the plasma from the charged body, and this layer is referred as to \emph{the plasma sheath}. The plasma sheath is known to appear when \emph{the Bohm criterion} holds. 
For a more detailed discussion on the physicality of the Bohm criterion and the formation of plasma sheaths, we refer to \cite{DB1, BT1, La29, Ri91}.

In order to mathematically investigate the solutions related to plasma sheaths, 
we consider the system \eqref{Vlasov1}--\eqref{Poisson1} 
in a half space, 
with physically relevant  initial and boundary conditions
\begin{gather}
f (0,x,\xi)= f_{0}(x,\xi) \geq 0, \quad x>0 , \ \xi \in \mathbb R^{3},
\label{ini1} \\
f (t,0,\xi) = 0, \quad t>0, \ \xi_{1}>0,
\label{bc1} \\
\phi(t, 0)=\phi_{b}>0, \quad t>0, 
\label{bc3} \\
\lim_{x \to\infty} f (t,x,\xi) =  f_{\infty}(\xi) \geq 0, 
\quad t>0, \ \xi \in \mathbb R^{3},
\label{bc2} \\
\lim_{x \to\infty} \phi (t,x) =  0, \quad t>0.
\label{bc4} 
\end{gather}
\end{subequations}
The end state $f_{\infty}=f_{\infty}(\xi)$ is a nonnegative function with the quasi-natural condition
\begin{gather}
\int_{\mathbb R^{3}} f_{\infty} d\xi =1.
\label{qn1}
\end{gather}
Note that \eqref{qn1} turns out to be  a necessary condition for the solvability of the problem \eqref{VP1}, 
which is discussed in \cite{STZ1}.

Physically speaking, the boundary condition \eqref{bc1} is referred as to  the {\it completely absorbing boundary condition}.
As mentioned above, the wall is negatively charged when the sheath is formed, which is 
the case $\phi_{b}>0$ in \eqref{bc3} 
since $-\phi$ is the electrostatic potential.
The boundary condition  \eqref{bc4} indicates that the reference point of the potential is located at $x=\infty$.

  Due to 
the presence of a significant number of charge carriers (electrons), a plasma is electrically conductive, which makes any charges in plasmas can be readily neutralized. In many physical situations, a plasma can be treated as electrically neutral overall, while at smaller scales, the charges may give rise to a charged region. This characteristic of a plasma is referred to as the \emph{quasi-neutrality}. The scale at which the neutrality breaks down is often characterized by a parameter, called the Debye length $\lambda_D$. 
 In usual physical laboratory settings,  $\lambda_D$ is known to be a significantly small number, e.g., $\lambda_D\approx 10^{-4} {\rm m}$.

Our main purpose of this work is to mathematically justify the quasi-neutrality of plasma sheath.
 It is an intriguing question if \eqref{VP1} is qualitatively robust with respect to {  the small parameter $\ve= \sqrt{C\lambda_D}$} for some $C>0$, especially in the presence of a sharp boundary layer. In fact,  the problem becomes significantly difficult when the solution bears a sharp transition layer, which is the plasma sheath in our setting. To formulate the problem mathematically, we consider under  \emph{the Bohm criterion},  the corresponding limit problem ($\ve \to 0$), which is often called the quasi-neutral limit problem.

To study the quasi-neutral limit problem for \eqref{VP1}, we first consider the  limiting equations obtained by formally 
letting $\ve \to 0$ for \eqref{Vlasov1} and \eqref{Poisson1}:
\begin{subequations}\lb{zero-out}
\begin{gather}\label{lim_eq}
\partial_{t} f^0 + \xi_1 \partial_{x} f^0 +  \partial_{x} \phi^0 \partial_{\xi_{1}} f^0 = 0,\\
-\int_{\mathbb R^{3}}  f^0 d\xi + e^{-\phi^0} = 0.
\label{lim_eq1}
\end{gather}
For this system, the initial and boundary conditions are given by 
\begin{gather}
f^0 (0,x,\xi)= f_{0}(x,\xi) \geq 0, \quad x>0 , \ \xi \in \mathbb R^{3}, \\
f^{0} (t,0,\xi) = 0, \quad t>0, \ \xi_{1}>0,\\ 
\lim_{x \to\infty} f^0 (t,x,\xi) =  f_{\infty}(\xi), \quad t>0, \ \xi \in \mathbb R^{3}. 
\end{gather}
 \end{subequations}
Throughout this paper, we assume  that the solution $(f^0, \phi^0)$ exists and satisfies the 
Bohm condition, i.e., 
\begin{align}\label{bohm2}
\inf_{t\in[0,T]} \left( \int_{\mathbb R^{3}}f^{0}(t,0,\xi) d\xi- \int_{\mathbb R^{3}}\xi_{1}^{-2}f^{0}(t,0,\xi) d\xi \right)>0.
\end{align}
It is also assumed that 
\begin{gather}
\inf_{t\in [0,T], \, x \in \overline{\mathbb R_{+}}} \int_{\mathbb R^{3}} f^{0}(t,x,\xi) d\xi >0,
\label{p1}\\
{\rm supp} f^{0}(t,x,\xi) \subset \left\{(t,x,\xi) \in [0,T] \times \overline{\mathbb R_{+}} \times \mathbb R^{3} \, \left| \, \xi_{1}  \leq -2c_{*}, \ |\xi| \leq \frac{1}{2}{C_{*}} \right.\right\},
\label{supp1}\\
\inf_{t\in [0,T]} (\phi_{b} - \phi^{0}(t,0))>0
\label{p2}
\end{gather}
for some constants $c_{*}$ and $C_{*}>0$ with $c_{*} < C_{*}/4$.
The condition \eqref{p1} implies that the number density is positive.
We use the condition \eqref{supp1} to ensure the compatibility condition of the zeroth order and the integrability of $f^{0}$ in \eqref{lim_eq1}.
The condition \eqref{p2} is required to avoid the situation that the sign of $\phi_{b} - \phi(t,0)$ is changing.
The well-posedness of the problem \eqref{zero-out} for fairly {\it general initial data} is one of our main interests for future work.
On the other hand, we remark that the problem \eqref{zero-out} has an explicit solution $(f^{0},\phi^{0})=(f_{0},0)$ for the homogeneous initial data, i.e., $f_{0}=f_{0}(\xi)$ (see subsection \ref{2.5}).

We remark that for the limiting problem \eqref{zero-out},  no boundary condition is imposed for $\phi^0$ at the boundary $x=0$, while for \eqref{VP1} the Dirichlet boundary condition for $\phi$ is prescribed as \er{bc3}. 
In general, these values do not agree and this discrepancy gives rise to a sharp transition layer near the boundary, related to the plasma sheath.
This makes the associated quasi-neutral limit problem \emph{singular}; the derivatives of such layer solutions blow up as $\ve\to0$.
To study the quasi-neutral limit problem \eqref{VP1}, the first step is to investigate the sharp transition layers near the boundary.
We refer the readers to  \cite{GHJT, GHT12, GK12, HK90, HMNW12, IP06, JPT16, Lio73, TW1998} for singular limit problems in various contexts. 
More closely related to our problems, at the macroscopic level, the quasi-neutral limit problem for the Euler--Poisson system in the presence of a boundary layer is studied in \cite{GHR2, JKS16, JKS20}. Especially, in a similar spirit to the present work,  the approximate equations for the Euler--Poisson system are systematically derived, and their well-posedness is discussed in \cite{JKS21} for an annular domain.   
On the other hand,  at the microscopic level, the quasi-neutral limit problems for the Vlasov--Poisson system are considered in \cite{ HI1,HI2, HR}, for which the formal limit is justified in the framework of the Wasserstein distance. Since the problems are considered in a periodic domain, no sharp boundary layers appear.  
%
However, in the presence of boundary layers, which are physically relevant to the plasma sheath, to our best knowledge, the detailed analysis addressing the derivation and solvability of the approximate inner and outer equations, and their properties and estimates, 
is not available up to date. 
The purpose of this paper is to resolve these issues and analyze them in great detail.

From a simple example of the reaction-diffusion equation in \cite[Figure 1]{JKS21},
one can expect that the solution of \er{VP1} becomes \emph{singular} near the boundary as $\ve\to 0$, for instance, its pointwise derivatives diverge. 
We will verify that this is the case by investigating the asymptotic behavior in the quasi-neutral limit, for which it is important to construct the \emph{good approximate solutions}.
To this end, 
we use the outer and inner expansions. 
The inner and outer solutions are commonly used tools that arise in the method of asymptotic expansions to construct accurate approximate solutions to   \emph{singularly} perturbed partial differential equations.
In what follows, we present a systematic way to approximate \eqref{VP1} with the initial and boundary conditions \eqref{ini1}--\eqref{bc4}, and study the resulting approximate equations with the associated initial and boundary conditions in Section~\ref{sec2}. Now we present our main result. 

\medskip

\newcommand{\fa}{f^{A}}

\noindent
{\sc \bf Main result.} 
{\it Let $m\ge1$, 
the order of approximation. 
 The approximate solutions $(f^{A}, \phi^{A})$ are constructed, defined in \eqref{app_exp_all1}.  
Then the following holds. 
\begin{enumerate}[(i)]
\item 
By substituting $(f, \phi) = (\fa, \phia)$ into the left-hand side of \eqref{VP1}, we set $\mathcal{R}_1$ and $\mathcal{R}_2$ as 
\begin{subequations}\label{VP-AP}
\begin{gather}
\D_{t} \fa + \xi_{1} \partial_{x} \fa+\partial_{x} \phia \partial_{\xi_{1}} \fa =: \mathcal{R}_1, 
\label{Vlasov-AP1}
\\
\ve \partial_{xx} \phia - \int_{\mathbb R^{3}} \fa d\xi + e^{-{\phia}}=: \mathcal{R}_2. 
\label{Poisson-AP1}
\end{gather}
\end{subequations}
For any $l, n \in\mathbb{N}_{0}$ and multi-index $\beta=(\beta_{1},\beta_{2},\beta_{3}) \in (\mathbb{N}_{0})^{3}$, 
\begin{subequations}\lb{Apro}
\begin{align} 
|\partial_{t}^l\partial_{x}^n\partial_{\xi}^{\beta}\mathcal{R}_1| &\leq C \ep^{(m+1)/2} +  C \ep^{(m-n)/2} e^{-cx/\sqrt{\ep}},
\lb{Apro1} \\
|\partial_{t}^l \partial_{x}^n\mathcal{R}_2| &\leq C \ep^{(m+1)/2} +  C \ep^{(m+1-n)/2} e^{-cx/\sqrt{\ep}}, 
\lb{Apro2}
\end{align}
\end{subequations}
where $c$ and $C$ are positive constants independent of $\ep$, $t$, $x$, and $\xi$.

\item For any $l,n \in\mathbb{N}_{0}$ and $\beta \in (\mathbb{N}_{0})^{3}$, it holds that 
\begin{align}
\sup_{t\in [0,T], \, \xi \in \mathbb R^{3}}| \partial_t^l \partial_x^{n} \partial_\xi^{\beta} f^{A} (t,x,\xi) | & \leq 
C\left(\ve^{-n/2} e^{- c x / \sqrt{\ve}} + 1 \right),
 \lb{app-prop-1} \\
\sup_{t\in [0,T]}|  \partial_t^l \partial_x^{n} \phi^{A} (t,x) | & \leq
C\left(\ve^{-n/2} e^{- c x / \sqrt{\ve}} + 1 \right),
 \lb{app-prop-2} 
\end{align}
where $c$ and $C$ are positive constants independent of $\ep$ and $x$.

\item We numerically verify that $(f^{A}, \phi^{A})$ converges to $(f,\phi)$ in $L^\infty$ as $\ve\to0$. 
\end{enumerate}
}

\medskip

More specifically, for the assertion $(iii)$, we provide various numerical experiments demonstrating  the approximate solutions converge to the exact ones as $\ep \rightarrow 0$ in Section \ref{sec5}.
Our numerical tests indicate that the solutions of the original problem \eqref{VP1} and the approximate solutions both exhibit sharp transition layers near the boundary after some time for fairly general initial data, see Figure~\ref{ex2_vari_eps_f},   Figures~\ref{ex3_f}--\ref{ex3_phi}, and Figures~\ref{ex4_f}--\ref{ex4_phi}. In fact, for all cases, the outer solutions are getting flatter as time goes on, which is exactly what is expected for the outer solutions to behave. 
We also provide numerical evidence of the convergence of the approximate solutions to the original ones,  see Figure~\ref{ex2_vari_eps_err}. 
Various numerical experiments demonstrate that the approximate equations derived by asymptotic expansion are good approximation to the original ones in the presence of sharp boundary layers. 

The assertion $(ii)$ indicates that our approximate solutions stay bounded away from the boundary and get singular near the boundary. More precisely, the thickness of the boundary layer is a scale of $\ve^{1/2}$, the rescaled Debye length, in which  the derivatives of the solutions become unbounded for small $\ve>0$.  This is exactly what our numerical experiments demonstrate, where the sharp transition layers near the boundary are the plasma sheath. This provides an evidence of the quasi-neutrality of plasma sheath, that is formally claimed in the context of physics.

The proofs of $(i)$ and $(ii)$  in our main result are given in Sections~\ref{sec3} and \ref{sec4} under some reasonable conditions (see (3.1), (3.8), and (3.20)).
One of the main difficulties to prove the assertion $(ii)$ lies on the fact that the inner equations that are derived by the asymptotic expansion are not ODEs, but they are still given as PDEs contrast to those arising in most singular limit problems, for instance see the Euler--Possion system in \cite{GHR2,JKS21}.
To resolve this, we reduce the inner equations to ODEs only for the electrostatic potential by applying the characteristics method as well as introducing a new coordinate. 
Here, the Bohm criterion plays an essential role in deriving the estimates of the inner solutions.
In the proof of $(i)$, it is difficult to handle the the nonlinearity $e^{-\phi}$, 
for which we make use of the Talyor expansion techniques developed in \cite{JKS21}.

The paper is organized as follows: In Section~\ref{sec2}, we derive the approximate equations. 
Section~\ref{sec3} is devoted to the estimates of the inner solutions. 
The estimates immediately lead to the assertion $(ii)$ in the main result.
In Section~\ref{sec4}, we prove the assertion $(i)$ by using the assertion $(ii)$.
In Section~\ref{sec5}, we provide some numerical experiments 
demonstrating the convergence of the approximate solutions to the exact  ones, and the properties of both approximate and exact ones. 

\section{Approximate solutions by asymptotic expansion}\label{sec2}
 
 In this section, we derive the approximate equations by an asymptotic expansion method. 
To efficiently handle the nonlinear terms in the original problem \eqref{VP1}, 
we introduce a new variable $w = e^{-\phi}$.
Then, we write (\ref{VP1}) as
 \begin{subequations}\label{eqs}
   \begin{gather}
\D_{t} f + \xi_{1} \partial_{x} f+\partial_{x} \phi \partial_{\xi_{1}} f =0, \quad t>0, \ x>0 , \ \xi \in \mathbb R^{3},
\label{eq1}
\\
\ve \partial_{xx} \phi 
= \int_{\mathbb R^{3}} f d\xi - w, \quad t>0, \ x>0, \label{eq2} \\
w = e^{-\phi}, \label{eq3}
   \end{gather}
supplemented with the initial and boundary conditions
\begin{gather}
f (0,x,\xi)= f_{0}(x,\xi) \geq 0, \quad x>0 , \ \xi \in \mathbb R^{3},
\label{sini1} \\
f (t,0,\xi) = 0, \quad t>0, \ \xi_{1}>0,
\label{sbc1} \\
\phi(t,0)=\phi_{b},
\label{sbc3} \\
\lim_{x \to\infty} f (t,x,\xi) =  f_{\infty}(\xi), 
\quad t>0, \ \xi \in \mathbb R^{3},
\label{sbc2} \\
\lim_{x \to\infty} \phi (t,x) =  0, \quad t>0.
\label{sbc4} 
\end{gather}
\end{subequations}

Subsections \ref{2.1}--\ref{2.4} provide the outer and inner expansions with respect to the small parameter $\ep$.
In subsection \ref{2.5}, we study the well-posedness of the approximate solutions for the homogeneous initial data $f_{0}$.
To this end, we introduce the multi-index notations and multinomial expansion 
\begin{align*}
\alpha &= (\alpha_1,\ldots,\alpha_l,\ldots),\\
\left( \begin{array}{cc} k \\ \alpha \end{array} \right) &= \frac{k!}{\alpha_{1} ! \cdots \alpha_{l} ! \cdots },\\
|\alpha| &= \alpha_1+\alpha_2 + \cdots+\alpha_l+\cdots,\\
\|\alpha\| &= \alpha_{1} + 2 \alpha_{2} + \cdots + l \alpha_{l} + \cdots, \\
u^\alpha &= \big(u^1\big)^{\alpha_1}\cdots \big(u^l\big)^{\alpha_l}\cdots, 
\end{align*}
and we recall the following statement developed in Appendix 6.2 of \cite{JKS21}:
\begin{lem}[\cite{JKS21}]\label{Taylor1}
Let $g$ be an analytic function.  There holds formally that 
\begin{gather*}
g\bigg( \sum_{j=0}^\infty \ep^{j/2} u^j \bigg) = g(u^0) + \sum_{j=1}^\infty \ep^{j/2}g'(u^0)u^j + \sum_{j=2}^\infty \ep^{j/2}\mathcal{J}^{j-1}_g(u),
\\
\mathcal{J}^{j-1}_g(u) := \sum_{k=1}^j \frac{g^{(k)}(u^0)}{k!}\sum_{|\alpha| = k,\ \|\alpha\|=j, \ \alpha_j=0 } \left( \begin{array}{cc} k\\ \alpha \end{array} \right)  u^{\alpha}, \quad  j\geq 2,
\end{gather*}
where  $\mathcal{J}^{j-1}_g(u)$ is independent of $u^{j}$ and determined only by $u^0,u^1,\ldots,u^{j-1}$.
\end{lem}

\subsection{Equations of outer solutions}\lb{2.1}
To investigate the solutions away from the boundary, we introduce the outer expansion, 
\begin{align}\label{outer_exp}
(f, \phi,w)(t,x,\xi) = \sum_{j=0}^\infty \ve^{j/2} (f^j, \phi^j,w^j)(t,x,\xi).
\end{align}
We substitute the expansion (\ref{outer_exp}) into (\ref{eq1})--(\ref{eq3})
to find that
\begin{subequations}\lb{outer_exp0}
\begin{gather}
\sum_{j=0}^\infty \ve^{j/2} \partial_{t} f^j  + \xi_1 \sum_{j=0}^\infty \ve^{j/2} \partial_{x} f^j
+ \sum_{j=0}^\infty \ve^{j/2}\sum_{l+k=j, \ l,k \geq 0}\partial_{x} \phi^l \partial_{\xi_{1}} f^k  = 0,
\label{outer_exp1}\\
\sum_{j=0}^\infty \ep^{(j+2)/2} \partial_{xx} \phi^j - \sum_{j=0}^\infty \ve^{j/2} \int_{\mathbb R^{3}}  f^j d\xi + \sum_{j=0}^\infty \ve^{j/2} w^j=0.
\label{outer_exp2}
\end{gather}
\end{subequations}
At the order of $\ep^{j/2}$, $j \geq 0$, in (\ref{outer_exp1}), we obtain the outer equations:
\begin{subequations}\label{outer_eqs}
\begin{gather}
\partial_{t} f^0 + \xi_1 \partial_{x} f^0 + \partial_{x} \phi^0 \partial_{\xi_{1}} f^0 = 0,\\
\partial_{t} f^1 + \xi_1 \partial_{x} f^1 + \partial_{x} \phi^1 \partial_{\xi_{1}} f^0 + \partial_{x} \phi^0 \partial_{\xi_{1}} f^1 =  0,\\
\partial_{t} f^j + \xi_1 \partial_{x} f^j + \partial_{x} \phi^j \partial_{\xi_{1}} f^0 + \partial_{x} \phi^0 \partial_{\xi_{1}} f^j   =  -\sum_{l=1}^{j-1} \partial_{x} \phi^l \partial_{\xi_{1}} f^{j-l},\quad  j \geq 2.
\end{gather}
From \er{outer_exp2}, we similarly obtain the outer equations at the order of $\ve^{j/2}$, $j \geq 0$, respectively:
\begin{gather}
\int_{\mathbb R^{3}}  f^j d\xi - w^j = 0, \quad j=0,1,\\
\int_{\mathbb R^{3}}  f^j d\xi - w^j = \partial_{xx} \phi^{j-2},\quad j \geq 2.
\end{gather}
\end{subequations}
From (\ref{eq3}), using Lemma \ref{Taylor1} with $g(\phi)=e^{-\phi}$, we write
\begin{subequations}\label{expp}
\begin{align}
\begin{split}
&w^0 = e^{-\phi^0},\\
&w^j = -e^{-\phi^0}\phi^j + \mathcal{J}^{j-1}_{\exp}(\phi),\ j \geq 1,
\end{split}
\end{align}
where
\begin{gather}
\mathcal{J}^{j-1}_{\exp}(\phi) := 
\left\{
\begin{array}{ll}
0, & j=1,
\\
\displaystyle \sum_{k=1}^j\frac{(-1)^{k}}{k!} e^{-\phi^0} \sum_{|\alpha| = k,\ \|\alpha\| = j, \ \alpha_j=0} \left( \begin{array}{cc} k\\ \alpha \end{array} \right)  \phi^{\alpha}, & j\geq2.
\end{array}
\right.
\end{gather}
\end{subequations}

From (\ref{outer_eqs}) with $j=0$, we have the limiting equations,
\begin{subequations}\label{outer_sol_eq0}
\begin{gather}
\partial_{t} f^0 + \xi_1 \partial_{x} f^0 + \partial_{x} \phi^0 \partial_{\xi_{1}} f^0 = 0,
\label{outer_eq-1}\\
\int_{\mathbb R^{3}}  f^0 d\xi - e^{-\phi^0} = 0.
\label{outer_eq0}
\end{gather}
The initial and boundary conditions for these equations are given by
\begin{gather}
f^0 (0,x,\xi)= f_{0}(x,\xi), \quad x>0 , \ \xi \in \mathbb R^{3}, \\
f^0 (t,0,\xi) = 0, \quad t>0, \ \xi_{1}>0,\\
\lim_{x \to\infty} f^0 (t,x,\xi) =  f_{\infty}(\xi), \quad t>0, \ \xi \in \mathbb R^{3}.
\end{gather}
\end{subequations}
We also obtain the outer equations at $j=1$,
\begin{subequations}\label{outer_sol_eqj}
\begin{gather}
\partial_{t} f^1 + \xi_1 \partial_{x} f^1 + \partial_{x} \phi^1 \partial_{\xi_{1}} f^0 + \partial_{x} \phi^0 \partial_{\xi_{1}} f^1 =  0,
\label{outer_eq1}\\
\int_{\mathbb R^{3}}  f^1 d\xi + e^{-\phi^0}\phi^1 = 0,
\label{outer_eq1'}
\end{gather}
and the outer equations at $j \geq 2$, 
\begin{gather}
\partial_{t} f^j + \xi_1 \partial_{x} f^j + \partial_{x} \phi^j \partial_{\xi_{1}} f^0 + \partial_{x} \phi^0 \partial_{\xi_{1}} f^j  =  -\sum_{l=1}^{j-1}\partial_{x} \phi^l \partial_{\xi_{1}} f^{j-l},
\label{outer_eqj}\\
\int_{\mathbb R^{3}}  f^j d\xi + e^{-\phi^0}\phi^j = \partial_{xx} \phi^{j-2} + \mathcal{J}^{j-1}_{\exp}(\phi).
\label{outer_eqj'}
\end{gather}
We remark that the right-hand sides of (\ref{outer_eqj}) and (\ref{outer_eqj'}) are all determined only by the previous terms $(f^l,\phi^l)$, $l=0,1,\ldots,j-1$, and that the equations in \eqref{outer_sol_eqj} are all linear.
The initial and boundary conditions for the outer equations for $j \geq 1$ are given by
\begin{gather}
f^j (0,x,\xi)= 0, \quad x>0 , \ \xi \in \mathbb R^{3}, \\
f^j (t,0,\xi) = 0, \quad t>0, \ \xi_{1}>0, \\
\lim_{x \to\infty} f^j (t,x,\xi) =  0, \quad t>0, \ \xi \in \mathbb R^{3}. 
\end{gather}
\end{subequations}

\subsection{Equations of inner solutions}\lb{2.2}

To study the behaviors of solutions near the boundary $\{x=0\}$, we introduce the inner expansions,
\begin{align}\label{outer_inner_exp_app}
(f, \phi,w)(t, x, \xi) 
=  \sum_{j=0}^\infty \ve^{j/2} (f^j+F^j, \phi^j+\Phi^j,w^j+W^j)(t, x, \xi),
\end{align}
where $(f^{j},\phi^{j},w^{j})$ are the outer expansions discussed in (\ref{outer_exp}), and $(F^{j},\Phi^{j},W^{j})$ are the inner expansions we now discuss.

Substituting (\ref{outer_inner_exp_app}) into (\ref{eq1})--(\ref{eq3}) and subtracting (\ref{outer_exp0}) from the resutling equations respectively, we obtain
\begin{subequations}\label{exp0_int}
\begin{align}
&\sum_{j=0}^\infty \ve^{j/2} \partial_{t} F^j + \xi_1\sum_{j=0}^\infty \ve^{j/2} \partial_{x} F^j 
\notag \\
&\quad + \sum_{j=0}^\infty \ve^{j/2}\!\sum_{\substack{l+k=j, \\ l,k\geq 0}}\!\!\big(\partial_{x} \phi^l \partial_{\xi_{1}}F^k + \partial_{x} \Phi^l \partial_{\xi_{1}} f^k + \partial_{x} \Phi^l \partial_{\xi_{1}} F^k \big)= 0,\label{exp1_int}\\
&\sum_{j=0}^\infty \ep^{(j+2)/2} \partial_{xx} \Phi^j  - \sum_{j=0}^\infty \ve^{j/2}\int_{\mathbb R^{3}}  F^j d\xi + \sum_{j=0}^\infty \ve^{j/2} W^j=0.\label{exp2_int}
\end{align}
\end{subequations}
To construct the inner equations near the boundary $\{x=0\}$,
we introduce the stretched variable $\bar{x}$ as $$\bar{x}=\frac{x}{\sqrt{\ep}}.$$
Then it is easy to see that   the outer solutions verify 
\begin{align}\label{obs1}
|f^j(t,x,\xi)-f^j(t,0,\xi)| \leq C \ep^{1/2} \bar{x}\|\partial_{x} f^j(t,\cdot,\xi)\|_{L^\infty}.
\end{align}
Namely, we use $\bar{x}$ for $(F^{j},\Phi^{j})$ and \eqref{obs1} for $(\partial_{\xi_{1}}f^{j},\partial_{x}\phi^{j})$.

Noting that $\partial^n_{x} = \ep^{-n/2}\partial^n_{\bar{x}}$, we rewrite the equations \er{exp1_int} and \er{exp2_int} as
\begin{align}\label{formal_exp_inner}
\sum_{j=0}^\infty\ep^{(j-1)/2}\mathcal{R}_{1,j}(F,\Phi) = \sum_{j=0}^\infty\ep^{j/2}\mathcal{R}_{2,j}(F,\Phi,W) = 0,
\end{align}
where
\begin{align*}
&\mathcal{R}_{1,j}(F,\Phi) := \\
&\left\{
\begin{array}{ll}
\displaystyle \xi_1 \partial_{\bar{x}} F^0  + \partial_{\bar{x}} \Phi^0 \partial_{\xi_{1}} F^0 + \partial_{\bar{x}} \Phi^0 \partial_{\xi_{1}} f^0(t,0,\xi), & j=0,
\\[15pt]
\displaystyle \xi_1 \partial_{\bar{x}} F^1 + \partial_{\bar{x}} \Phi^0 \ep^{-1/2} (\partial_{\xi_{1}} f^0 - \partial_{\xi_{1}} f^0(t,0,\xi))  &
\\
\displaystyle  \quad + \sum_{\substack{l+k=1, \\ l,k\geq 0}} \partial_{\bar{x}} \Phi^l \partial_{\xi_{1}} F^k + \partial_{x}\phi^0(t,0)\partial_{\xi_{1}} F^0 + \sum_{\substack{l+k=1, \\ l,k\geq 0}} \partial_{\bar{x}} \Phi^l \partial_{\xi_{1}} f^k(t,0,\xi) +\partial_{t} F^{0}, & j= 1,
\\[30pt]
\displaystyle \xi_1 \partial_{\bar{x}} F^j + \!\!\sum_{\substack{l+k=j-2, \\ l,k\geq 0}}\!\!\ \ep^{-1/2} (\partial_{x} \phi^l - \partial_{x} \phi^l(t,0)) \partial_{\xi_{1}}F^k 
\\
\displaystyle  \quad
+ \sum_{\substack{l+k=j-1, \\ l,k\geq 0}} \partial_{\bar{x}} \Phi^l \ep^{-1/2} (\partial_{\xi_{1}}f^k - \partial_{\xi_{1}}f^k(t,0,\xi))  
+ \sum_{\substack{l+k=j, \\ l,k\geq 0}} \partial_{\bar{x}} \Phi^l \partial_{\xi_{1}} F^k 
\\
\displaystyle  \quad 
+ \sum_{\substack{l+k=j-1, \\ l,k\geq 0}} \partial_{x} \phi^l (t,0) \partial_{\xi_{1}} F^k + \sum_{\substack{l+k=j, \\ l,k\geq 0}} \partial_{\bar{x}} \Phi^l \partial_{\xi_{1}} f^k(t,0,\xi) +\partial_t F^{j-1}, & j\geq 2,
\end{array}
\right.
\\[10pt]
&\mathcal{R}_{2,j}(F,\Phi,W) := -\Phi^j_{\bar{x}\bar{x}} + \int_{\mathbb R^{3}}  F^j d\xi -  W^j,\quad j \geq 0.
\end{align*}
Here, the functions $W^j$ are determined as follows.
As we did in (\ref{expp}), we similarly find that
\begin{subequations}\label{exp45}
\begin{align}
\begin{split}
&\sum_{j=0}^\infty \ve^{j/2}W^j = \exp\bigg(-\sum_{j=0}^\infty\ve^{j/2}(\phi^j+\Phi^j)\bigg) - \exp\bigg(-\sum_{j=0}^\infty\ve^{j/2}\phi^j\bigg)\\
&\mspace{85mu} = e^{-\phi^0-\Phi^0} - e^{-\phi^0} + \sum_{j=1}^\infty  \ve^{j/2}\big(\mathcal{I}^{j}(\Phi) + \mathcal{J}^{j-1}(\Phi)\big),
\end{split}
\end{align}
where for $j \geq 1$,
\begin{align}
\mathcal{I}^{j}(\Phi) &:= -e^{-\phi^0-\Phi^0} (\phi^j+\Phi^j) + e^{-\phi^0} \phi^j, \lb{IandI2}\\
\mathcal{J}^{j-1}(\Phi) &:= \mathcal{J}^{j-1}_{\exp}(\phi+\Phi) - \mathcal{J}^{j-1}_{\exp}(\phi).\lb{IandI21}
\end{align}
\end{subequations}

In the same manner of the derivation of \er{formal_exp_inner}, using the observation (\ref{obs1})  we see from \er{exp45} that
\begin{subequations}\label{ws}
\begin{align}
\begin{split}
&W^0 =  e^{-\phi^0(t,0)-\Phi^0} - e^{-\phi^0(t,0)},\\
&W^1 =  \mathcal{I}^1_0(\Phi) +  \ep^{-1/2}\big( e^{-\phi^0-\Phi^0} - e^{-\phi^0} - W^0\big),\\
&W^j =  \mathcal{I}^j_0(\Phi) + \mathcal{J}^{j-1}(\Phi) + \ep^{-1/2}(\mathcal{I}^{j-1}(\Phi) - \mathcal{I}^{j-1}_{0}(\Phi) ),\ j \geq 2,
\end{split}
\end{align}
where
\begin{align}
&\mathcal{I}^{j}_{0}(\Phi) := -e^{-\phi^0(t,0)-\Phi^0} (\phi^j(t,0)+\Phi^j) + e^{-\phi^0(t,0)} \phi^j(t,0).
\end{align}
\end{subequations}


To obtain the inner equations, at each order in (\ref{formal_exp_inner}), we set 
\begin{align}\label{inner_eq_at_each_ep}
\mathcal{R}_{1,j}(F,\Phi) = \mathcal{R}_{2,j}(F,\Phi,W) = 0.
\end{align}
In (\ref{inner_eq_at_each_ep}), keeping only the $j$th terms $(F^j,\Phi^j)$ called {\it the inner solutions} or  {\it correctors}, we obtain the inner equations as follows. 
At $j=0$, the zeroth order corrector $(F^0,\Phi^0)$ satisfies the following nonlinear PDEs:
\begin{subequations}\label{inner_eqs0}
\begin{gather}
\xi_1 \partial_{\bar{x}} F^0  + \partial_{\bar{x}} \Phi^0 \partial_{\xi_{1}} F^0 + \partial_{\bar{x}} \Phi^0 \partial_{\xi_{1}} f^0(t,0,\xi) = 0,\label{inner_01}\\
-\partial_{\bar{x}\bar{x}}\Phi^0 + \int_{\mathbb R^{3}}  F^0 d\xi -  W^0 = 0.\label{inner_02}
\end{gather}
The boundary conditions for these equations are given by
\begin{gather}
F^{0}(t,0,\xi) = 0,\quad t>0, \ \xi_1>0, 
\label{bc01} \\
\lim_{\bar{x} \to\infty} {F}^{0}(t,\bar{x},\xi)=0,  \quad t>0, \ \xi_1>0, 
\label{bc02} \\
\Phi^{0}(t,0) = \phi_{b}-\phi^0(t,0), \quad t>0,
\label{bc03} \\
\lim_{\bar{x} \to\infty} \Phi^{0} (t,\bar{x}) = 0, \quad t>0.
\label{bc04}
\end{gather}
\end{subequations}

At $j=1$, we have
\begin{subequations}\label{inner_eqsj}
\begin{gather}
\xi_1 \partial_{\bar{x}} F^1  + \partial_{\bar{x}} \Phi^0 \partial_{\xi_{1}} F^1 + \partial_{\bar{x}} \Phi^1 \partial_{\xi_{1}}F^0  + \partial_{\bar{x}} \Phi^1 \partial_{\xi_{1}} f^0(t,0,\xi) = - {\mathcal F}^{0}(F,\Phi), 
\label{inner_eqsj1}\\
-\partial_{\bar{x}\bar{x}} \Phi^1 + \int_{\mathbb R^{3}}  F^1 d\xi -  W^1 = 0, 
\end{gather}
where
\begin{align}
{\mathcal F}^{0}(F,\Phi) & := \partial_{\bar{x}} \Phi^0 \ep^{-1/2} (\partial_{\xi_{1}} f^0 - \partial_{\xi_{1}} f^0(t,0,\xi)) +\partial_{x} \phi^0(t,0)\partial_{\xi_{1}}F^0 
\notag\\
& \quad + \partial_{\bar{x}} \Phi^0 \partial_{\xi_{1}} f^1(t,0,\xi) + \partial_{t} F^{0}.
\end{align}
At $j\geq 2$, the $j$th order inner solution $(F^j,\Phi^j)$
satisfies 
\begin{gather}
\xi_1 \partial_{\bar{x}} F^j + \partial_{\bar{x}} \Phi^0 \partial_{\xi_{1}} F^j + \partial_{\bar{x}} \Phi^j \partial_{\xi_{1}} F^0  + \partial_{\bar{x}} \Phi^j \partial_{\xi_{1}} f^0(t,0,\xi) 
=- {\mathcal F}^{j-1}(F,\Phi),
\label{inner_eqsj4}\\
-\partial_{\bar{x}\bar{x}}\Phi^j + \int_{\mathbb R^{3}}  F^j d\xi -  W^j = 0,
\end{gather}
where
\begin{align*}
\begin{split}
&{\mathcal F}^{j-1}(F,\Phi) := 
\\
& \sum_{\substack{l+k=j-2, \\ l,k\geq 0}}\!\!\ \ep^{-1/2} (\partial_{x} \phi^l - \partial_{x}\phi^l(t,0)) \partial_{\xi_{1}}F^k + \!\!\sum_{\substack{l+k=j-1, \\ l,k\geq 0}} \partial_{\bar{x}} \Phi^l \ep^{-1/2} (\partial_{\xi_{1}} f^k - \partial_{\xi_{1}} f^k(t,0,\xi))
\\
& + \sum_{l=1}^{j-1}\partial_{\bar{x}}\Phi^l\partial_{\xi_{1}}F^{j-l} + \sum_{\substack{l+k=j-1, \\ l,k\geq 0}} \partial_{x}\phi^l(t,0)\partial_{\xi_{1}}F^k +  \sum_{l=0}^{j-1} \partial_{\bar{x}} \Phi^l \partial_{\xi_{1}} f^{j-l}(t,0,\xi) + \partial_{t} F^{j-1}.
\end{split}
\end{align*}
We note that ${\mathcal F}^{j-1}(F,\Phi) $ are all determined by the outer solutions $(f^{l},\phi^{l})$ for $l \leq j$ and the previous inner solutions $(F^{l},\Phi^{l})$ for $l \leq j-1$.
Furthermore, we see that  the inner equations for $j \geq 1$ are all linear.
The boundary conditions for the inner equations for $j \geq 1$ are given by
\begin{gather}
F^{j}(t,0,\xi) = 0,\quad t>0, \ \xi_1>0, 
\label{bcj1} \\
\lim_{\bar{x} \to\infty} {F}^{j}(t,\bar{x},\xi)=0,  \quad t>0, \ \xi_1>0, 
\label{bcj2} \\
\Phi^{j}(t,0) = 0, \quad t>0,
\label{bcj3} \\
\lim_{\bar{x} \to\infty} \Phi^{j} (t,\bar{x}) = 0, \quad t>0.
\label{bcj4}
\end{gather}
\end{subequations}

\subsection{The explicit expressions of finite expansions}\label{2.4}
Let us now define the approximate solutions $(f^{A},\phi^{A})$ of the initial--boundary value problem (\ref{VP1}) by
\begin{equation}\label{app_exp_all1}
(f^{A},\phi^{A})=(f^{A,m},\phi^{A,m})(t, x,\xi)  :=  \sum_{j=0}^m \ve^{j/2} \big\{ (f^j, \phi^j)(t, x,\xi) + (F^j, \Phi^j)(t, \bar{x},\xi)\big\}.
\end{equation}
Under the initial and boundary conditions given for the outer and inner solutions, the approximate solution 
$(\fa, \phia)$ satisfies the same conditions as in \eqref{ini1}--\eqref{bc4}.

\subsection{Homogeneous initial data}\lb{2.5}
In this subsection, we consider homogeneous initial data $f_{0}$ with \eqref{bohm2}--\eqref{supp1} and the following:
\begin{subequations}\label{inicond0}
\begin{gather}
\text{$f_{0}=f_{0}(\xi) \geq 0$ is { homogeneous, i.e., independent of $x$}},
\\
\int_{\mathbb R^{3}} f_{0}(\xi) \,d\xi= \int_{\mathbb R^{3}}  f_{\infty}(\xi) d\xi = 1.
\end{gather}
\end{subequations}
The second condition comes from the quasi-neutral condition \eqref{qn1}.  
For this initial data $f_{0}$, the inner solutions of \eqref{outer_sol_eq0} and \eqref{outer_sol_eqj} can be written as follows:
\begin{gather}\label{simple_sol}
(f^{0},\phi^{0})=(f_{0}(\xi),0), \quad (f^{j},\phi^{j},w^{j})=(0,0,0), \quad j \geq 1.
\end{gather}

Then the expressions for $W^j$ can then be simplified as
\begin{align*}
\begin{split}
&W^0 =  e^{-\Phi^0} - 1,\\
&W^j =  -e^{-\Phi^0} \Phi^j+ \mathcal{J}^{j-1}_{\exp}(\Phi),\ j \geq 1,
\end{split}
\end{align*}
where $\mathcal{J}^{j-1}_{\exp}(\Phi)$ is defined in \eqref{expp}.
From (\ref{simple_sol}), we see that the inner equations in (\ref{inner_eqs0}) are independent  of $t$,
and we have 
\begin{subequations}\label{0th-inner}
\begin{gather}
\xi_1 \partial_{\bar{x}} F^0 + \partial_{\bar{x}} \Phi^0 \partial_{\xi_{1}} F^0 + \partial_{\bar{x}} \Phi^0 \partial_{\xi_{1}} f_{0} (\xi) = 0,\\
-\partial_{\bar{x}\bar{x}}\Phi^0 + \int_{\mathbb R^{3}}  F^0 d\xi -  (e^{-\Phi^0} - 1)= 0
\end{gather}
with the boundary conditions
\begin{gather}
F^0(0,\xi) = 0,\quad \xi_1>0, \\
\Phi^0(0) = \phi_{b}, \\
\lim_{\bar{x} \to\infty} F^0(\bar{x},\xi) = 0, \quad \xi \in \mathbb R^{3}, \\
\lim_{\bar{x} \to\infty} \Phi^0(\bar{x})=0.
\end{gather}
\end{subequations}
For $j\geq 1$, we have
\begin{subequations}\label{jth-inner}
\begin{gather}
\xi_1 \partial_{\bar{x}} F^j + \partial_{\bar{x}} \Phi^0 \partial_{\xi_{1}} F^j + \partial_{\bar{x}} \Phi^j \partial_{\xi_{1}} F^0 + \partial_{\bar{x}} \Phi^j \partial_{\xi_{1}} f_{0}(\xi) 
=- {\mathcal F}^{j-1}(F,\Phi),\\
-\partial_{\bar{x}\bar{x}}\Phi^j + \int_{\mathbb R^{3}}  F^j d\xi + e^{-\Phi^0} \Phi^j = \mathcal{J}^{j-1}_{\exp}(\Phi)
\end{gather}
with the boundary conditions
\begin{gather}
F^j(0,\xi) = 0,\quad \xi_1>0, \\
\Phi^j(0) = 0, \\
\lim_{\bar{x} \to\infty} F^j(\bar{x},\xi) = 0, \quad \xi \in \mathbb R^{3}, \\
\lim_{\bar{x} \to\infty} \Phi^j(\bar{x})=0.
\end{gather}
\end{subequations}
The expressions for ${\mathcal F}^{j-1}(F,\Phi)$, $j \geq 1$, are also simplified as
\begin{align*}
{\mathcal F}^{0}(F,\Phi) &= \partial_{\bar{x}} \Phi^0 \ep^{-1/2} (\partial_{\xi_{1}} f^0 -\partial_{\xi_{1}} f^0 (t,0,\xi)) + \partial_{t}F^{0}=0,\\
{\mathcal F}^{j-1}(F,\Phi) &= \partial_{\bar{x}} \Phi^{j-1} \ep^{-1/2} (\partial_{\xi_{1}} f^0 - \partial_{\xi_{1}}f^0(t,0,\xi)) +\sum_{l=1}^{j-1}\partial_{\bar{x}}\Phi^l \partial_{\xi_{1}} F^{j-l} + \partial_{t} F^{j-1} \notag \\
&=\!\!\sum_{l=1}^{j-1} \partial_{\bar{x}}\Phi^l \partial_{\xi_{1}} F^{j-l},\quad j\geq 2. 
\end{align*}
Then it is easy to see that $(F^{j},\Phi^{j}) =(0,0)$ solve the problem \eqref{jth-inner}.

Eventually, for all $m \geq 0$, the approximate solutions $(f^{A},\phi^{A})$ can be written by
\begin{gather*}\label{app_exp_all1H}
(f^{A},\phi^{A})=(f^{A,m},\phi^{A,m})(t, x,\xi)  = (f_{0}(\xi)+F^{0}(x,\xi),\Phi^{0}(x)).
\end{gather*}
The well-definedness of $(F^{0},\Phi^{0})$ for \eqref{0th-inner} can be shown by introducing the new unknown function ${\tilde{F}^{0}}:=F^{0}+f_{0}(0,\xi)$ and just applying Theorem 2.2 in \cite{MM2}.
Furthermore, it is also seen that $\partial_{x} \Phi^{0}<0$ holds, and $F^0$ is written as
\begin{gather*}
F^{0}(\bar{x},\xi) =  f_{0}(-\sqrt{\xi_{1}^{2}-2\Phi^{0}(\bar{x})},\xi')\chi(\xi_{1}^{2}-2\Phi^{0}(\bar{x}))\chi(-\xi_{1}) -  f_{0}(\xi), 
\end{gather*}
where $\chi(s)$ is the one dimensional indicator function of the set $\{s>0\}$.
Hence, the approximate solution is well-defined.

\section{Estimates of the inner solutions}\label{sec3}


In this section, we show the assertion $(ii)$ in the main result. 
We begin from establishing the estimates for the inner solutions, 
where we treat the general initial data $f_{0}=f_{0}(x,\xi)$.
To this end, we assume that the initial--boundary value problems \er{outer_sol_eq0} and \eqref{outer_sol_eqj} have smooth  local-in-time solutions  $(f^j,\phi^j)$ which satisfy the following:
\begin{subequations}\label{OE0}
\begin{gather}
\sup_{t\in[0,T], \ x\in \overline{\mathbb R_{+}},  \ \xi\in \mathbb R^{3} }
|\partial_t^l\partial_x^n\partial_{\xi}^{\beta}(f^j,\phi^j)(t,x,\xi)| \leq C_j,\quad  j \geq 0,
\lb{OE1}\\
{\rm supp} f^{j} (t,x,\xi) \subset \{(t, x,\xi) \in [0,T] \times \overline{\mathbb R_{+}} \times \mathbb R^{3} \, | \, \xi_{1} \leq -c_{*}, \ |\xi| \leq C_{*}\}, \quad j \geq 0,
\lb{OE2}
\end{gather}
where $C_{j}$ is a positive constant independent of $\ve$, and $c_{*}$ and $C_{*}$ are the same constants in \eqref{supp1}.
We also assume that the boundary value problems \eqref{inner_eqs0} and \eqref{inner_eqsj}
have smooth solutions. 
We remark that all the assumptions are true if we choose the homogeneous initial data $f_{0}=f_{0}(\xi)$ as discussed in subsection \ref{2.5}. For simplicity, we suppose that
\begin{gather}\label{OE3}
\phi_{b} - \phi^{0}(t,0) < \frac{c_{*}^{2}}{2}.
\end{gather}
\end{subequations}

Subsections \ref{3.1} and \ref{3.2} provide the estimates of the zeroth and $j$th  order inner solutions, respectively. 
Those estimates are summarized in Lemmas~\ref{left_corr_est_lem} and \ref{left_corr_est_lem_j} below.
The assertion $(ii)$ in the main result immediately follows from these lemmas and the definition of our approximate solutions 
$(f^{A},\phi^{A})$ in  \eqref{app_exp_all1}.

\subsection{Zeroth order inner solutions}\label{3.1}

Suppose that the inner solution $(F^0, \Phi^0)$ solves the problem \eqref{inner_eqs0}.
Set 
\begin{gather}\label{newf1}
{\tilde{F}^{0}}={\tilde{F}^{0}}(t,\bar{x},\xi):=F^{0}(t,\bar{x},\xi)+f^{0}(t,0,\xi).
\end{gather}
Using \eqref{outer_eq0}, we see that $({\tilde{F}^{0}},\Phi^{0})$ solves
\begin{subequations}\label{inner_sol_eq00}
\begin{gather}
\xi_1 \partial_{\bar{x}} \tilde{F}^{0} + \partial_{\bar{x}} \Phi^0 \partial_{\xi_{1}} \tilde{F}^{0}  = 0,\label{inner_001}\\
-\partial_{\bar{x}\bar{x}}\Phi^0 + \int_{\mathbb R^{3}}  \tilde{F}^{0} d\xi - e^{-\phi^0(t,0)} e^{-\Phi^0}  = 0
\label{inner_002}
\end{gather}
with the boundary conditions
\begin{gather}
{\tilde{F}^{0}}(t,0,\xi) = 0,\quad  t>0, \ \xi_1>0, 
\label{bc001} \\
\lim_{\bar{x} \to\infty} {\tilde{F}^{0}}(t,\bar{x},\xi)=f^{0}(t,0,\xi), \quad t>0,\ \xi\in\mathbb R^{3},
\label{bc002} \\
\Phi^{0}(t,0) = \phi_{b}-\phi^0(t,0), \quad t>0,
\label{bc003} \\
\lim_{\bar{x} \to\infty} \Phi^{0} (t,\bar{x}) = 0, \quad t>0.
\label{bc004}
\end{gather}
\end{subequations}
Owing to the assumption \eqref{p2}, we see that $\phi_{b}-\phi^0(t,0)>0$ for any $t>0$.

First we regard $\Phi^{0}$ as a given decreasing function with respect to $\bar{x}$, and then apply the characteristics method to obtain the following formula of $\tilde{F}^{0}$:
\begin{align}\label{fform00}
\tilde{F}^{0}(t,\bar{x},\xi)
&=f_{0}(t,0,-\sqrt{\xi_{1}^{2}-2\Phi^{0}(t,\bar{x})},\xi')\chi(\xi_{1}^{2}-2\Phi^{0}(t,\bar{x}))\chi(-\xi_{1})
\notag\\
&=f_{0}(t,0,-\sqrt{\xi_{1}^{2}-2\Phi^{0}(t,\bar{x})},\xi')\chi(-\xi_{1}-c_{*}),
\end{align}
where $\chi(s)$ is the one dimensional indicator function of the set $\{s>0\}$, 
and we have used the condition \eqref{OE0} in deriving the last equality.
It is seen from \eqref{OE2} and \eqref{fform00} that the $\xi$-support of $\tilde{F}^{0}(t,x,\xi)$ is bounded uniformly in $(t,x)$. 


Next we reduce the problem \eqref{inner_sol_eq00} to an ODE only for $\Phi^{0}$. 
Integrating \eqref{fform00} over $\mathbb R^{3}$, 
and using \eqref{supp1} and the change of variable $\sqrt{\xi_{1}^{2}-2\Phi^{0}}=-\zeta_{1}$, 
we see that
\begin{gather*}
\int_{\mathbb R^{3}} \tilde{F}^{0}(t,\bar{x},\xi) \,d\xi = \int_{\mathbb R^{3}} f_{0}(t,0,\xi)\frac{-\xi_{1}}{\sqrt{\xi_{1}^{2}+2\Phi^{0}(t,\bar{x})}} d\xi.
\label{rho0'}
\end{gather*}
Substituting this into \er{inner_002} yields the following ODE for $\Phi^{0}_{0}$:
\begin{equation}\lb{phieq1} 
-\partial_{\bar{x}\bar{x}}\Phi^0 + \int_{\mathbb R^{3}} f_{0}(t,0,\xi)\frac{-\xi_{1}}{\sqrt{\xi_{1}^{2}+2\Phi^{0}}} d\xi- e^{-\phi^0(t,0)} e^{-\Phi^0}  = 0.
\end{equation}

By the mean value theorem, we can rewrite \er{phieq1} as
\begin{align}\label{eq_phi00}
-\partial_{\bar{x}\bar{x}}\Phi^0 + S(t,\Phi^0)\Phi^0 = 0,
\end{align}
where \begin{align*}
S(t,\Phi^0) = \int^1_0 \partial_\Phi \Big(\int_{\mathbb R^{3}} f_{0}(t,0,\xi)\frac{-\xi_{1}}{\sqrt{\xi_{1}^{2}+2\Phi}} d\xi \Big) \Big|_{\Phi=\theta\Phi^{0}}+ e^{-\phi^0(t,0)}e^{-\theta\Phi^0} d\theta.
\end{align*}
Let us discuss the property of $S(t,\Phi^0)$.
We see from \er{bohm2} and \eqref{outer_eq0} that
\begin{align}\label{s0}
S(t,0) = -\int_{\mathbb R^{3}} \xi_{1}^{-2}f_{0}(t,0,\xi) d\xi + \int_{\mathbb R^{3}} f_{0}(t,0,\xi) d\xi
\geq c > 0, \quad t \in [0,T].
\end{align}
From the maximum principle,  the solution $\Phi^0$ is found to be monotonic with respect to $\bar{x}$ such that $|\Phi^0(t,\bar{x})| \leq |\phi_{b}-\phi^0(t,0)|$. 
Assuming $|\phi_{b}-\phi^0(t,0)| \ll 1$, we see from \er{s0} and the continuity of $S$ that $S(t,\Phi)  \geq c_0> 0$ for $\Phi \in [0,\phi_{b}-\phi^0(t,0)]$. Now $S(t,\Phi^0(t,\bar{x})) \geq c_0 > 0$ holds for all $\bar{x} \in [0,\infty)$. Thus there is a positive lower bound $c_0>0$, independent of $t$, such that
\begin{align}\label{decay_s1}
0 < c_0 \leq S(t,\Phi^0) \leq C_0,\quad  t \in [0,T].
\end{align}

We are now in a position to derive the estimate of $(F^{0},\Phi^{0})$.

\begin{lem}\label{left_corr_est_lem}
Let $l,n \in\mathbb{N}_{0}$ and $\beta \in (\mathbb{N}_{0})^{3}$.
Assume that \eqref{bohm2}--\eqref{p2}, \er{OE0}, and \er{decay_s1} hold.
Then there exist positive constants  $c$ and  $C$ independent of $\ep$, $ \phi_{b}$, and $x$  such that
\begin{align}
\sup_{t\in [0,T]} \left|\partial^l_t \partial_x^n \Phi^0\left(t,\frac{x}{\sqrt{\ve}}\right)\right| &\leq C\ep^{-n/2}\exp\left(-c\frac{x}{\sqrt{\ve}}\right),
\label{pt_phi0_dir_est1}
\\
\sup_{t\in [0,T], \, \xi \in \mathbb R^{3}}\left|\partial^l_t \partial_x^n \partial_\xi^\beta F^0\left(t,\frac{x}{\sqrt{\ve}},\xi\right)\right| &\leq C\ep^{-n/2}\exp\left(-c\frac{x}{\sqrt{\ve}}\right).
\label{pt_phi0_dir_est2} 
\end{align}
In particular, we have
\begin{align}\label{pt_est_phi_00}
\sup_{t\in [0,T]}\left|\Phi^0\left(t,\frac{x}{\sqrt{\ve}}\right)\right| &\leq |\phi_{b}-\phi^0(t,0)| \exp\left(-c\frac{x}{\sqrt{\ve}}\right),
\\
\label{pt_est_nu00}
\sup_{t\in [0,T], \, \xi \in \mathbb R^{3}}\left|F^0\left(t,\frac{x}{\sqrt{\ve}},\xi\right)\right| &\leq  C |\phi_{b}-\phi^0(t,0)| \exp\left(-c\frac{x}{\sqrt{\ve}}\right).
\end{align}
\end{lem}

\begin{proof}
It is sufficient to show the estimates of $\Phi^{0}$ in \eqref{pt_phi0_dir_est1} and \eqref{pt_est_phi_00}, since  the estimates of $F^{0}$ in \eqref{pt_phi0_dir_est2} and \eqref{pt_est_nu00} immediately follow from \eqref{OE0}, \eqref{newf1}, and \eqref{fform00}.

We prove the estimate of $\Phi^{0}$ by induction on $l$.
For $l=0$, applying Lemma 6.1 in \cite{JKS21} to the boundary value problem (\ref{eq_phi00}), \eqref{bc003}, and \eqref{bc004} with \er{decay_s1}, we immediately have \eqref{pt_est_phi_00} and
\begin{align}\label{pt_est_phi0}
|\partial_{\bar{x}}^n\Phi^0| \leq C\exp\big(-c\bar{x}\big),\  n \geq 0,
\end{align}
where $c$ and $C$ are positive constants independent of $\ep$, $t$, $x$, and $\xi$.

For $l \geq 1$, we now assume that the estimates (\ref{pt_phi0_dir_est1}) hold at the orders $0,1,\ldots,l-1$ with any $n \geq 0$. 
Applying $\partial_t^l$ to (\ref{eq_phi00}), we have
\begin{align}
\begin{split}\label{eq_phi0_t1_l}
&-\partial_{\bar{x}\bar{x}}(\partial^l_t\Phi^0) + \big(S(t,\Phi^0) +   \partial_\Phi S(t,\Phi^0)\Phi^0 \big)\partial^l_t\Phi^0 \\
&= \sum_{\substack{m_1+\cdots+m_l \leq l,\\ 1\leq m_i < l}} ( \text{partial derivatives of $S$ w.r.t. $t,\Phi$ with orders $\leq l$} ) \ \partial^{m_1}_t\Phi^0\cdots \partial^{m_l}_t\Phi^0.
\end{split}
\end{align}
Owing to \er{pt_est_phi0}, we can choose $M>0$, independent of $t$, sufficiently large so that
$S(t,\Phi^0) + S_\Phi(t,\Phi^0)\Phi^0 \geq c_0/2$ for all $\bar{x} \in [M,\infty)$. 
By our assumption for the induction argument, the right-hand side of (\ref{eq_phi0_t1_l}) satisfies the decay condition in Lemma 6.1 in \cite{JKS21}. 
Since $\partial_{\bar{x}}^n\partial^l_t\Phi^0$ is smooth and hence bounded on $[0,M]$, 
by applying Lemma 6.1 in \cite{JKS21} to the equation (\ref{eq_phi0_t1_l}) on $(M,\infty)$, we conclude that $|\partial_{\bar{x}}^n\partial^l_t\Phi^0| \leq C\exp\big(-c(\bar{x}-M)\big)$ for $n \geq 0$ and  $\bar{x}\in(M,\infty)$. 
Thus, the estimate of $\Phi^{0}$ in (\ref{pt_phi0_dir_est1}) holds on $(0,\infty)$. 
The proof is complete.
\end{proof}

\subsection{$j$th order inner solutions}\label{3.2}

Suppose that the inner solution $(F^j, \Phi^j)$ for $j \geq 1$ solves the problem \eqref{inner_eqsj}, i.e., 
\begin{subequations}\label{inner_sol_eqjj}
\begin{gather}
\xi_1 \partial_{\bar{x}} F^j + \partial_{\bar{x}} \Phi^0 \partial_{\xi_{1}} F^j + \partial_{\bar{x}} \Phi^j \partial_{\xi_{1}} \tilde{F}^0 
=- {\mathcal F}^{j-1}(F,\Phi),
\label{inner_jj1}\\
-\partial_{\bar{x}\bar{x}}\Phi^j + \int_{\mathbb R^{3}}  F^j d\xi 
+e^{-\phi^0(t,0)-\Phi^0} \Phi^j = - {\mathcal G}^{j-1}(\Phi), 
\label{inner_jj2}\\
F^{j}(t,0,\xi) = 0,\quad t>0, \ \xi_1>0, 
\label{bcjj1} \\
\lim_{\bar{x} \to\infty} {F}^{j}(t,\bar{x},\xi)=0,  \quad t>0, \ \xi \in \mathbb R^{3}, 
\label{bcjj2} \\
\Phi^{j}(t,0) = 0, \quad t>0,
\label{bcjj3} \\
\lim_{\bar{x} \to\infty} \Phi^{j} (t,\bar{x}) = 0, \quad t>0,
\label{bcjj4}
\end{gather}
where $\tilde{F}^0$ is defined in \eqref{newf1} and
\begin{gather}
{\mathcal G}^{j-1}(\Phi):=\left\{
\begin{array}{ll}
\ep^{-1/2}\big( e^{-\phi^0-\Phi^0} - e^{-\phi^0} - e^{-\phi^0(t,0)-\Phi^0} + e^{-\phi^0(t,0)}\big)
\\
-(e^{-\phi^0(t,0)-\Phi^0}-  e^{-\phi^0(t,0)}) \phi^1(t,0), & j=1,
\\[10pt]
\mathcal{J}^{j-1}(\Phi) + \ep^{-1/2}(\mathcal{I}^{j-1}(\Phi) - \mathcal{I}^{j-1}_{0}(\Phi) )
\\
-(e^{-\phi^0(t,0)-\Phi^0}-  e^{-\phi^0(t,0)}) \phi^j(t,0), & j\geq2.
\end{array}
\right.
\end{gather}
\end{subequations}

First we track the support of $F^{j}$.
At $j=1$, by using \eqref{OE0} and \eqref{fform00},
it is seen that ${\mathcal F}^{0}(F,\Phi)={\mathcal F}^{0}(F,\Phi)\chi(-\xi_{1}-c_{*})$ holds, and also
the $\xi$-support of ${\mathcal F}^{0}(F,\Phi)$ is bounded uniformly in $(t,x)$. 
Then regarding $\Phi^{j}$ as a given function and applying the characteristics method,
we see that ${F}^{1}={F}^{1}\chi(-\xi_{1}-c_{*})$ holds, and also
the $\xi$-support of ${F}^{1}$ is bounded uniformly in $(t,x)$. 
Inductively, one can conclude that the same properties hold for $j \geq 2$.
In particular, it is enough to consider $F^{j}$ and the equation \eqref{inner_jj1} only in $\{\xi_{1} \leq -c_{*}\}$.

Next we find a formula of $F^{j}$.
Using the change of variable $\zeta_{1}=-\sqrt{\xi_{1}^{2}-2\Phi^{0}(t,\bar{x})}$ and $\zeta'=\xi'$ for the equation \eqref{inner_jj1}  with the fact $\tilde{F}^0(t,\bar{x},-\sqrt{\zeta_{1}^{2}+2\Phi^{0}(t,\bar{x})},\zeta')=f^{0}(t,0,\zeta)$, we have
\begin{gather*}
\zeta_{1} \partial_{\bar{x}} \left( F^j (t,\bar{x},-\sqrt{\zeta_{1}^{2}+2\Phi^{0}(t,\bar{x})},\zeta')\right)+ \partial_{\bar{x}} \Phi^j \partial_{\zeta_{1}} f^{0}(t,0,\zeta)
= \frac{\zeta_{1}}{\sqrt{\zeta_{1}^{2}+2\Phi^{0}(t,\bar{x})}} {\mathcal F}^{j-1}(F,\Phi).
\end{gather*}
Integrating this over $(\bar{x},\infty)$ gives the formula
\begin{align}
&\zeta_{1} F^j(t,\bar{x},-\sqrt{\zeta_{1}^{2}+2\Phi^{0}(t,\bar{x})},\zeta')
\notag \\
&= - \Phi^j \partial_{\zeta_{1}} f^{0}(t,0,\zeta)
-\int_{\bar{x}}^{\infty} \frac{\zeta_{1}}{\sqrt{\zeta_{1}^{2}+2\Phi^{0}(t,y)}} {\mathcal F}^{j-1}(F,\Phi) dy.
\label{cv_eq1}
\end{align}

Let reduce the problem \eqref{inner_sol_eqjj} to an ODE only for $\Phi^{j}$.
Using the above change of variable again, we observe that
\begin{align}
\int_{\mathbb R^{3}}  F^j(t,\bar{x},\xi)  d\xi 
&= \int_{\mathbb R^{3}}  F^j(t,\bar{x},\xi) \chi(-\xi_{1}-c_{*}) d\xi 
\notag\\
&= \int_{-\infty}^{0}  \int_{\mathbb R^{2}}  \frac{-\zeta_{1}F^j(t,\bar{x},-\sqrt{\zeta_{1}^{2}+2\Phi^{0}(t,\bar{x})},\zeta')}{\sqrt{\zeta_{1}^{2}+2\Phi^{0}(t,\bar{x})}} d\zeta_{1}d\zeta'.
\label{cv_int1}
\end{align}
Substituting \eqref{cv_eq1} and \eqref{cv_int1} into \eqref{inner_jj2} leads to 
\begin{align}\lb{corr_phi_scalar_1}
-\partial_{\bar{x}\bar{x}} \Phi^j + G(t,\bar{x}) \Phi^j = {\mathcal G}^{j-1}(\Phi)+{\mathcal H}^{j-1}(t,\bar{x}),
\end{align}
where
\begin{align*}
G(t,\bar{x}) &:= e^{-\phi^0(t,0)-\Phi^0} 
+ \int_{-\infty}^{0}  \int_{\mathbb R^{2}}   \partial_{\zeta_{1}}f^{0}(t,0,\zeta) \frac{1}{\sqrt{\zeta_{1}^{2}+2\Phi^{0}(t,\bar{x})}} d\zeta_{1}d\zeta' \notag \\
&= e^{-\phi^0(t,0)-\Phi^0} 
- \int_{\mathbb R^{3}}   f^{0}(t,0,\xi) \frac{-\xi_{1}}{(\xi_{1}^{2}+2\Phi^{0}(t,\bar{x}))^{3/2}} d\xi, 
\\
{\mathcal H}^{j-1}(t,\bar{x}) &:=  
-\int_{-\infty}^{0}  \int_{\mathbb R^{2}} 
\left(\int_{\bar{x}}^{\infty} \frac{\zeta_{1}}{\sqrt{\zeta_{1}^{2}+2\Phi^{0}(t,y)}} {\mathcal F}^{j-1}(F,\Phi) dy\right)
 \frac{1}{\sqrt{\zeta_{1}^{2}+2\Phi^{0}(t,\bar{x})}} d\zeta_{1}d\zeta'.
\end{align*}
We discuss the property of $G(t,\bar{x})$.
If $\Phi^{0}(t,\bar{x})=0$, i.e., $\bar{x}=\infty$, it follows from \er{bohm2} and \eqref{outer_sol_eq0} that
\begin{align}\label{g0}
G(t,\infty) = -\int_{\mathbb R^{3}} \xi^{-2}f^{0}(t,0,\xi) d\xi + \int_{\mathbb R^{3}} f^{0}(t,0,\xi) d\xi
\geq c > 0,\ t \in [0,T].
\end{align}
Assuming $\phi_{b}-\phi^{0}(t,0) \ll 1$, we see from \er{g0} and $0 \leq \Phi^{0}(t,\bar{x}) \leq \Phi^{0}(t,0)=\phi_{b}-\phi^{0}(t,0)$ that $G(t,\bar{x}) \geq c_0 > 0$ for all $\bar{x} \in [0,\infty)$. 
Thus there is a positive lower bound $c_0>0$, independent of $t$, such that
\begin{align}\label{ass_g}
0 < c_0 \leq G(t,\bar{x}) \leq C_0,\  t \in [0,T], \ \bar{x} \in \overline{\mathbb R_{+}}.
\end{align}

We now estimate the inner solution $(F^{j},\Phi^{j})$ for $j\geq 1$.
\begin{lem}\label{left_corr_est_lem_j}
Let $j\geq 1$, $l,n \in\mathbb{N}_{0}$, and $\beta \in (\mathbb{N}_{0})^{3}$. Assume that \eqref{bohm2}--\eqref{p2}, \er{OE0},  and  \er{ass_g}  hold.
Then there exist positive constants $C$  and $c$ independent of $\ep$ and $x$ such that
\begin{align}
\sup_{t\in [0,T]} \left|\partial^l_t\partial_x^n \Phi^{j}\left(t,\frac{x}{\sqrt{\ve}}\right)\right| &\leq C\ep^{-{n}/{2}}\exp\left(-c\frac{x}{\sqrt{\ve}}\right),
\label{pt_dir_est1} \\
\sup_{t\in [0,T], \, \xi \in \mathbb R^{3}} \left|\partial^l_t\partial_x^n\partial^\beta_\xi F^{j}\left(t,\frac{x}{\sqrt{\ve}},\xi\right)\right| &\leq C\ep^{-{n}/{2}}\exp\left(-c\frac{x}{\sqrt{\ve}}\right).
\label{pt_dir_est2} 
\end{align}
\end{lem}

\begin{proof} 
It is sufficient to show the estimates of $\Phi^{j}$ in \eqref{pt_dir_est1}, since  the estimates of $F^{j}$ in \eqref{pt_dir_est2} immediately follows from \eqref{cv_eq1}.

We prove the estimate of $\Phi^{j}$ in (\ref{pt_dir_est1}) by induction on  $j \geq 1$ and $l \geq 0$. 
Recall that $\Phi^j$ satisfies \eqref{corr_phi_scalar_1}, \eqref{bcjj3}, and \eqref{bcjj4}.
First we treat the case $j=1$ and $l=0$.
Owing to \eqref{OE0} and Lemma \ref{left_corr_est_lem},
the right-hand side of \eqref{corr_phi_scalar_1} satisfies the decay condition in Lemma 6.1 in \cite{JKS21}. 
Thus the lemma gives 
\begin{align*}
|\partial_{\bar{x}}^n\Phi^1(t,\bar{x})| \leq C\exp\big(-c\bar{x}\big),\quad  n \geq 0.
\end{align*}

Next, for $j=1$ and $l \geq 1$, we assume that  the estimate of $\Phi^{1}$ in (\ref{pt_dir_est1}) is true for the orders $0,1,\ldots,l-1$ with any $n \geq 0$. 
Applying $\partial_t^l$ to (\ref{corr_phi_scalar_1}), we obtain
\begin{align*}
-\partial_{_{\bar{x}\bar{x}}}(\partial^l_t\Phi^1) + G\partial^l_t \Phi^1 
= \sum_{i=0}^{l-1}\left(
\begin{array}{c}
l\\
i\\
\end{array}
\right)
 \partial^{l-i}_tG \partial^i_t\Phi^1 + \partial^l_t ({\mathcal G}^{j-1}+{\mathcal H}^{j-1}).
\end{align*}
It is seen from the assumption of induction
that the right-hand side satisfies the decay condition in Lemma 6.1 in \cite{JKS21}. 
Thus the lemma gives (\ref{pt_dir_est1}). 

For any $j\ge2$, by the induction argument, we can similarly obtain the estimate of $\Phi^{j}$ in (\ref{pt_dir_est1}) for all   $l,n \geq 0$. The proof is complete.
\end{proof}

\section{Estimates of the errors $({\mathcal R}_{1},{\mathcal R}_{2})$} \label{sec4}


In this section, we show the assertion $(i)$ in the main result.
We recall the definition \eqref{app_exp_all1} of our approximate solution, i.e.,
\begin{equation*}
(f^A,\phi^A) =(f^{A,m},\phi^{A,m})=  \sum_{j=0}^m \ve^{j/2} \big\{ (f^j, \phi^j)(t, x,\xi) + (F^j, \Phi^j)(t, \bar{x},\xi)\big\}.
\end{equation*}

\begin{lem}
Assume \eqref{bohm2}--\eqref{p2}, \eqref{OE0}, \eqref{decay_s1}, and \eqref{ass_g} hold.
Then there exist positive constants $c$ and $C$ independent of $\ve$, $t$, $x$, and $\xi$ such that \eqref{Apro} holds.
\end{lem}
\begin{proof}
We first show the estimate \eqref{Apro1}. 
To do so, we rewrite $\mathcal{R}_1$ defined in \eqref{VP-AP} using $f^A = f^{A,m+1} - \ve^{(m+1)/2}(f^{m+1}+F^{m+1})$ as follows:
\begin{align*}
\mathcal{R}_1&=  \partial_{t} f^{A,m+1} + \xi_1 \partial_{x} f^{A,m+1} +  \partial_{x} \phi^{A,m+1} \partial_{\xi_{1}} f^{A,m+1} + \ve^{(m+1)/2} \mathcal{T}_1, 
\end{align*}
where
\begin{align*}
\begin{split}
\mathcal{T}_1 &= - \partial_t (f^{m+1}+F^{m+1}) - \xi_1 \partial_{x} (f^{m+1}+F^{m+1}) 
- \big(\partial_{x} \phi^{m+1} + \partial_{x} \Phi^{m+1} \big)\partial_{\xi_{1}} f^{A,m+1} \\
&\quad  - \partial_{x} \phi^{A,m+1} \partial_{\xi_{1}} (f^{m+1}+F^{m+1})
+ \ve^{(m+1)/2} \big(\partial_{x} \phi^{m+1} + \partial_{x} \Phi^{m+1} \big) \partial_{\xi_{1}} (f^{m+1}+F^{m+1}).
\end{split}
\end{align*}
From (\ref{OE0}) and Lemmas \ref{left_corr_est_lem} and \ref{left_corr_est_lem_j}, we see that $\ve^{(m+1)/2}\mathcal{T}_1$, along with their derivatives, are bounded by the right-hand side of (\ref{Apro1}). 
The thing to note here is that the lowest power terms with respect to $\ep$ originate from $\mathcal{T}_1$.
Thus we focus ourself on estimating the other terms in $\mathcal{R}_1$.

Now we can rewrite $\mathcal{R}_1$ as 
\begin{align*}
\mathcal{R}_1 = r_1(f,\phi) + \mathcal{R}^*_1(F,\Phi) + \mathcal{R}^*_1 + \ve^{(m+1)/2}\mathcal{T}_1,
\end{align*}
where 
\begin{align*}
\begin{split}
r_1(f,\phi) &:= \sum_{j=0}^{m+1}\ep^{j/2}(\partial_t f^j + \xi_1 \partial_x f^j) + \sum_{j=0}^{m+1}\ep^{j/2}\sum_{\substack{l+k=j,\\ l,k \geq 0}} \partial_x\phi^l \partial_{\xi_1} f^k,\\
\mathcal{R}^*_1(F,\Phi) &:= \sum_{j=0}^{m+1}\ep^{j/2}\Big\{\partial_t F^j + 
\xi_1 \partial_{x} F^j + \!\!\sum_{\substack{l+k=j, \\ l,k\geq 0}}\!\!\ (\partial_{x} \phi^l - \partial_{x} \phi^l(t,0)) \partial_{\xi_{1}}F^k \\
&\quad + \sum_{\substack{l+k=j, \\ l,k\geq 0}} \partial_{x} \Phi^l (\partial_{\xi_{1}}f^k - \partial_{\xi_{1}}f^k(t,0,\xi))  
+ \sum_{\substack{l+k=j, \\ l,k\geq 0}} \partial_{x} \Phi^l \partial_{\xi_{1}} F^k \\
&\quad + \sum_{\substack{l+k=j, \\ l,k\geq 0}} \partial_{x} \phi^l (t,0) \partial_{\xi_{1}} F^k + \sum_{\substack{l+k=j, \\ l,k\geq 0}} \partial_{x} \Phi^l \partial_{\xi_{1}} f^k(t,0,\xi)
\Big\},
\end{split}\\
\begin{split}
\mathcal{R}^*_1 &:= \sum_{j=m+2}^{2(m+1)} \ep^{j/2} \Big\{\sum_{\substack{l+k=j,\\ 1\leq l,k \leq m+1}} \partial_x\phi^l \partial_{\xi_1} f^k \\
& \qquad \qquad \qquad \qquad
+ \sum_{\substack{l+k=j,\\ 1\leq l,k \leq m+1}} (\partial_x\phi^l \partial_{\xi_1}F^k + \partial_x\Phi^l \partial_{\xi_1}f^k + \partial_x\Phi^l \partial_{\xi_1}F^k )\Big\}.
\end{split}
\end{align*}
From (\ref{outer_eq-1}), (\ref{outer_eq1}), and (\ref{outer_eqj}), we find that
\begin{align*}
r_1(f,\phi) &= 0.
\end{align*}
Furthermore, it follows from (\ref{inner_eq_at_each_ep}), i.e., $\mathcal{R}_{1,j}(F,\Phi) = 0$, $j=0,1,\ldots,m$, in (\ref{formal_exp_inner}) that
\begin{align*}
\mathcal{R}^*_1(F,\Phi) &=  \ep^{(m+1)/2}\partial_t F^{m+1} + 
\ep^{m/2}\!\!\sum_{\substack{l+k=m, \\ l,k\geq 0}}\!\!\ (\partial_{x} \phi^l - \partial_{x} \phi^l(t,0)) \partial_{\xi_{1}}F^k \\
&\quad + \ep^{(m+1)/2}\!\!\sum_{\substack{l+k=m+1, \\ l,k\geq 0}}\!\!\ (\partial_{x} \phi^l - \partial_{x} \phi^l(t,0)) \partial_{\xi_{1}}F^k\\
&\quad + \ep^{m/2}\sum_{\substack{l+k=m+1, \\ l,k\geq 0}} \partial_{x} \Phi^l (\partial_{\xi_{1}}f^k - \partial_{\xi_{1}}f^k(t,0,\xi)) \\
& \quad+ \ep^{(m+1)/2}\sum_{\substack{l+k=m+1, \\ l,k\geq 0}} \partial_{x} \phi^l (t,0) \partial_{\xi_{1}} F^k.
\end{align*}
Using (\ref{obs1}), (\ref{OE0}), and Lemmas \ref{left_corr_est_lem} and \ref{left_corr_est_lem_j} to estimate $\mathcal{R}^*_1(F,\Phi)$ and $\mathcal{R}^*_1$,
we arrive at \eqref{Apro1}. Indeed, e.g., we can estimate as follows: 
\begin{align*}
\big| \partial^l_t\partial_x^n\partial^\beta_\xi \{(\partial_{x} \phi^l - \partial_{x} \phi^l(t,0)) F^{j}\}\big| 
& \leq C_1\ep^{(1-n)/{2}}\frac{x}{\sqrt{\ve}}\exp\left(-c_1\frac{x}{\sqrt{\ve}}\right) \\
& \leq C\ep^{(1-n)/{2}}\exp\left(-c\frac{x}{\sqrt{\ve}}\right), \quad \text{where $0 < c < c_1$}.
\end{align*}

Next we show the estimate \eqref{Apro2} on $\mathcal{R}_2$. Similarly as $\mathcal{R}_1$, we observe that
\begin{align*}
\begin{split}
\mathcal{R}_2
&= r_2(f,\phi) + \mathcal{R}^*_{2}(F,\Phi) + \widetilde{\mathcal{R}}_2+\ve^{(m+1)/2} \mathcal{T}_2,
\end{split}
\end{align*} 
where 
\begin{align*}
r_2(f,\phi) &:= \ep \sum_{j=0}^m\ep^{j/2}\partial_{xx} \phi^j
-  \sum_{j=0}^m\ep^{j/2} \int_{\mathbb R^{3}} f^{j} d\xi + \sum_{j=0}^m\ep^{j/2} w^j,\\
\mathcal{R}^*_2(F,\Phi) &:= \sum_{j=0}^m\ep^{j/2} \Big(\ep \partial_{xx}\Phi^j - \int_{\mathbb R^{3}} F^{j} d\xi  +  W^j \Big),\notag\\
\widetilde{\mathcal{R}}_2 &:= e^{- \phi^{A,m}} -  \sum_{j=0}^{m+1}\ep^{j/2} (w^j + W^j),
\\
\mathcal{T}_2 & := w^{m+1} + W^{m+1}.
\end{align*}
It immediately follows from (\ref{OE0}) and Lemmas \ref{left_corr_est_lem} and \ref{left_corr_est_lem_j} 
that $\ve^{(m+1)/2}\mathcal{T}_2$, along with their derivatives, are bounded by the right-hand side of (\ref{Apro2}). 
We focus ourself on estimating the other terms in $\mathcal{R}_2$.

From (\ref{expp}), (\ref{outer_eq0}), (\ref{outer_eq1'}), and (\ref{outer_eqj'}),
we find that
\begin{align*}
r_2(f,\phi) &= \ep^{(m+2)/2} \partial_{xx} \phi^m  + \ep^{(m+1)/2} \partial_{xx}\phi^{m-1},
\end{align*}
and thus $r_2(f,\phi)$ and its derivatives are bounded by the right-hand side of (\ref{Apro2}).
From (\ref{inner_eq_at_each_ep}), i.e., $\mathcal{R}_{2,j}(F,\Phi) = 0$, $j=0,1,\ldots,m$, in (\ref{formal_exp_inner}), we find that
\begin{align*}
\mathcal{R}^*_2(F,\Phi) = 0.
\end{align*}

We also rewrite $\widetilde{\mathcal{R}}_2$ as follows.
From (\ref{expp}), (\ref{exp45}), and  (\ref{ws}), we observe that
\begin{align*}
\sum_{j=0}^{m+1}\ep^{j/2} (w^j + W^j) 
&= e^{-\phi^0-\Phi^0} + \ep^{1/2} \big(- e^{-\phi^0}\phi^1 + \mathcal{J}_{\exp}^0(\phi) \big)\\
&\quad + \sum_{j=2}^{m+1}\ep^{j/2} \big(-e^{-\phi^0}\phi^j + \ep^{-1/2}\mathcal{I}^{j-1}(\Phi)\big) \\
&\quad + \sum_{j=2}^{m+1}\ep^{j/2} \big(\mathcal{J}^{j-1}_{\exp}(\phi)
+ \mathcal{J}^{j-1}(\Phi)\big) + \ep^{(m+1)/2}\mathcal{I}_0^{m+1}(\Phi)\\
&=\mathcal{K} + \ep^{(
m+1)/2}\big(-e^{-\phi^0}\phi^{m+1}+\mathcal{J}^{m}_{\exp}(\phi+\Phi) + \mathcal{I}_0^{m+1}(\Phi)\big),
\end{align*}
where
\begin{align*}
\mathcal{K} := e^{-\phi^0-\Phi^0} + \sum_{j=1}^{m}\ep^{j/2} \big( -e^{-\phi^0-\Phi^0}(\phi^j+\Phi^j)  + \mathcal{J}^{j-1}_{\exp}(\phi+\Phi) \big).
\end{align*}
On the other hand, it follows from Lemma 6.3 in \cite{JKS21}, which gives the estimate of
the truncation error of the formal expansion as in Lemma \ref{Taylor1}, that
\begin{align*}
e^{- \phi^{A,m}} = \exp\bigg( -\sum_{j=0}^m \ep^{j/2} (\phi^j+\Phi^j) \bigg) = \mathcal{K} + \ep^{(m+1)/2}\mathcal{L}_{m}(\phi+\Phi),
\end{align*}
where $\mathcal{L}_{m}(\phi+\Phi)$ can be estimated by using (\ref{OE0}) and Lemmas \ref{left_corr_est_lem} and \ref{left_corr_est_lem_j} as 
\begin{align*}
|\partial_{t}^l \partial_{x}^n \mathcal{L}_{m}(\phi+\Phi) |
\leq C  +  C \ep^{-n/2} e^{-cx/\sqrt{\ep}}.
\end{align*}
Combining the above results, we deduce that 
\begin{align*}
\widetilde{\mathcal{R}}_2 &=\ep^{(m+1)/2}\mathcal{L}_{m}(\phi+\Phi)-\ep^{(m+1)/2}\big(-e^{-\phi^0}\phi^{m+1}+\mathcal{J}^{m}_{\exp}(\phi+\Phi) + \mathcal{I}_0^{m+1}(\Phi)\big).
\end{align*}
Finally, we find that $|\partial_{t}^l\partial_{r}^n\widetilde{\mathcal{R}}_2|$ is bounded by the right-hand side of \er{Apro2} 
and hence \eqref{Apro2} holds. The proof is complete.
\end{proof}

\section{Numerical experiments}\label{sec5}



In this section, we present numerical simulations for { the Vlasov--Poisson system in (\ref{VP1}), their limit equations in (\ref{zero-out}), and the associated inner (boundary layer) equations in (\ref{inner_eqs0}).} To facilitate our computations, we consider the case of one dimensional velocity $\xi \in \mathbb R$.
From the discussion on the support of $F^{0}$ in Section \ref{sec3}, 
we can suppose that the supports of $f$, $f^{0}$, and $F^{0}$ are on $\{(x,\xi) \in [0,1]\times[-4,0]\}$. The computational domain we consider is defined as $\{(x,\xi) \in [0,1]\times[-4,0]\}$.

The equations take the following forms:
\begin{itemize}
\begin{subequations}\label{vp}
\item The Vlasov--Poisson system
\begin{gather}
f_t + \xi f_x+ \phi_x f_{\xi} =0, \quad t>0, \ x\in(0,1), \ \xi \in (-4,0),\label{vp_time}\\
\ve \partial_{xx} \phi  = \int_{\mathbb R} f d\xi - e^{-\phi}, \quad t>0, \ x\in(0,1)
   \end{gather}
with the initial and boundary conditions
\begin{gather}
f (0,x,\xi)= f_{0}(x,\xi) \geq 0, \quad x\in(0,1), \ \xi \in (-4,0), \\
\phi(t,0)=\phi_{b}>0, \\
f (t,1,\xi) =  f_{\infty}(\xi) \geq 0, \quad t>0, \ \xi \in (-4,0), \label{vp_bc}\\
\phi (t,1) =  0, \quad t>0.\label{vp_time2}
\end{gather}
\end{subequations}
\item The limit equations
\begin{subequations}\label{vp2}
\begin{gather}
f^0_t + \xi f^0_x + \phi^0_x f^0_{\xi} = 0, \quad t>0, \ x\in(0,1), \ \xi \in (-4,0),\label{vp_lim_time}\\
\int_{\mathbb R}  f^0 d\xi - e^{-\phi^0} = 0, \quad t>0, \ x\in(0,1)
\end{gather}
with the initial and boundary conditions
\begin{gather}
f^0 (0,x,\xi)= f_{0}(x,\xi) \geq 0, \quad x\in(0,1), \ \xi \in (-4,0), \\
f^0 (t,1,\xi) =  f_{\infty}(\xi) \geq 0, \quad t>0, \ \xi \in (-4,0).\label{vp_lim_time2}
\end{gather}
\end{subequations}

\item The inner (boundary layer) equations
\begin{subequations}\label{vp3}
\begin{gather}
\xi F^0_{x}  + \Phi^0_{x}F^0_{\xi} + \Phi^0_{x} f^0_{\xi}(t,0,\xi) = 0, \quad t>0, \ x\in(0,1), \ \xi \in (-4,0), \label{bl_two_pt}\\
-\ep\Phi^0_{xx} + \int_{\mathbb R}  F^0 d\xi -  e^{-\phi^0(t,0)-\Phi^0} + e^{-\phi^0(t,0)} = 0, \quad t>0, \ x\in(0,1)
\end{gather}
with the boundary conditions
\begin{gather}
{F}^{0}(t,1,\xi)=0,  \quad t>0, \ \xi \in (-4,0), \\
\Phi^{0}(t,0) = \phi_{b} -\phi^0(t,0), \quad t>0,  \\
\Phi^{0} (t,1) = 0, \quad t>0. \label{bl_two_pt2}
\end{gather}
\end{subequations}

\end{itemize}


%
%

In the numerical experiments presented below, we work with the computational domain $(x,\xi) \in [0,1]\times[-4,0]$. 
 To impose the initial and boundary conditions, we use the function $\sigma(\xi)$ defined by 
\begin{align*}
\sigma(\xi)  := \frac{1}{d}\exp\left(-\frac{(\xi-c)^2}{10^{-0.8}}\right),
\end{align*}
where the value of the parameter $d$ is chosen such that the condition (\ref{qn1}) holds, i.e.,
\begin{align*}
\int_{-4}^0 \sigma(\xi) d\xi = 1.
\end{align*}

For the limit solutions, it is important to note that
\begin{align*}
\phi^0(t,x) = -\log\left(\int_{-4}^0  f^0(t,x,\xi) d\xi\right).
\end{align*}
Hence, it is reasonable to assume that
\begin{align}\label{ini_constr}
\inf_{x\in (0,1)}\int_{-4}^0  f^0(0,x,\xi) d\xi > 0.
\end{align}

\begin{figure}[h]
  \centering
  \begin{tabular}{m{2mm}m{34mm}m{34mm}m{34mm}m{34mm}}
& \hspace{10mm} $\ep=10^{-1.5}$ & \hspace{10mm} $\ep=10^{-2.5}$ & \hspace{10mm} $\ep=10^{-3}$ & \hspace{10mm} $\ep=10^{-3.5}$\\ 
   $f$ & \resizebox{38mm}{!}{\includegraphics{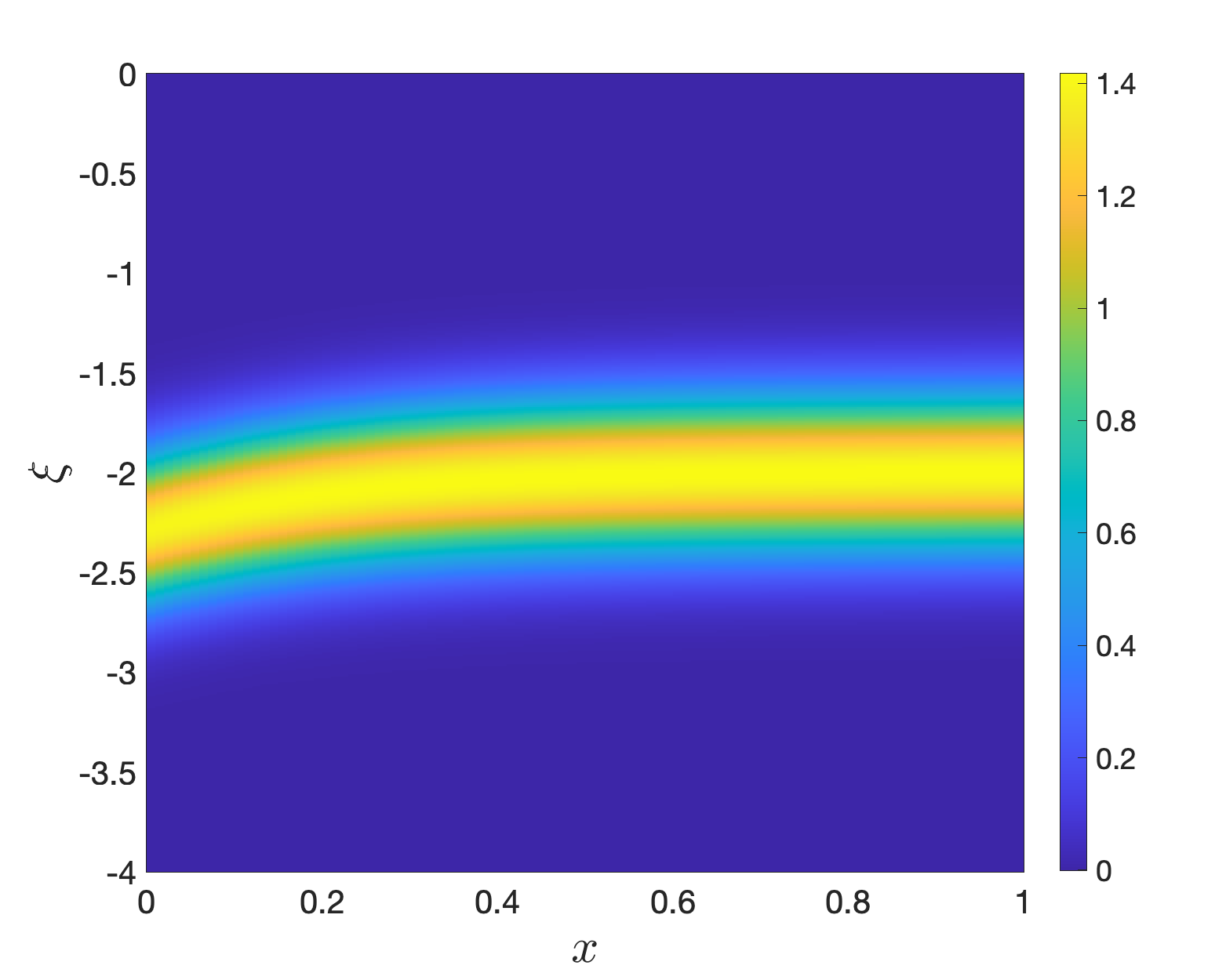}} & \resizebox{38mm}{!}{\includegraphics{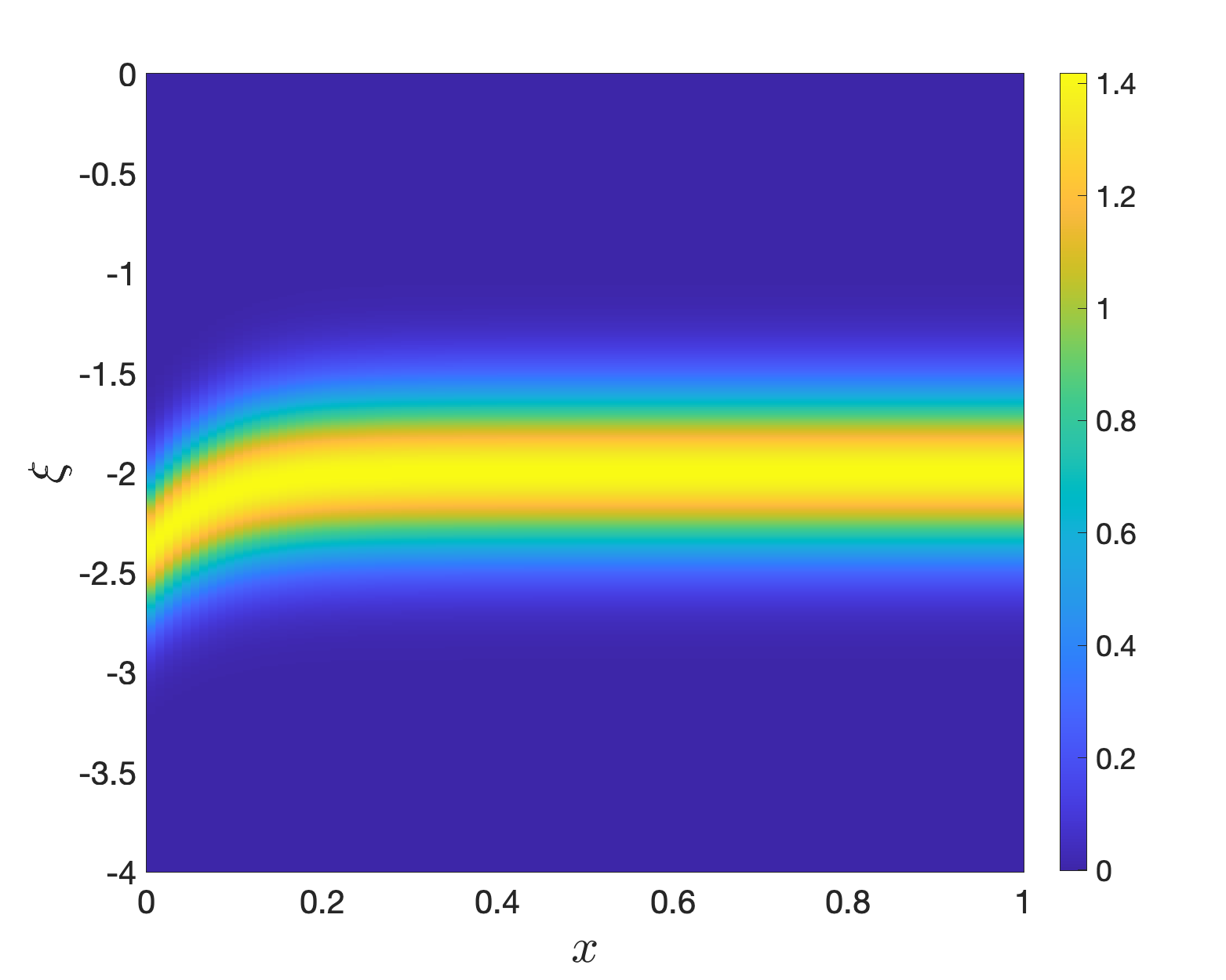}} & \resizebox{38mm}{!}{\includegraphics{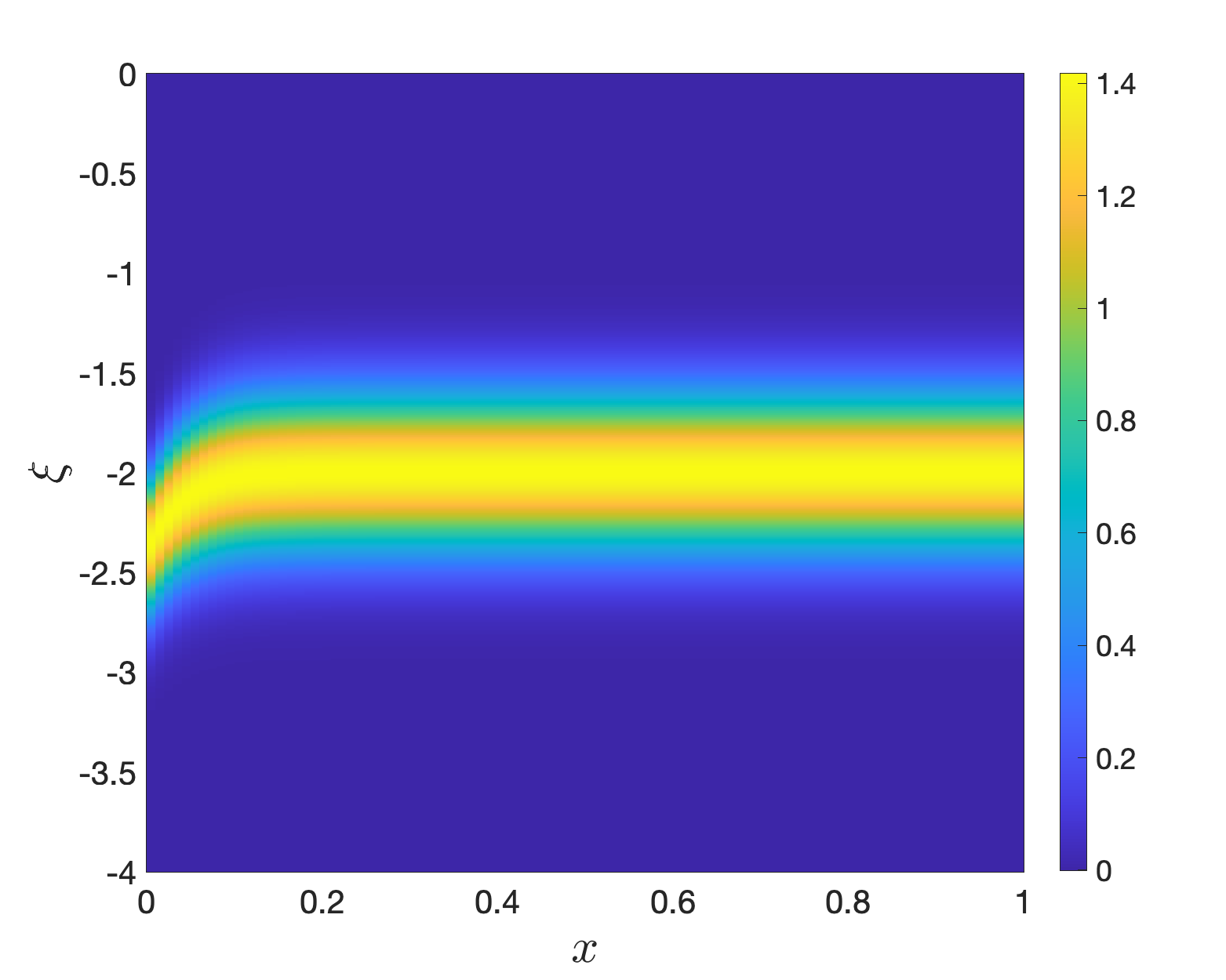}} & \resizebox{38mm}{!}{\includegraphics{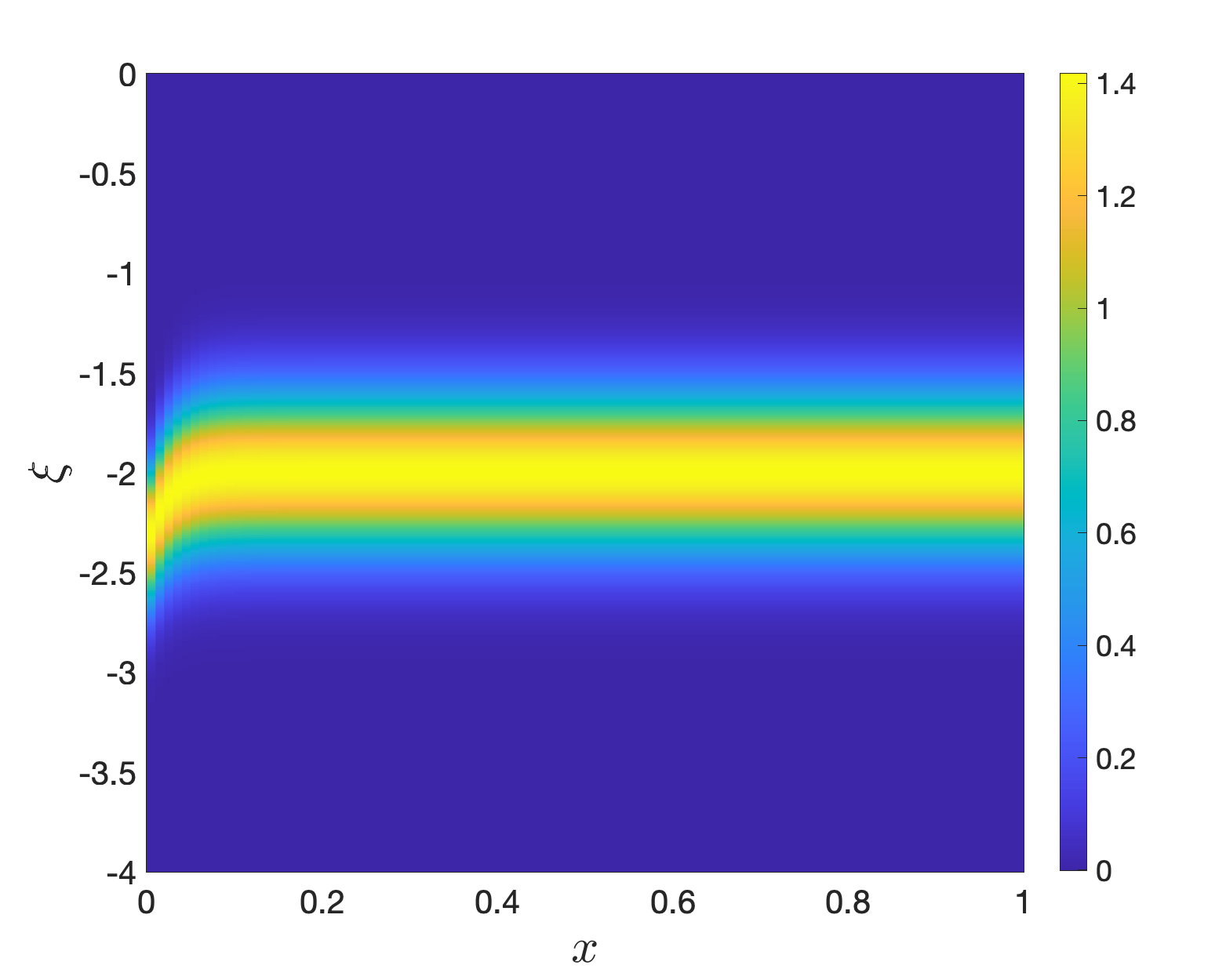}} \\
$F^0$ &     \resizebox{38mm}{!}{\includegraphics{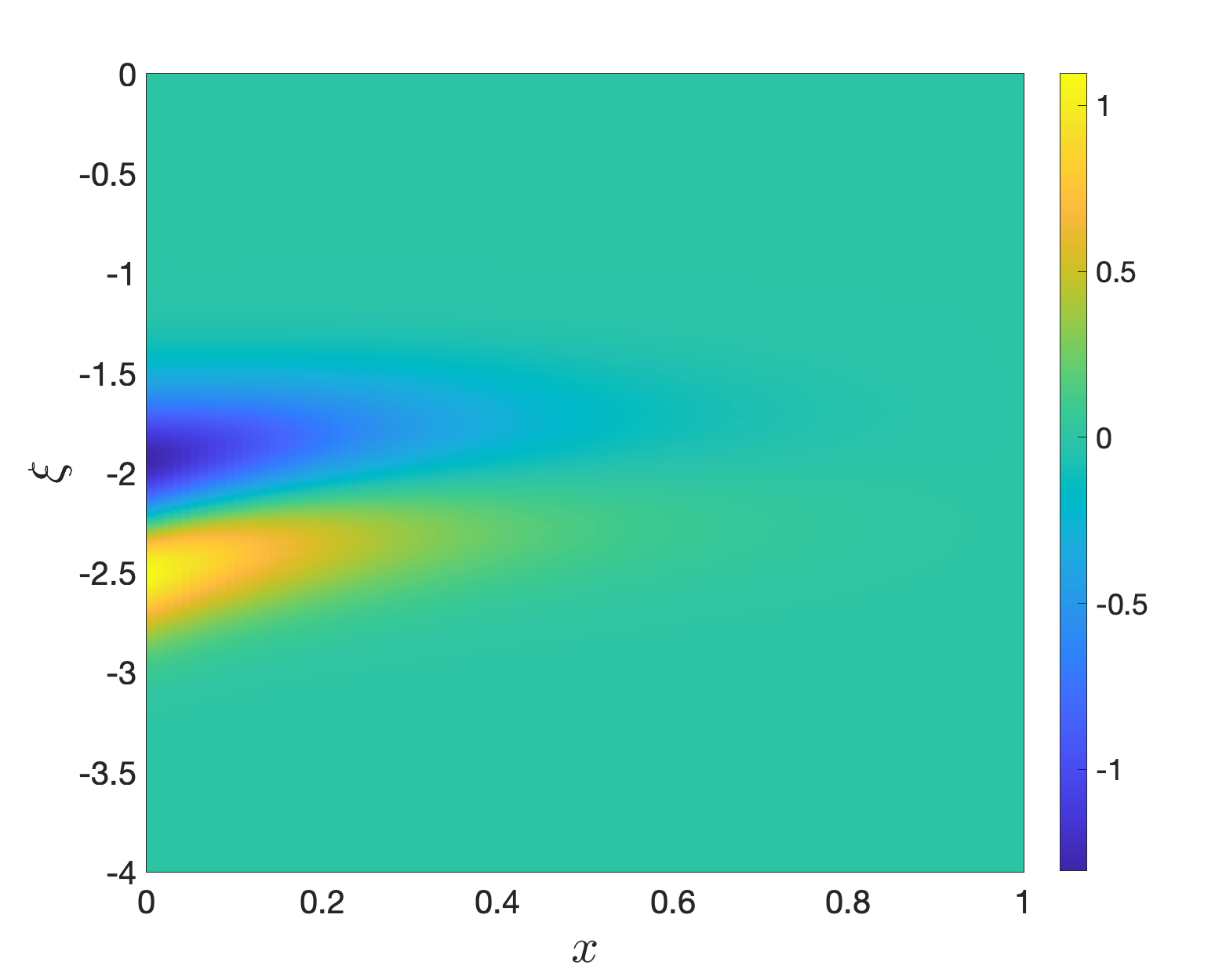}} & \resizebox{38mm}{!}{\includegraphics{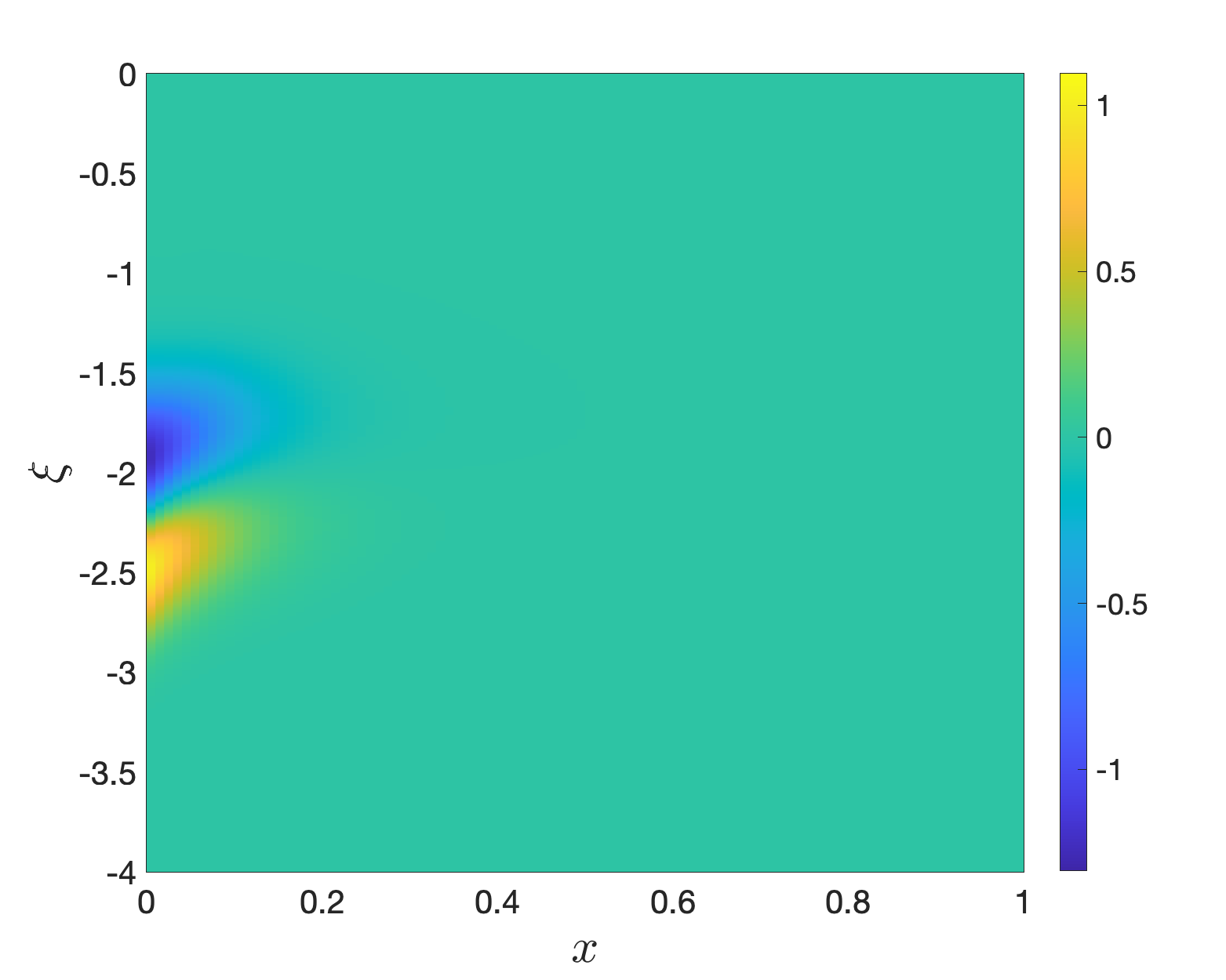}} & \resizebox{38mm}{!}{\includegraphics{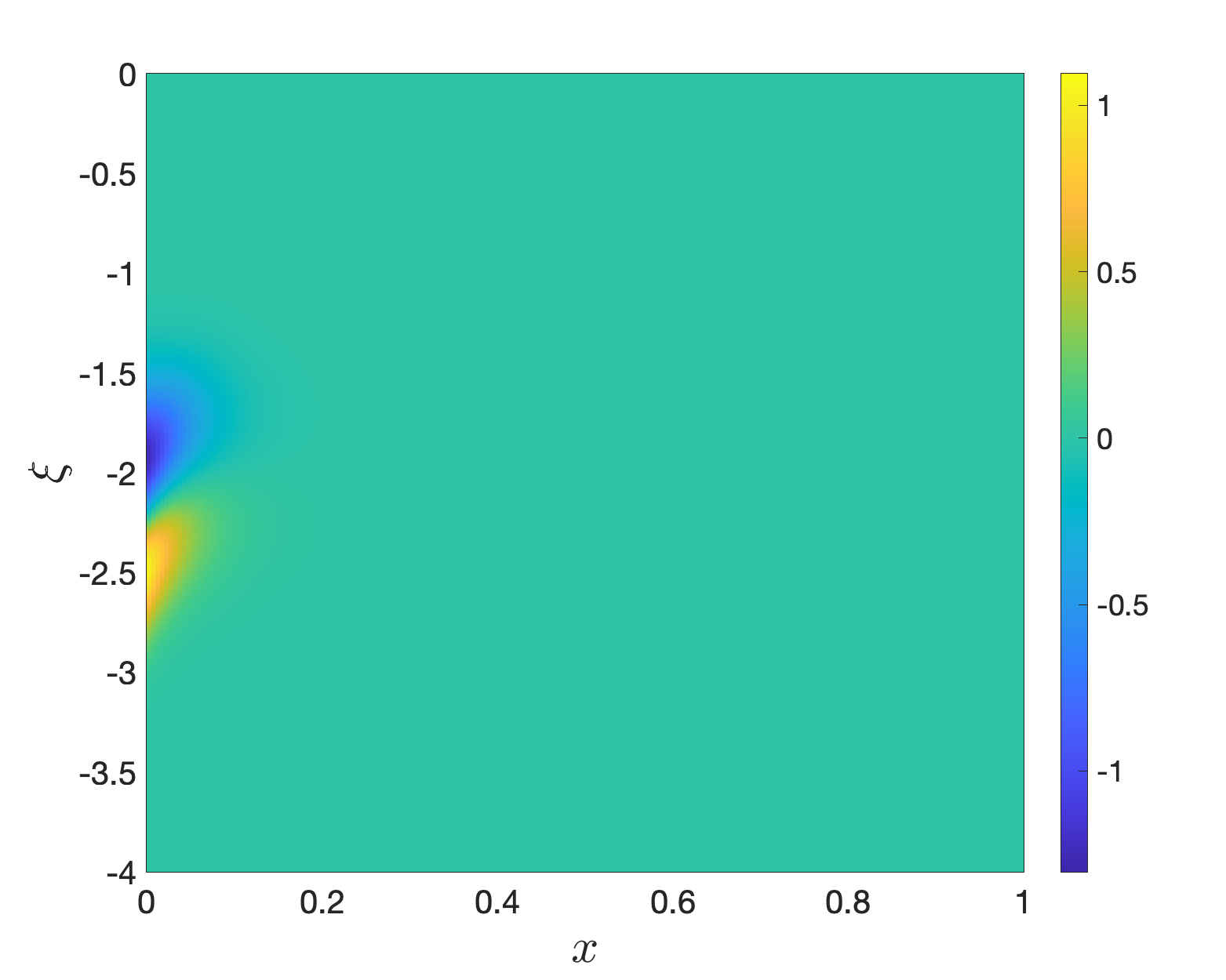}} & \resizebox{38mm}{!}{\includegraphics{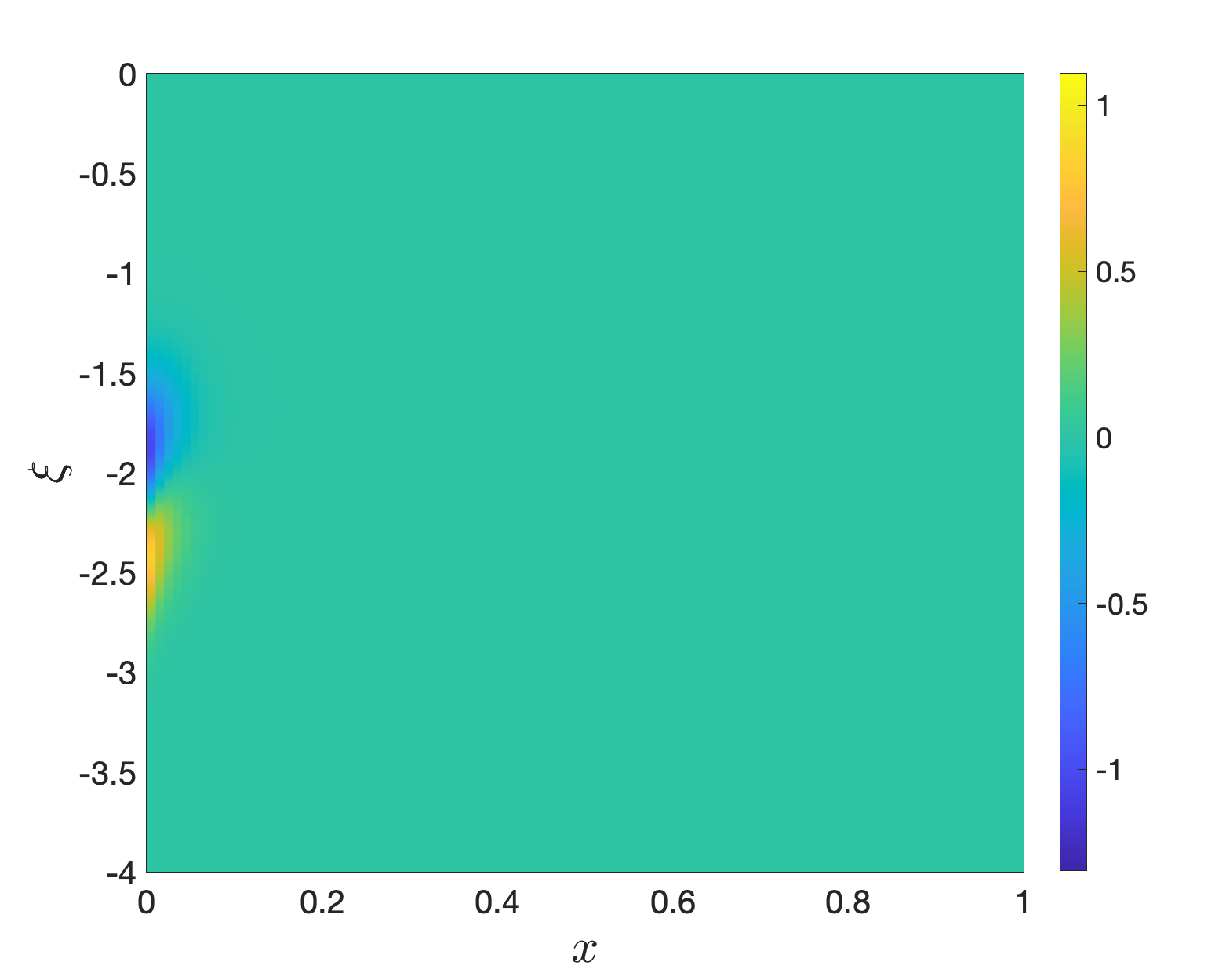}} \\
 $\substack{\phi\\ \\ \Phi^0 }$ & \resizebox{38mm}{!}{\includegraphics{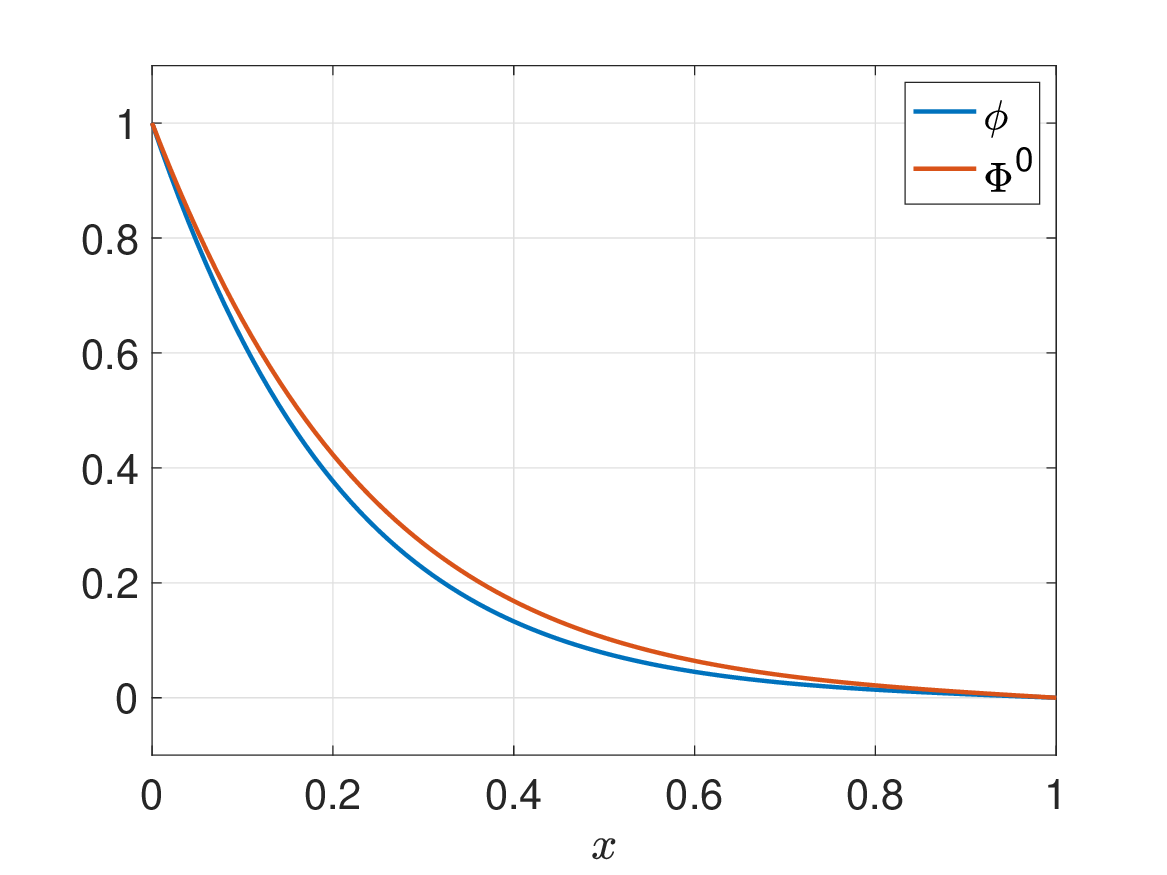}} & \resizebox{38mm}{!}{\includegraphics{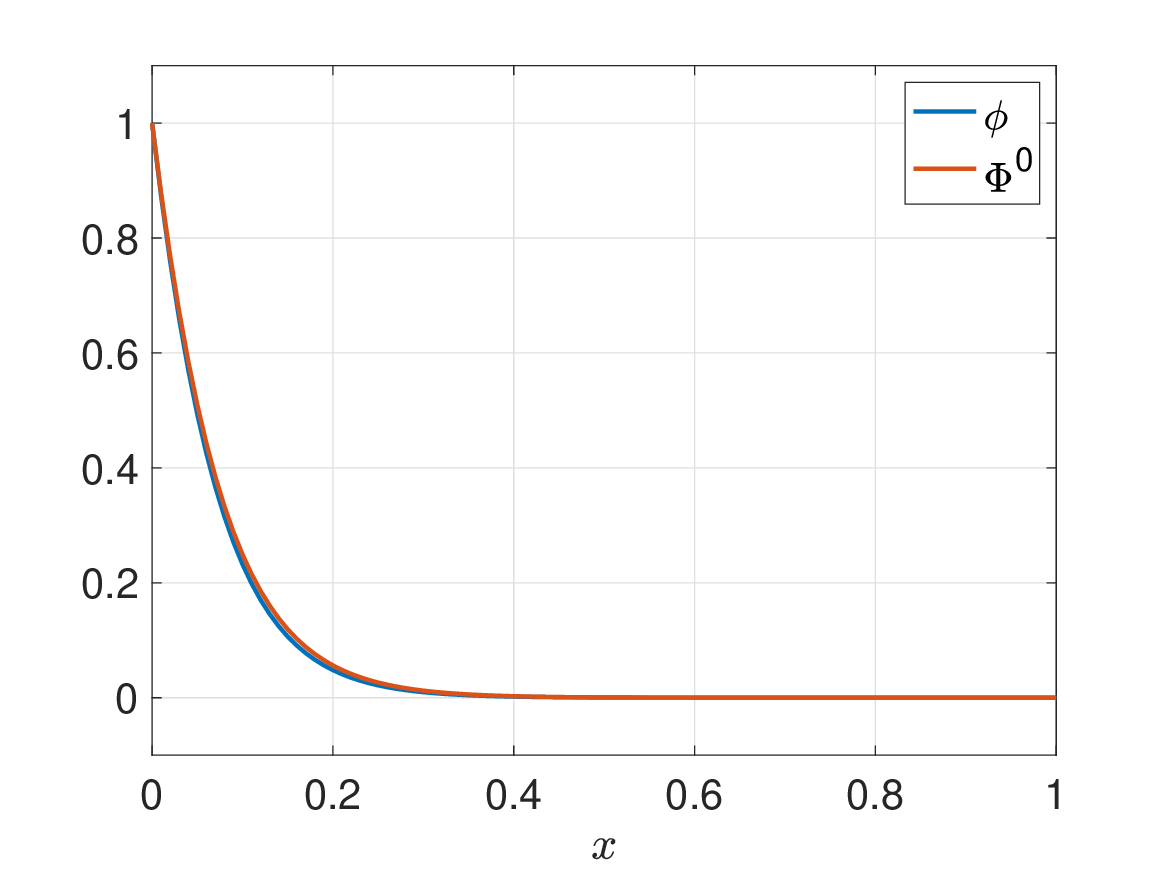}} & \resizebox{38mm}{!}{\includegraphics{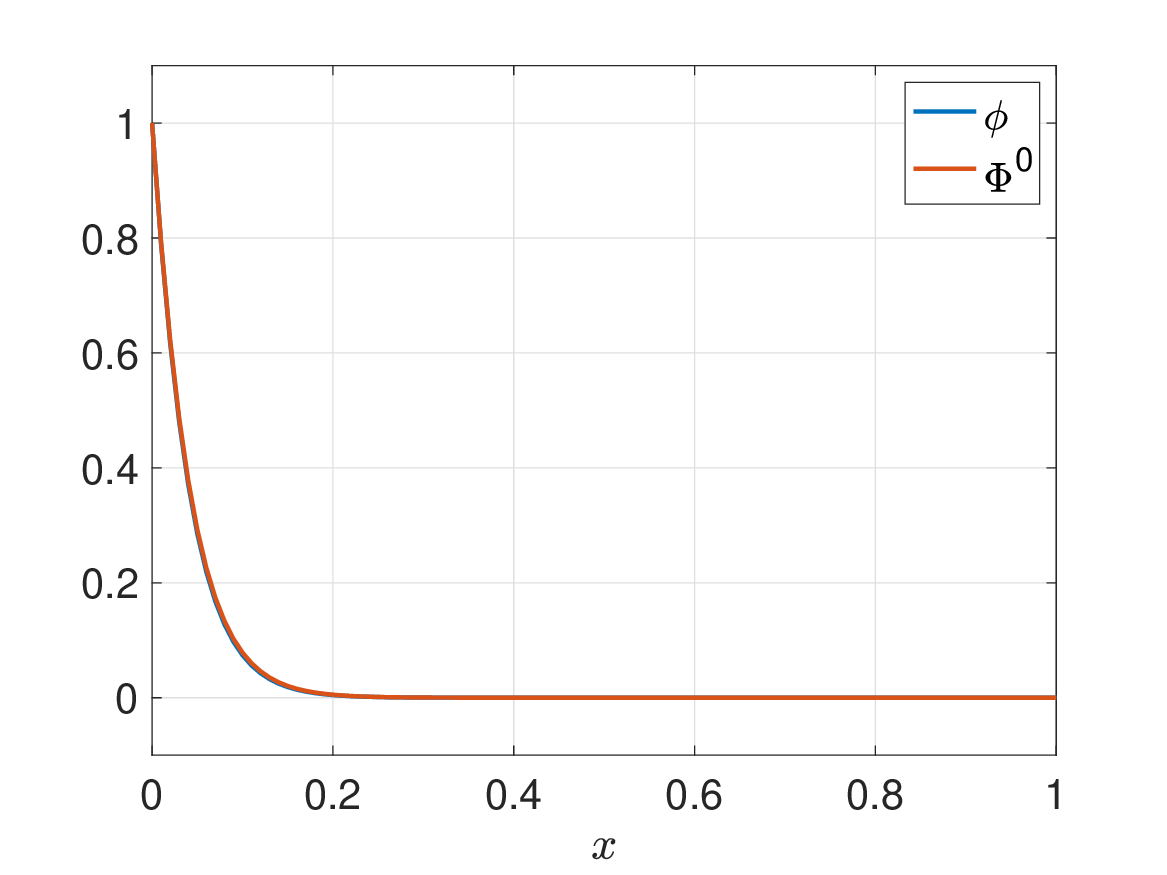}} & \resizebox{38mm}{!}{\includegraphics{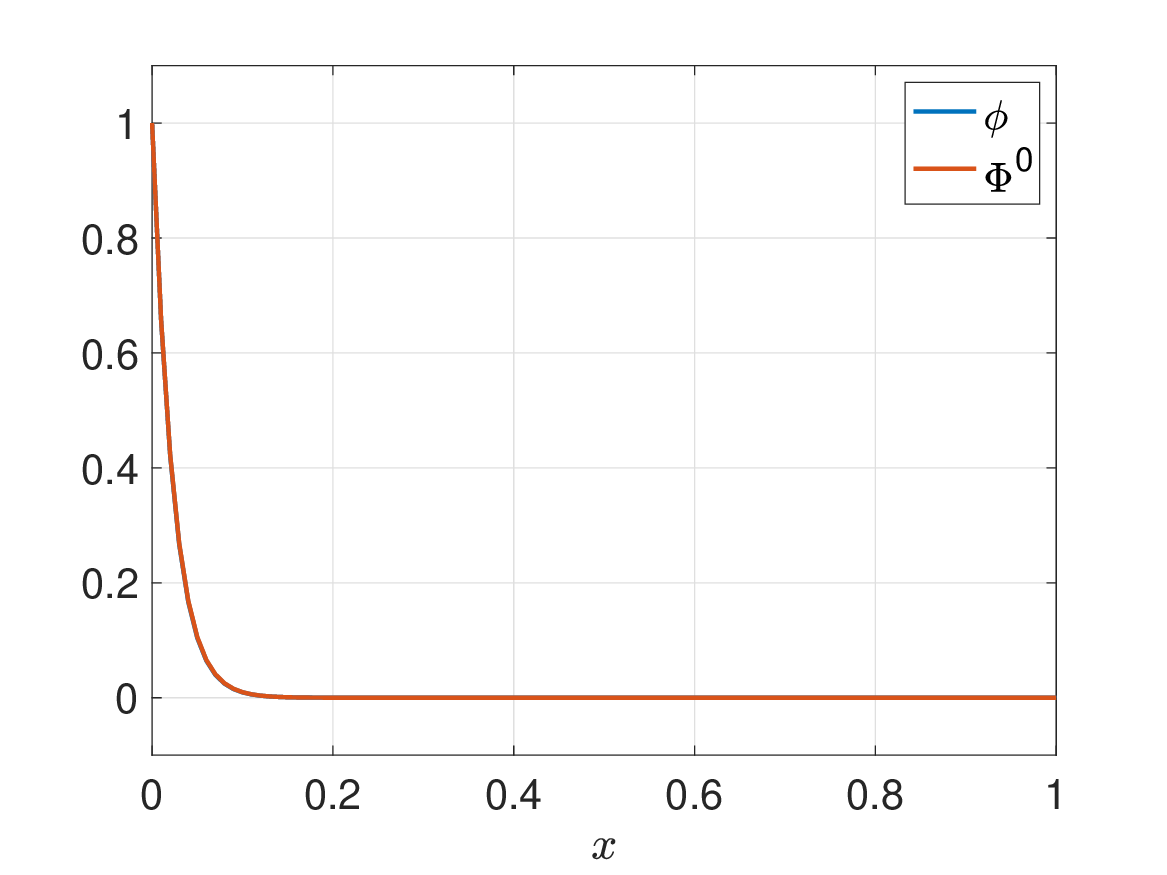}}
  \end{tabular}  
  \caption{\small{The plots of $f$, $F^0$, $\phi$, and $\Phi^0$ at $t=0.1$ for various $\ep$'s.}}\label{ex2_vari_eps_f}
\vspace{5mm}
  \begin{center}
  \begin{tabular}{m{15mm}m{65mm}m{65mm}}
& \hspace{30mm} (a) &  \hspace{30mm}  (b) \\
   &    \resizebox{60mm}{!}{\includegraphics{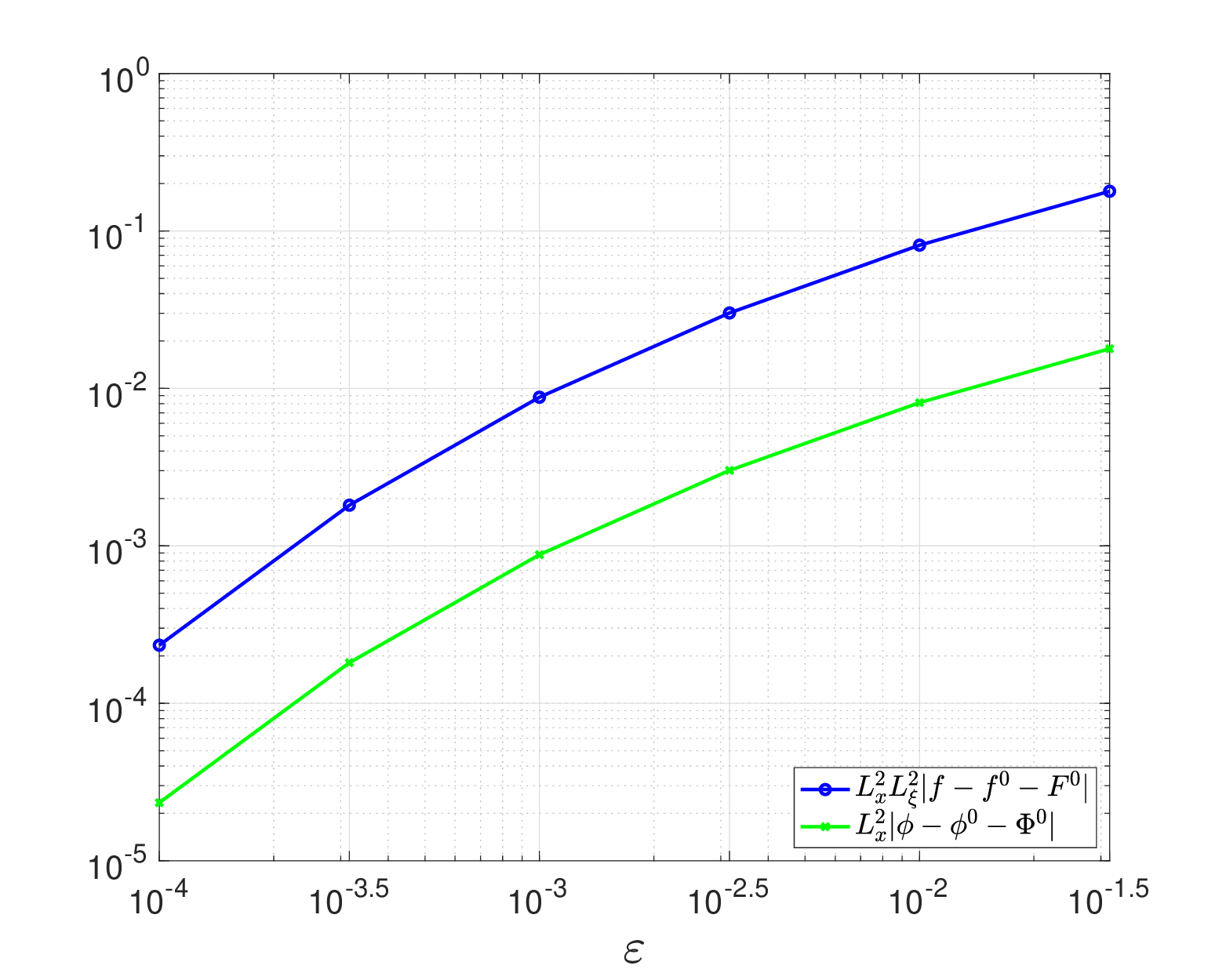}}  \hspace{-9mm} & 
       \resizebox{60mm}{!}{\includegraphics{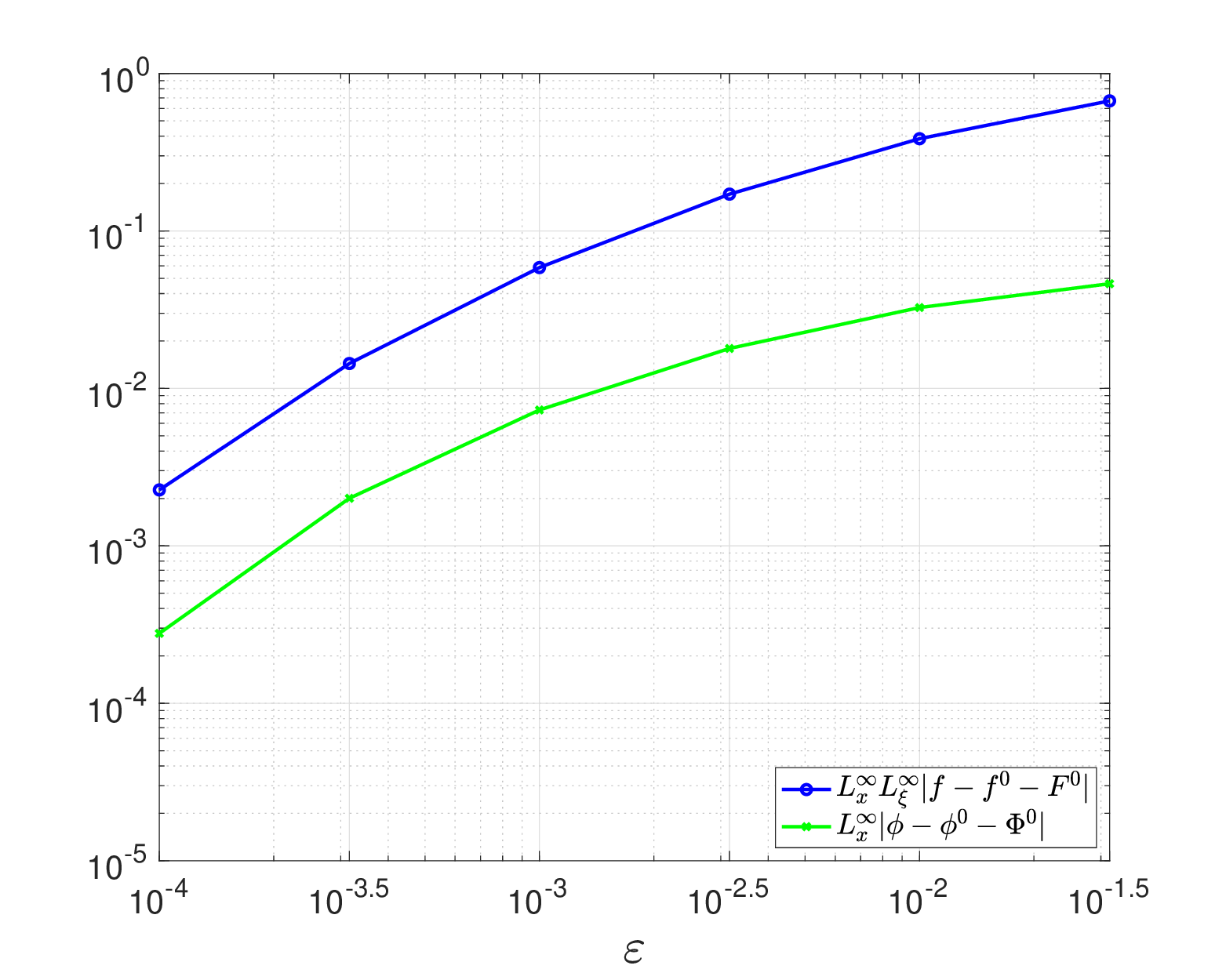}}\\
    \end{tabular}
    \caption{\small{Log-log plots of the errors $f-f^0-F^0$ and $\phi-\phi^0-\Phi^0$ at $t=0.1$, $(x,\xi) \in [0,1]\times[-4,0]$ in (a) $L^2((0,1)\times(-4,0))$, $L^2((0,1))$, and in (b) $L^\infty((0,1)\times(-4,0))$, $L^\infty((0,1))$.}}\label{ex2_vari_eps_err}
  \end{center}
\end{figure}

Now we will numerically assess the convergence of the asymptotic expansion $(f^0+F^0, \phi^0+\Phi^0)$ towards the solutions $(f,\phi) = (f^\ep,\phi^\ep)$ of the Vlasov--Poisson system in (\ref{vp}) as $\ep \rightarrow 0$. 
We adopt the initial condition $f_0(x,\xi) = \sigma(\xi)$ and boundary conditions $f_\infty(\xi) = \sigma(\xi)$, $\phi_b=1$, as described in (\ref{vp}). This leads to the following limit solution as discussed in subsection \ref{2.5}:
\begin{align*}
\phi^0 = 0, \quad f^0 = \sigma(\xi).
\end{align*}
Subsequently, we have the boundary layer equations:
\begin{subequations}
\begin{align*}
\xi F^0_{x}  + \Phi^0_{x}F^0_{\xi} + \Phi^0_{x} \sigma'(\xi) = 0,\\
-\ep\Phi^0_{xx} + \int_{\mathbb R}  F^0 d\xi -  e^{-\Phi^0} + 1 = 0
\end{align*}
with the boundary conditions
\begin{align*}
{F}^{0}(t,1,\xi)=0,\quad \Phi^{0}(t,0) = 1,\quad
\Phi^{0} (t,1) = 0. 
\end{align*}
\end{subequations}

To visualize the numerical solutions for various  values of $\ep$ at $t=0.1$, we present the plots for $f$, $F^0$, $\phi$, and $\Phi^0$, respectively in Figure \ref{ex2_vari_eps_f}.
Additionally, the error plots for $(f-f^0-F^0, \phi-\phi^0-\Phi^0)$ are given in Figure \ref{ex2_vari_eps_err}.



 
We also present  our numerical simulations with the following initial and boundary conditions in Cases (I) and (II):

\textbf{Case (I)}
\begin{itemize}
\item  Center of $\sigma(\xi)$: $c = -2$; 
\item  Initial condition: $f(0,x,\xi) = \sigma(\xi+1-x)$;
\item  Boundary conditions: $f(t,1,\xi) = \sigma(\xi)$, $\phi(t,0) = 1$, $\phi(t,1) = 0$;
\item  Initial conditions for limit solution: $f^0(0,x,\xi) = \sigma(\xi+1-x)$;
\item  Boundary conditions for limit solution: $f^0(t,1,\xi) = \sigma(\xi)$, $\phi^0(t,1) = 0$;
\item  Boundary conditions for boundary layers: $F^0(t,1,\xi) = 0$, $\Phi^0(t,0) = 1-\phi^0(t,0)$, $\Phi^0(t,1) = 0$.
\end{itemize}

\textbf{Case (II)}
\begin{itemize}
\item Center of $\sigma(\xi)$:  $c = -3$;
\item Initial condition: $f(0,x,\xi) = \sigma(\xi-1+x)$;
\item Boundary conditions: $f(t,1,\xi) = \sigma(\xi)$, $\phi(t,0) = 1$, $\phi(t,1) = 0$;
\item Initial conditions for limit solution: $f^0(0,x,\xi) = \sigma(\xi-1+x)$;
\item Boundary conditions for limit solution: $f^0(t,1,\xi) = \sigma(\xi)$, $\phi^0(t,1) = 0$;
\item Boundary conditions for boundary layers: $F^0(t,1,\xi) = 0$, $\Phi^0(t,0) = 1-\phi^0(t,0)$, $\Phi^0(t,1) = 0$.
\end{itemize}
The initial potentials $\phi(0,x)$ and $\phi^0(0,x)$ can be determined by using the equations
\begin{align*}
\ve \partial_{xx} \phi(0,x) = \int_{\mathbb R} f_0(x,\xi) d\xi - e^{-\phi(0,x)},\\
\phi^0(0,x) = -\log\left(\int_{-4}^0 f_0(x,\xi) d\xi\right).
\end{align*}
For the boundary layers $F^0(0,x,\xi)$ and $\Phi^0(0,x)$ at $t=0$, we need to solve the following equations:
\begin{subequations}
\begin{align*}
\xi F^0_{x}  + \Phi^0_{x}F^0_{\xi} + \Phi^0_{x} f^0_{\xi}(0,0,\xi) = 0,\\
-\varepsilon\Phi^0_{xx} + \int_{\mathbb R} F^0 d\xi -  e^{-\phi^0(0,0)-\Phi^0} + e^{-\phi^0(0,0)} = 0.
\end{align*}
\end{subequations}
The numerical results for  Case (I)  are given in Figures \ref{ex3_f}--\ref{ex3_phi}, while those for Case (II) can be found in Figures \ref{ex4_f}--\ref{ex4_phi}.

\subsection{Some discussion} 
In Figure \ref{ex3_ex4}, we provide numerical results for the zeroth moment $\int_{-4}^0 (f,f^0,F^0) d\xi$ and the first moment $\int_{-4}^0 \xi (f,f^0,F^0) d\xi$ at $t=0.1$ with $\ve=10^{-2}$. These represnet the density and momentum of the fluid at the macroscopic level. 
As we can see,  sharp transition layers of the density appear near the boundary. 
In the third column of Figure \ref{ex3_ex4}, it is worth noting that the momentum is trivial, i.e., $\int_{-4}^0\xi F^0 d\xi = 0$. This holds theoretically.  Indeed, by integrating the boundary layer equation 
\er{bl_two_pt} 
with respect to $\xi$, we see that
\begin{align*}
\left(\int_{-4}^0 \xi F^0 d\xi \right)_{x}  &= - \Phi^0_{x}(t,x)\Big(F^0(t,0) + f^0(t,0,0) - F^0(t,-4) - f^0(t,0,-4)\Big)
\\
&=0,
\end{align*}
where we have also used the fact that the $\xi$-supports of $F^0$ and $f^0$ are on $(-4,0)$ in deriving the last equality. 
Thus, the boundary condition $\xi F^0(t,1,\xi) = 0$ ensures that $\int_{-4}^0 \xi F^0 d\xi=0$ for all $x\in[0,1]$.

Furthermore, we present simulations for the cases that do not satisfy (\ref{ini_constr}), i.e, the case including the state of vacuum. These results are given in Figure \ref{ex10}, where $(\tilde{f}, \tilde{\phi})$ denotes a stationary solution. 
It is an interesting question to find a good approximate solution that takes account into the vacuum state.

\subsection{Numerical scheme} 
We shall briefly discuss the discretization  employed in our simulations. 
To address the entire set of equations in (\ref{vp})--(\ref{vp3}) and the associated stationary equations, we initiate the process by tackling the two-point boundary value problems.

For the interval $-L < \xi < 0$ with $L=4$, we utilize a specialized form of the Fourier series for $\tilde{f}$, expressed as
\begin{align*}
\tilde{f} &= \sum_{j=-M}^M a_j(x) \exp\left(2ij\pi\frac{\xi^2}{L^2}\right).
 \end{align*}
 Throughout the simulations, we employ the value $M=10$. Substituting this into the  stationary equation associated with \eqref{vp_time} and multiplying the resultant by $\exp(-2ik\pi\frac{\xi^2}{L^2})$, we have
\begin{align*}
\sum_{j=-M}^M a'_j(x) \xi \exp\left(2i(j-k)\pi\frac{\xi^2}{L^2}\right)  + \tilde{\phi}_x(x)\sum_{j=-M}^M a_j(x) \frac{4ij\pi}{L^2}\xi\exp\left(2i(j-k)\pi\frac{\xi^2}{L^2}\right) = 0.
 \end{align*}
Integrating over 
$(-L,0)$ in $\xi$, we then find that
\begin{subequations}\label{first_order}
\begin{align}
a_k(x)' &= -\tilde{\phi}_x(x)\frac{4 i k \pi}{L^2}a_k(x),\ k=-M,-M+1,\ldots,-1,0,1,\ldots,M, \\
\ve \partial_{xx} \tilde{\phi}(x) &= \sum_{j=-M}^M m_j a_j(x) - e^{-\tilde{\phi}(x)},
 \end{align}
where
\begin{align*}
m_j = \int_{-L}^0 \exp\left(2ij\pi\frac{\xi^2}{L^2}\right) d\xi = \left\{
\begin{array}{ll}
L & \text{for $j=0$,}\\
\frac{L}{2\sqrt{j}}\left({\sf fresnelc}(2\sqrt{j}) + i \cdot {\sf fresnels}(2\sqrt{j})\right)
& \text{for $j> 0$,}\\
\frac{L}{2\sqrt{|j|}}\left({\sf fresnelc}(2\sqrt{|j|}) - i \cdot {\sf fresnels}(2\sqrt{|j|})\right)
& \text{for $j< 0$.}
\end{array}
\right.
\end{align*}   
\end{subequations}
Here, {\sf fresnelc}($z$) and {\sf fresnels}($z$) are the Fresnel cosine and sine integrals of $z$, respectively.

Let $z_1 = \tilde{\phi}$, $z_2 = \tilde{\phi}_x$, $z_3=a_0$, $z_{2l+2} = a_{-l}$, and $z_{2l+3} = a_{l}$, where $l=1,2,\ldots M$.
The first-order system (\ref{first_order}) is then written as
\begin{subequations}\label{first_order_sys}
\begin{align}
\begin{split}
\frac{dz_1}{dx} &= z_2,\\
\frac{dz_2}{dx} &= \ep^{-1}\left(  m_0z_3 + \sum_{j=1}^M m_{-j} z_{2j+2} + \sum_{j=1}^M m_{j} z_{2j+3} - e^{-z_1} \right),\\
\frac{dz_3}{dx} &= 0,\\
\frac{dz_{2l+2}}{dx} &= \frac{4 i l \pi}{L^2}z_2z_{2l+2},\\
\frac{dz_{2l+3}}{dx} &= -\frac{4 i l \pi}{L^2}z_2z_{2l+3}.
\end{split}
\end{align}
The boundary conditions (\ref{vp_bc}) and (\ref{vp_time2}) are
\begin{align}\label{first_order_sys_bdry}
\begin{split}
&z_1(0) = 1,\ z_1(1) = 0,\\
&z_{2l+3}(1) =  -\frac{2}{L^2}\int_{-L}^0 \sigma(\xi)\exp\left(-2il\pi\frac{\xi^2}{L^2}\right)\xi d\xi,\ l=0,1,\ldots,M,\\
&z_{2l+2}(1) = -\frac{2}{L^2}\int_{-L}^0 \sigma(\xi)\exp\left(2il\pi\frac{\xi^2}{L^2}\right)\xi d\xi,\ l=1,\ldots,M.
\end{split}
\end{align}
\end{subequations}

For the boundary layer equations in (\ref{vp2}), we may use the first-order system (\ref{first_order_sys}) after some modifications. Note that the limit solutions $(f^0, \phi^0)$ of (\ref{vp3}) should be obtained in advance (see (\ref{limit_sys_discr}) below).

For the time-dependent problem (\ref{vp}), incorporating the implicit Euler discretizations in time for $f_t$ which offer larger stability regions compared to explicit ones, we may follow (\ref{first_order_sys}) with some modifications. Let $f^{n} = f^{n}(x,\xi) = f(t_n,x,\xi)$ and $\phi^{n}=\phi^{n}(x)=\phi(t_n,x)$, where $t_n = n\Delta t$, $n=1,2,\ldots$. Then, the problem is discretized as
\begin{subequations}
\begin{gather}
\frac{f^{n}-f^{n-1}}{\Delta t} + \xi f^{n}_x + \phi^n_x\partial_\xi f^{n} = 0, \quad  x\in(0,1), \ \xi \in (-4,0),\label{sta_eq_time}\\
\ve \partial_{xx} \phi^n  = \int_{\mathbb R} f^n d\xi - e^{-\phi^n}, \ x\in(0,1)
\end{gather}
with the boundary conditions 
\begin{align}\label{bdry_vp}
f^n(1,\xi) = \sigma(\xi),\quad \phi^n(0) = 1,\quad \phi^n(1) = 0.
\end{align}
\end{subequations}

\begin{figure}[h]
  \centering
  \begin{tabular}{m{1mm}m{34mm}m{34mm}m{34mm}m{34mm}}
\hspace{15mm} & \hspace{15mm} $t=0$ & \hspace{15mm} $t=0.1$ & \hspace{15mm} $t=0.4$ & \hspace{15mm} $t=1$\\ 
   {\small $f$} & \resizebox{38mm}{!}{\includegraphics{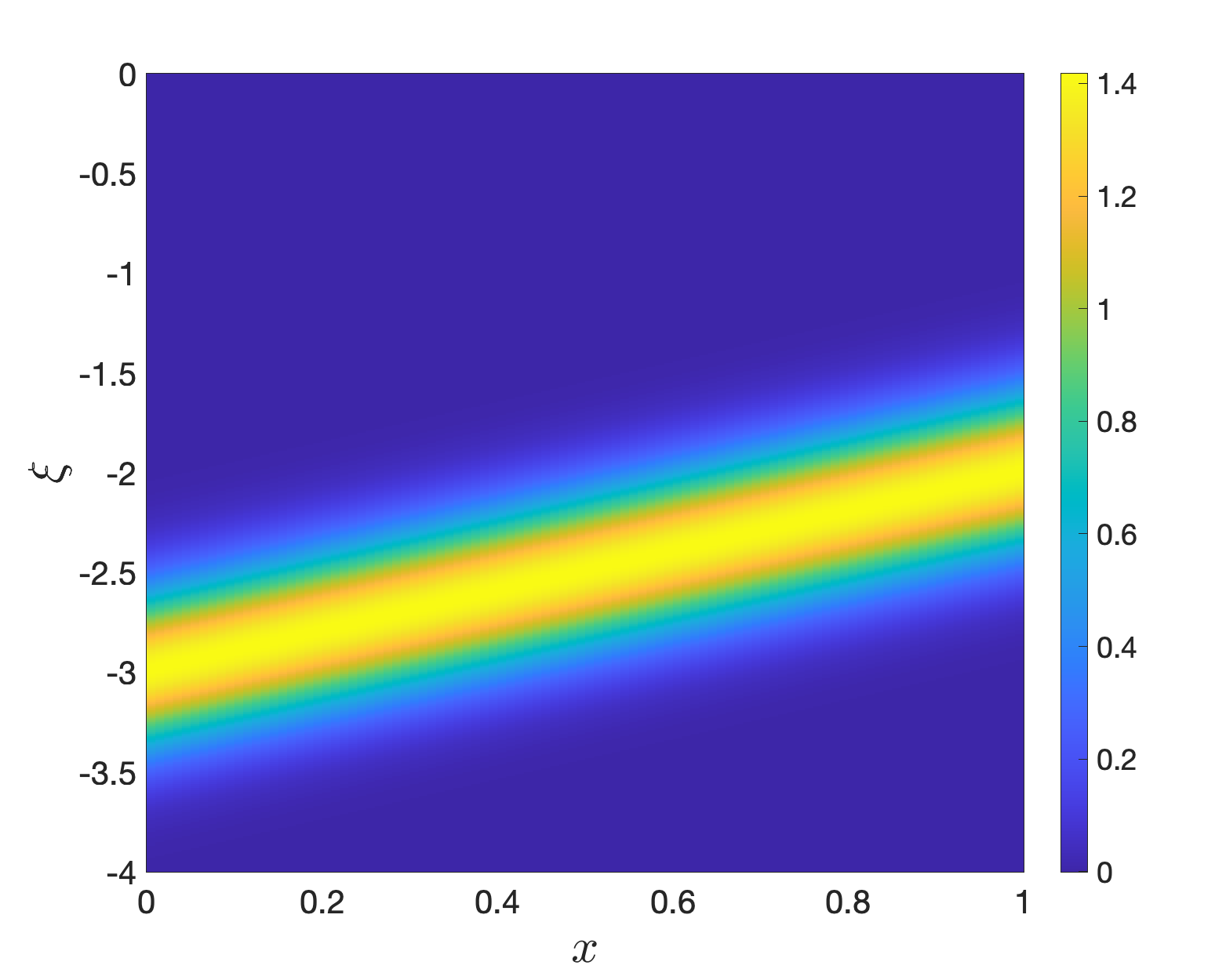}} & \resizebox{38mm}{!}{\includegraphics{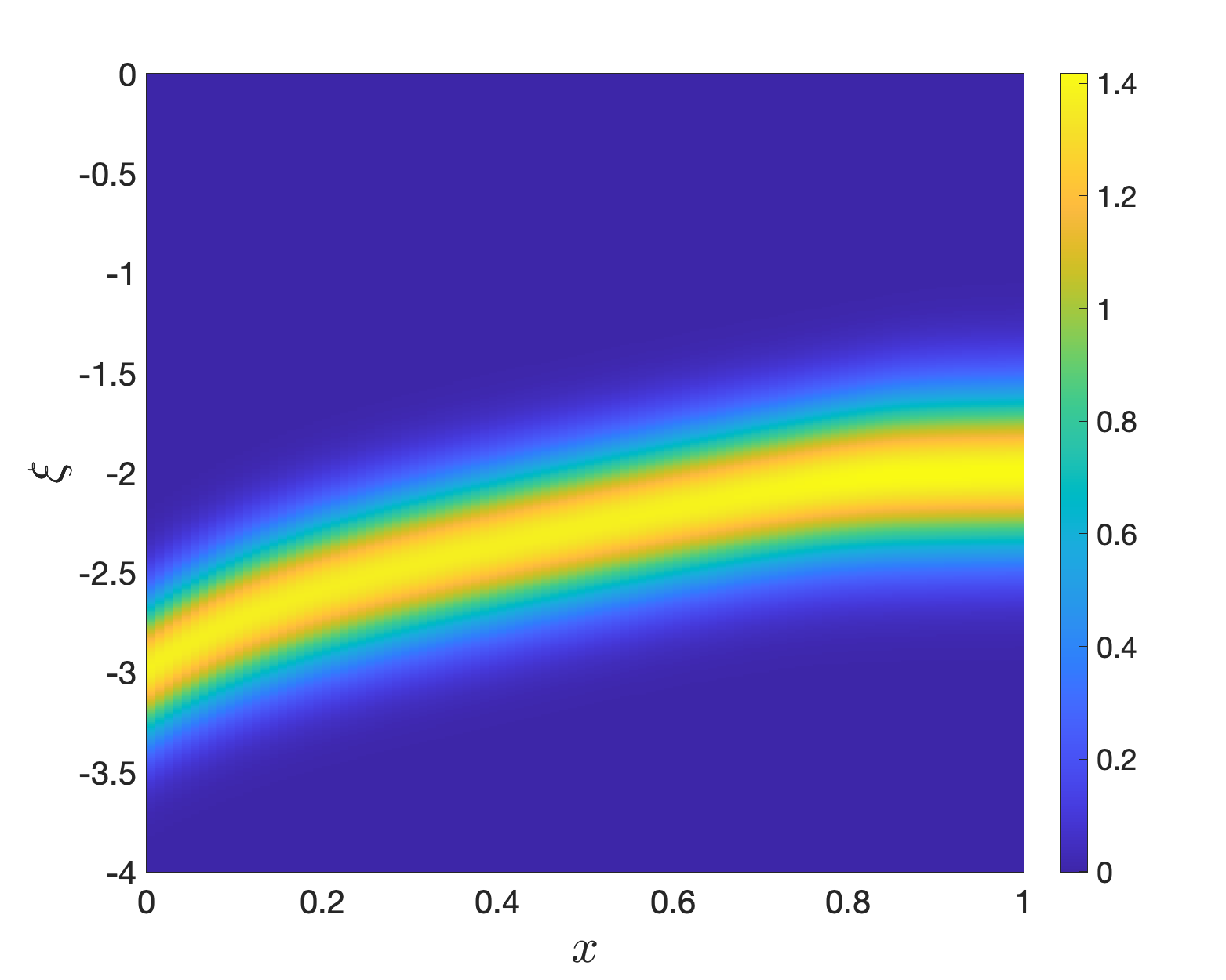}} & \resizebox{38mm}{!}{\includegraphics{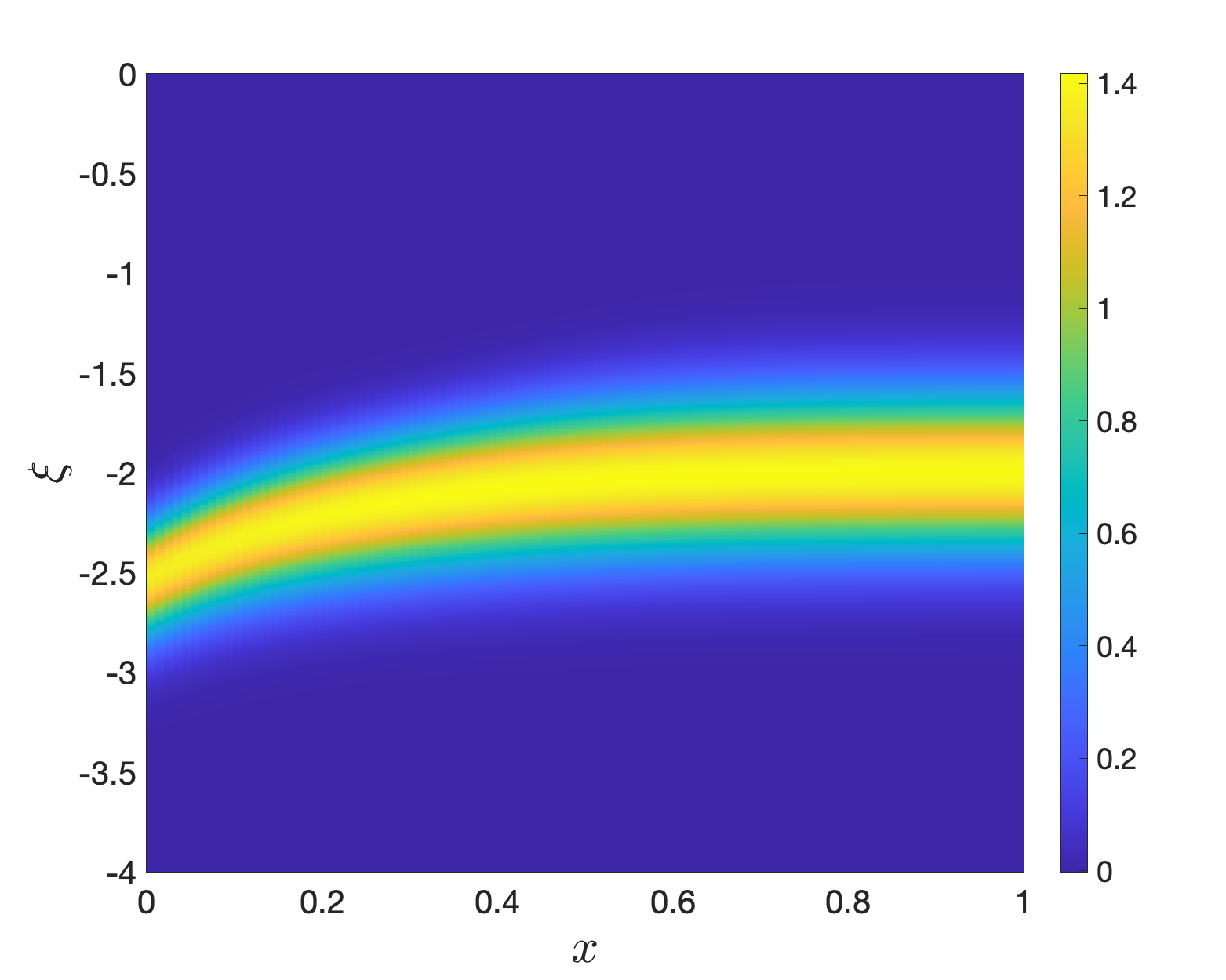}} & \resizebox{38mm}{!}{\includegraphics{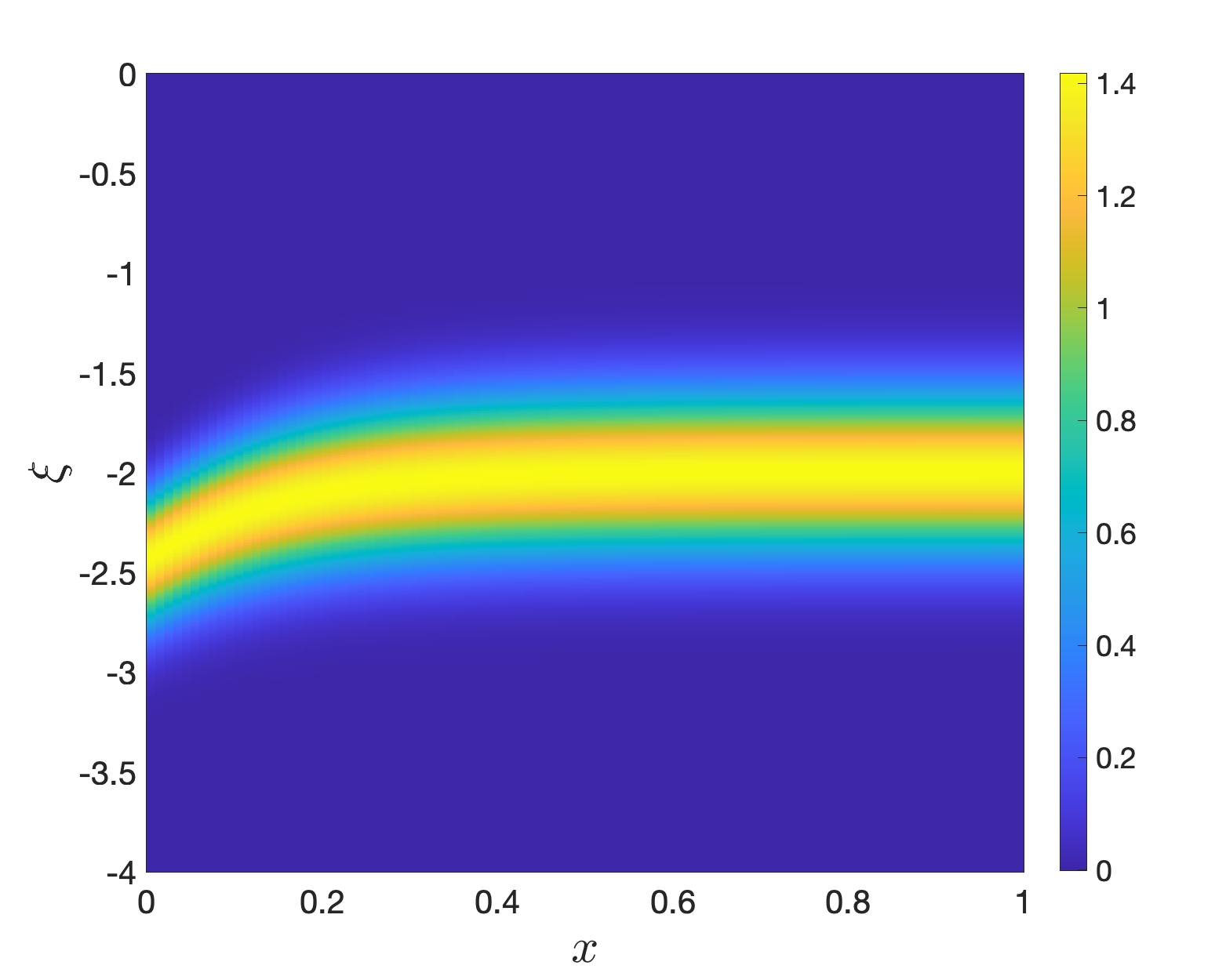}} \\
  {\small $f^0$} &   \resizebox{38mm}{!}{\includegraphics{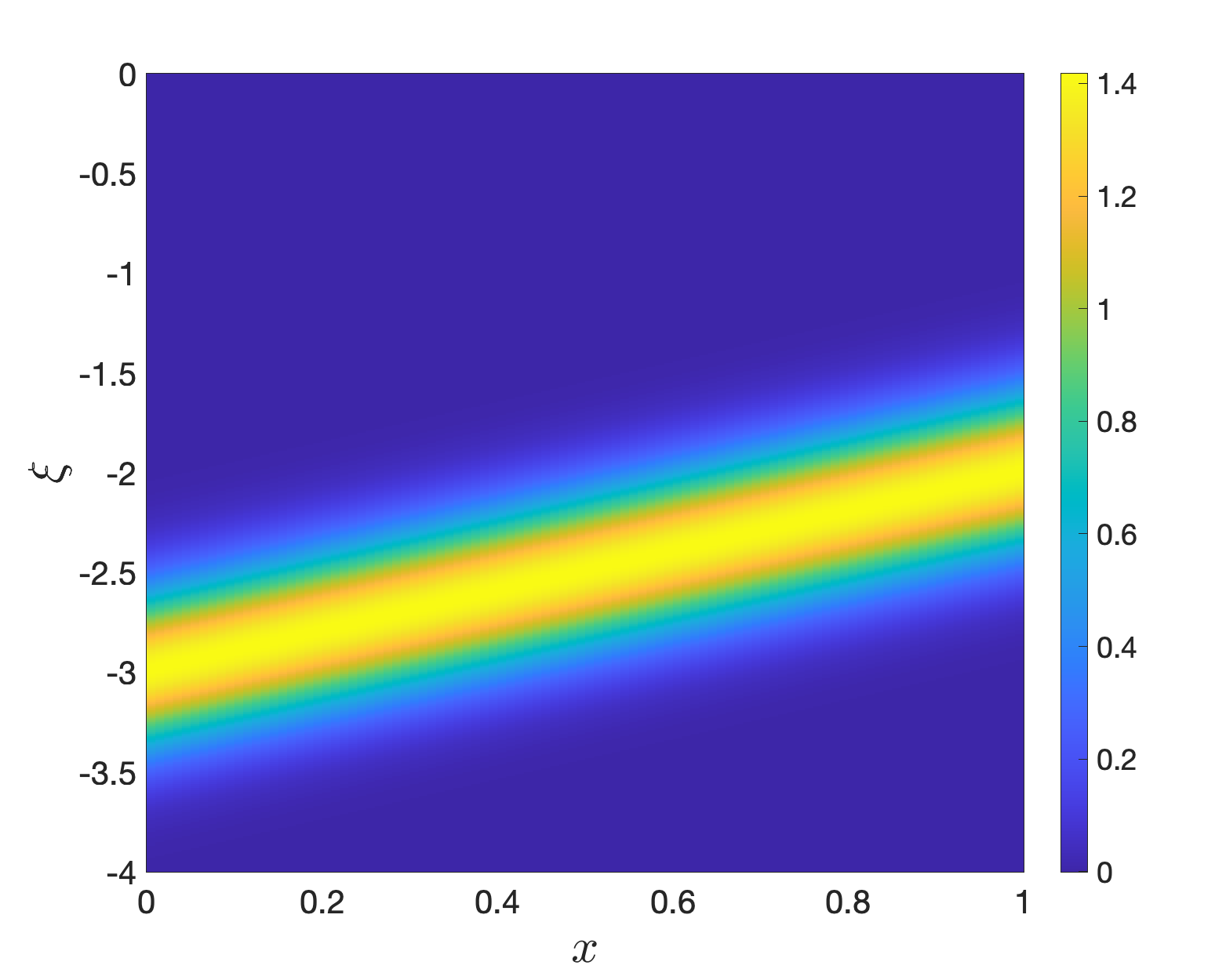}} & \resizebox{38mm}{!}{\includegraphics{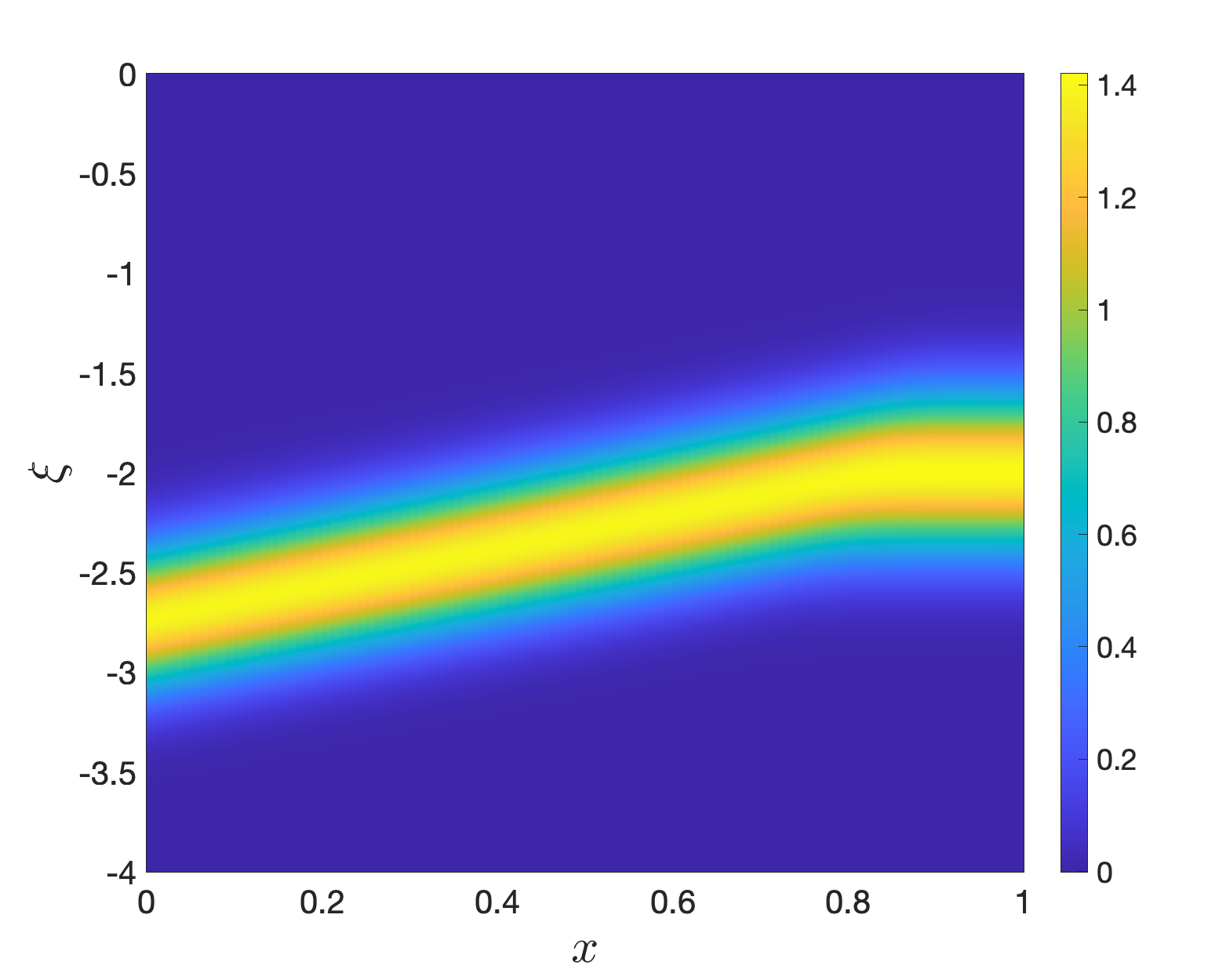}} & \resizebox{38mm}{!}{\includegraphics{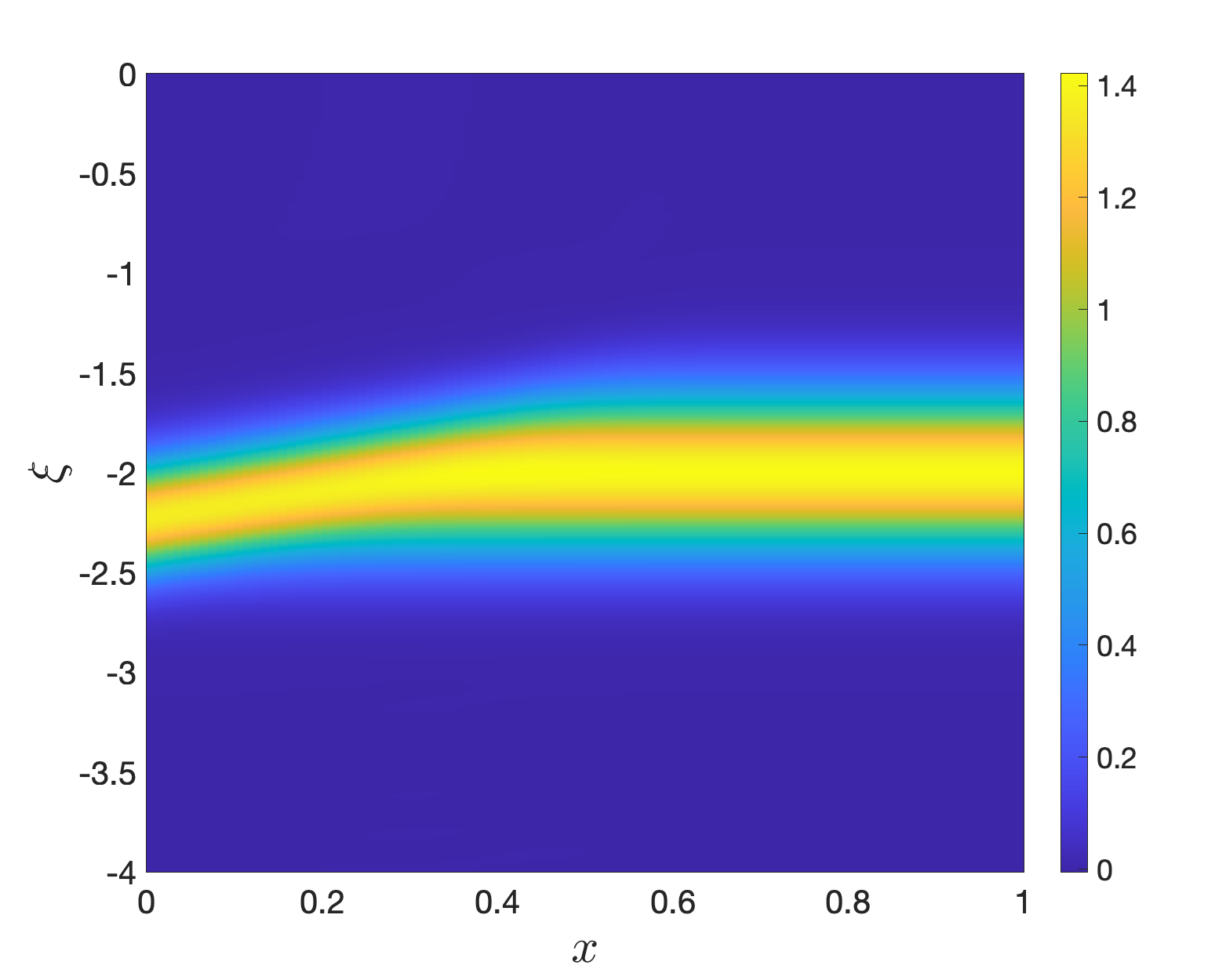}} & \resizebox{38mm}{!}{\includegraphics{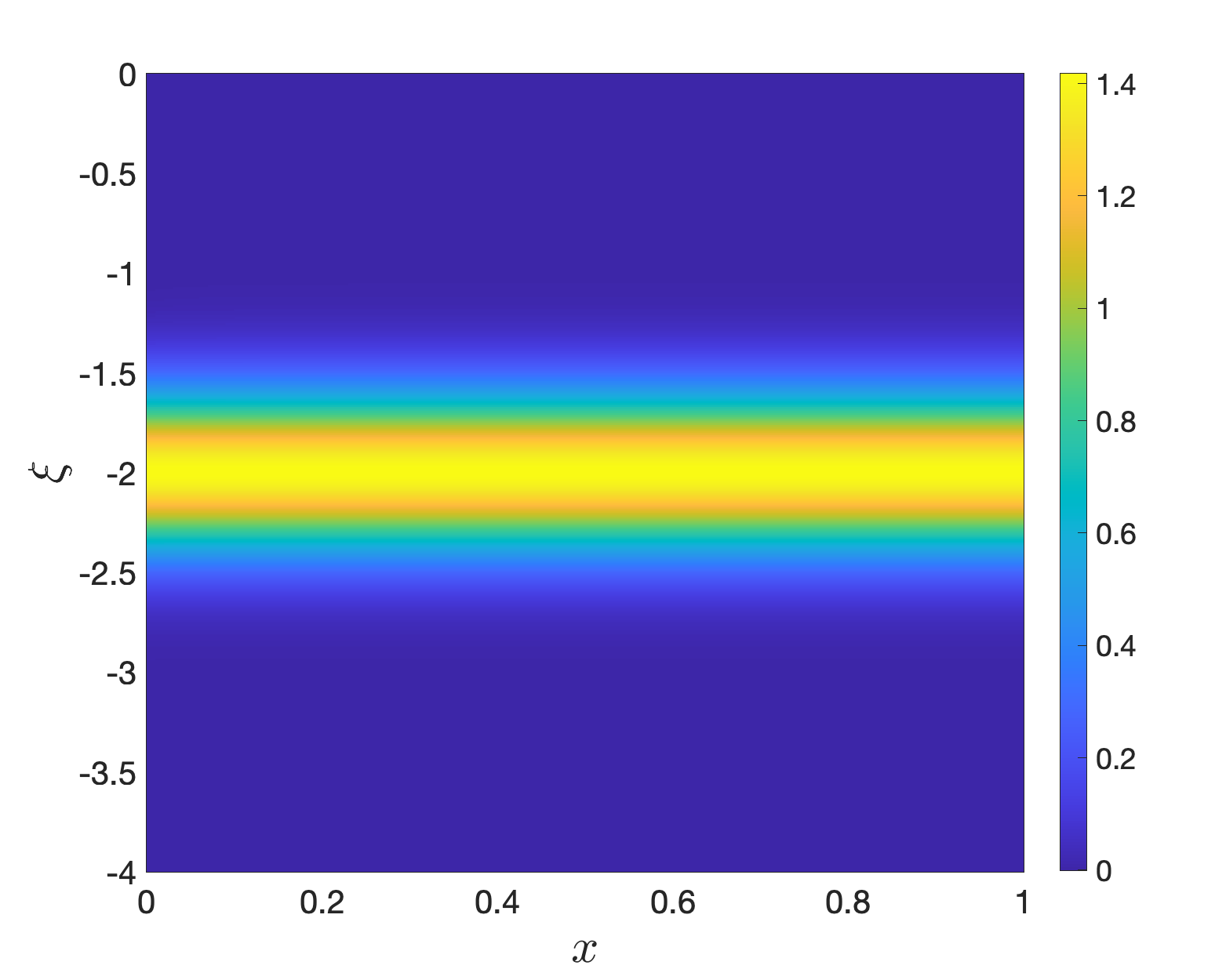}} \\
{\small $F^0$} &     \resizebox{38mm}{!}{\includegraphics{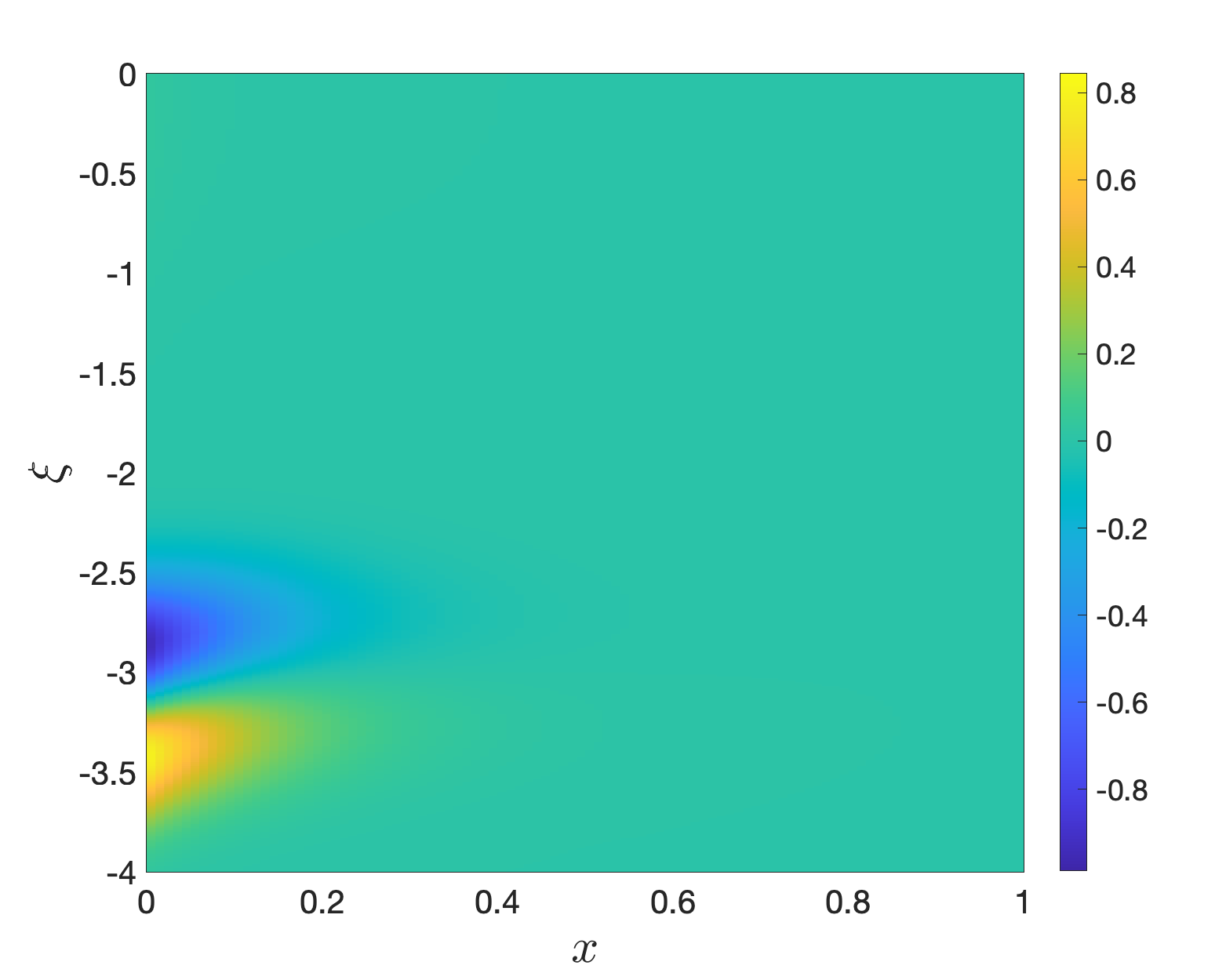}} & \resizebox{38mm}{!}{\includegraphics{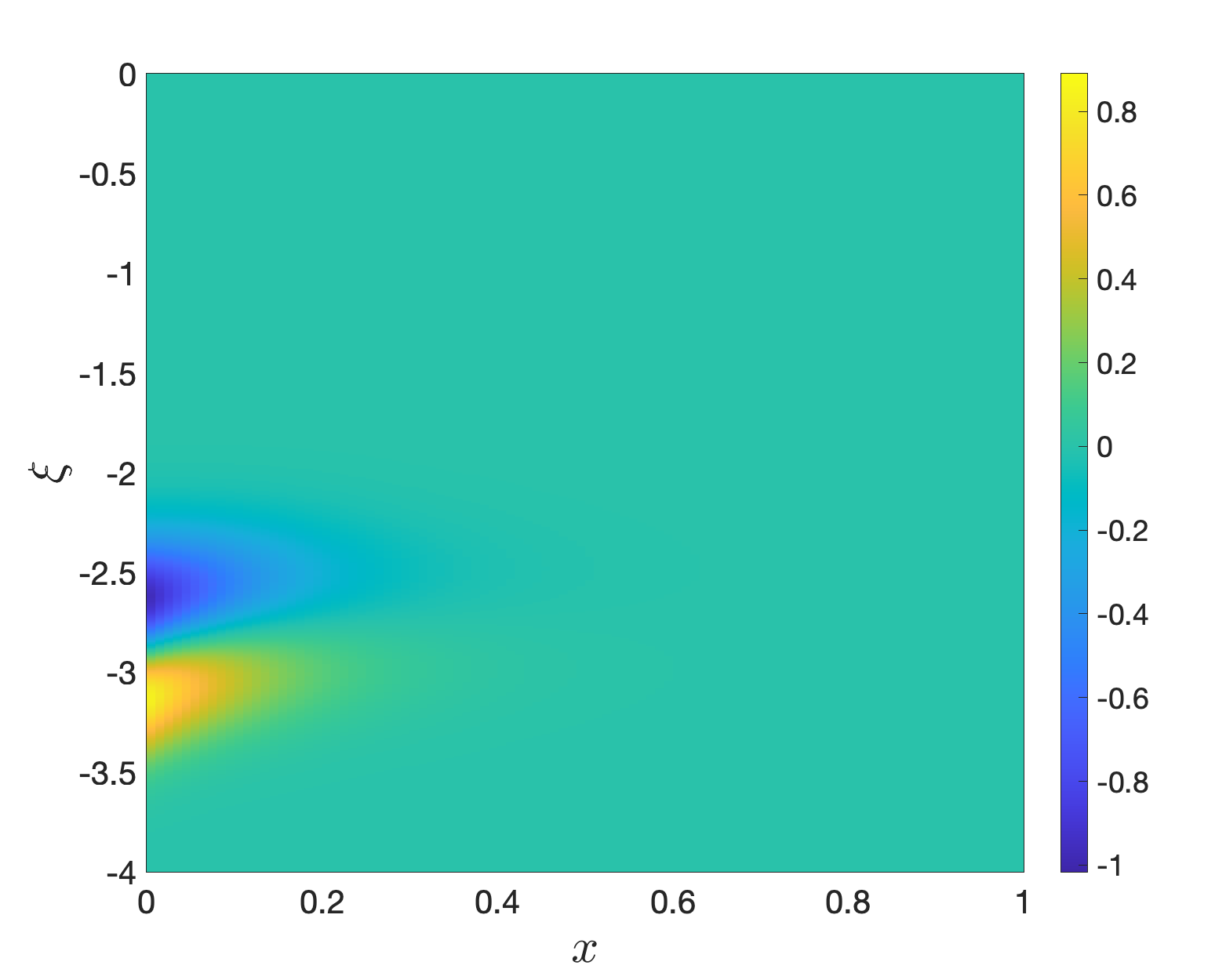}} & \resizebox{38mm}{!}{\includegraphics{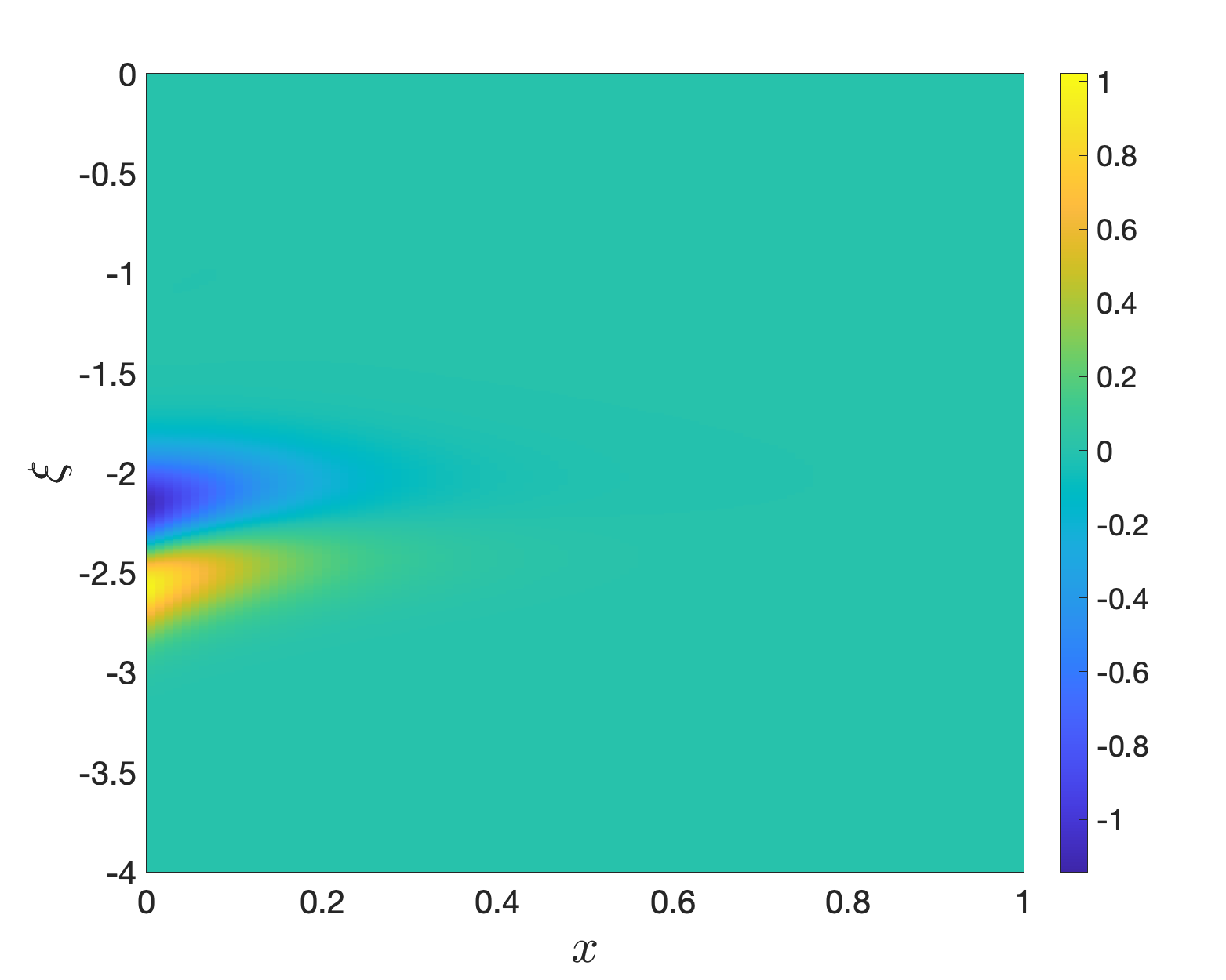}} & \resizebox{38mm}{!}{\includegraphics{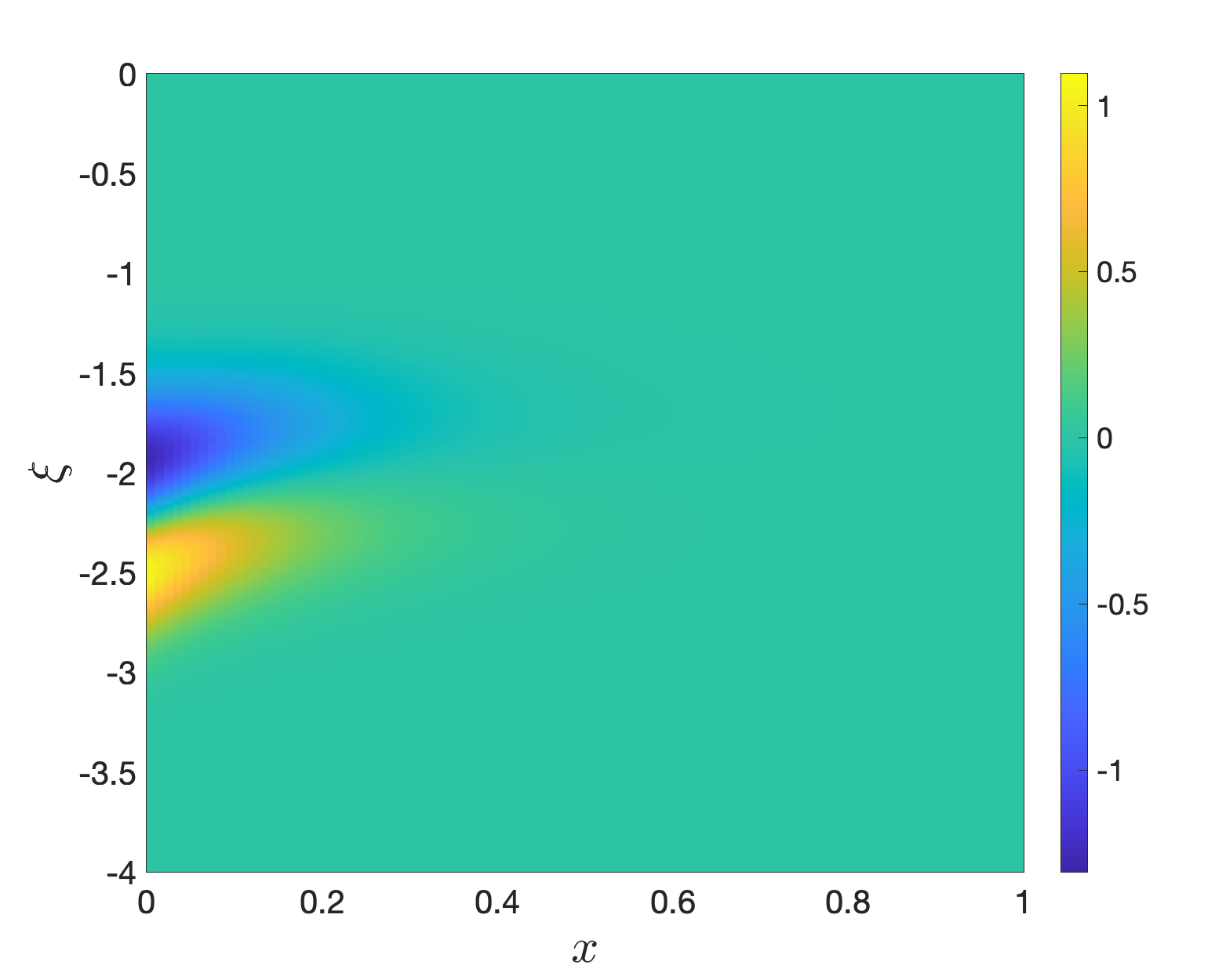}} 
  \end{tabular}  
  \caption{\small{The plots of $f$, $f^0$, and $F^0$ for $\ep = 10^{-2}$  with the initial and boundary conditions in Case (I).}}\label{ex3_f}
\vspace{5mm}
  \centering
  \begin{tabular}{m{1mm}m{34mm}m{34mm}m{34mm}m{34mm}}
\hspace{15mm} & \hspace{15mm} $t=0$ & \hspace{15mm} $t=0.1$ & \hspace{15mm} $t=0.4$ & \hspace{15mm} $t=1$\\ 
   {\small $\phi$} & \resizebox{38mm}{!}{\includegraphics{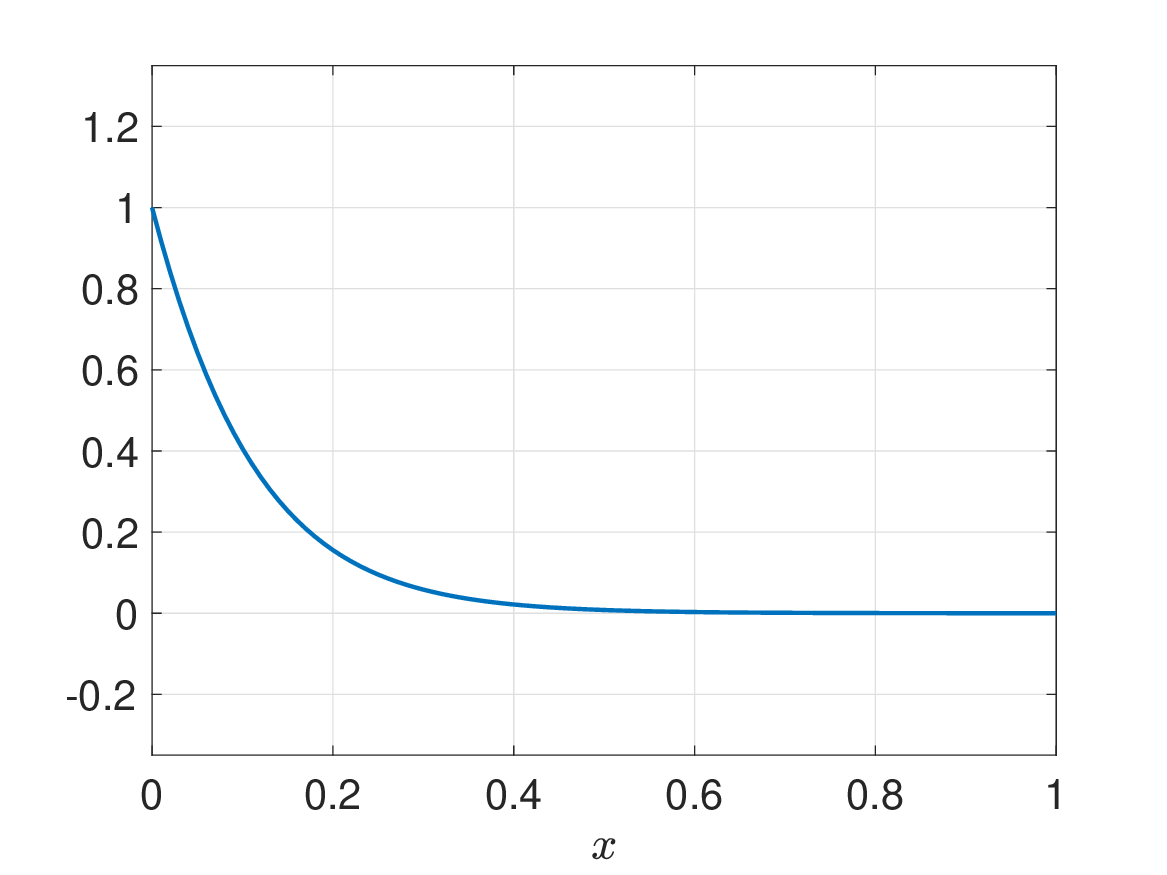}} & \resizebox{38mm}{!}{\includegraphics{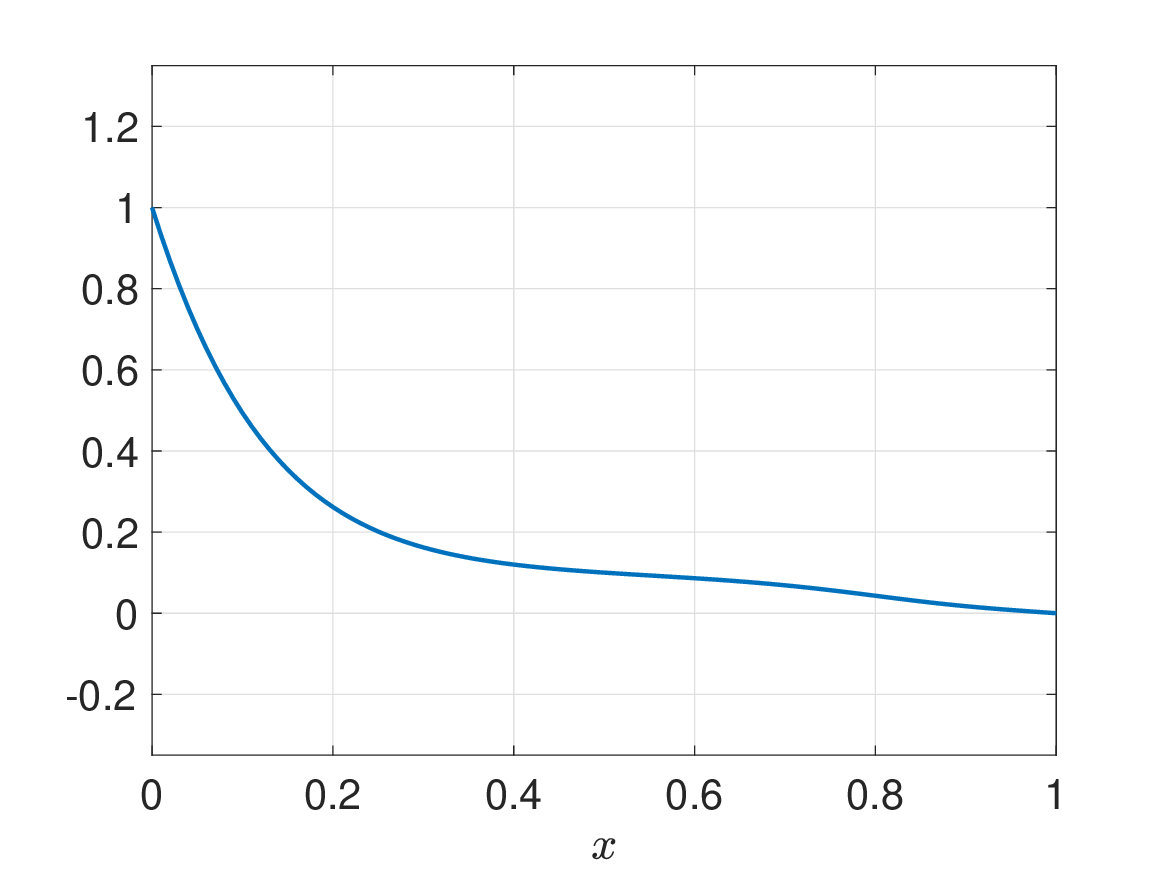}} & \resizebox{38mm}{!}{\includegraphics{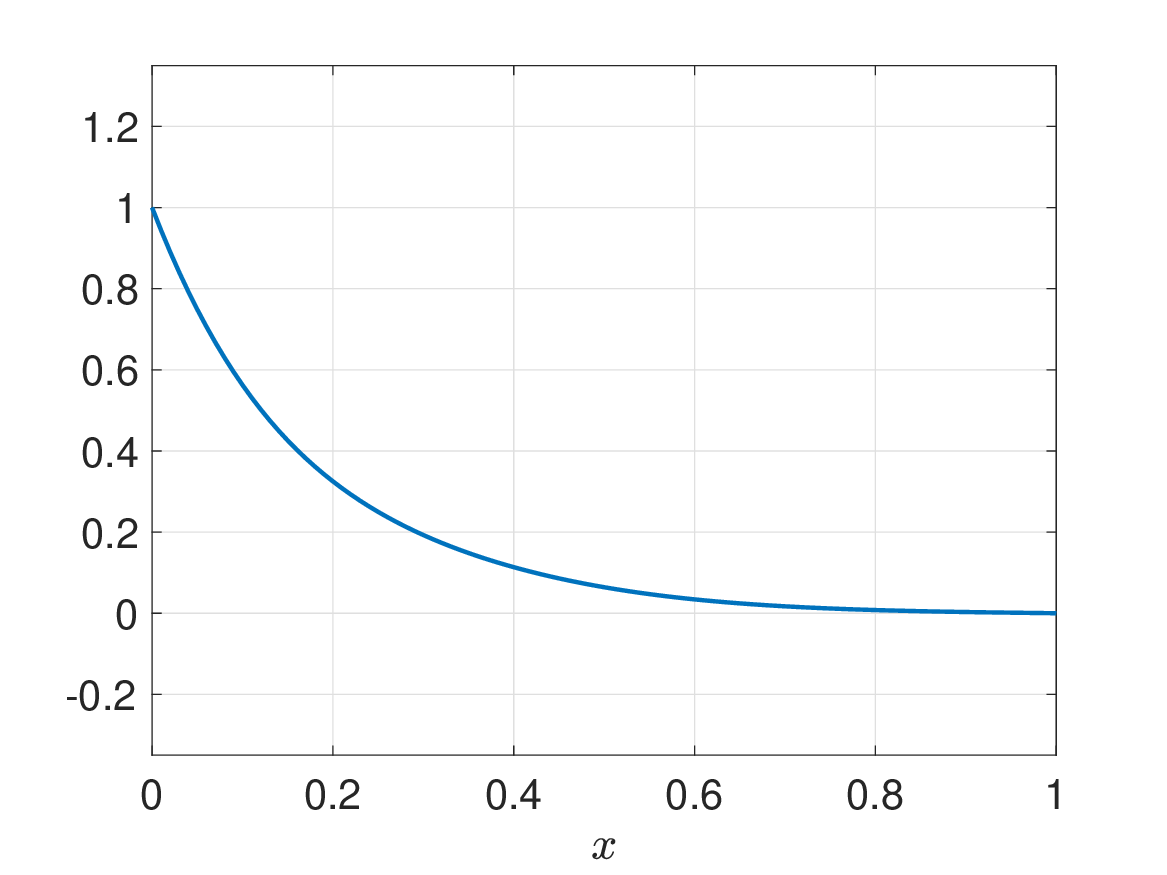}} & \resizebox{38mm}{!}{\includegraphics{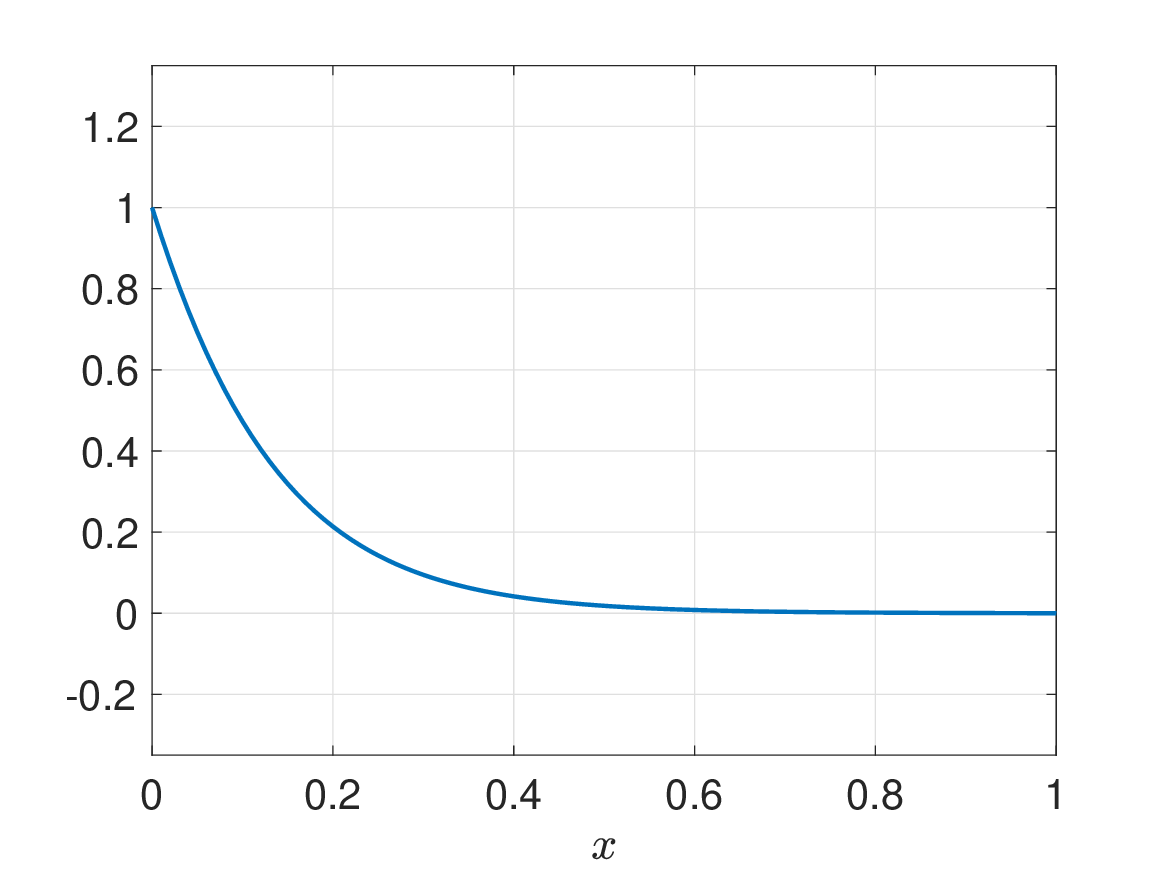}} \\
  {\small $\phi^0$} &   \resizebox{38mm}{!}{\includegraphics{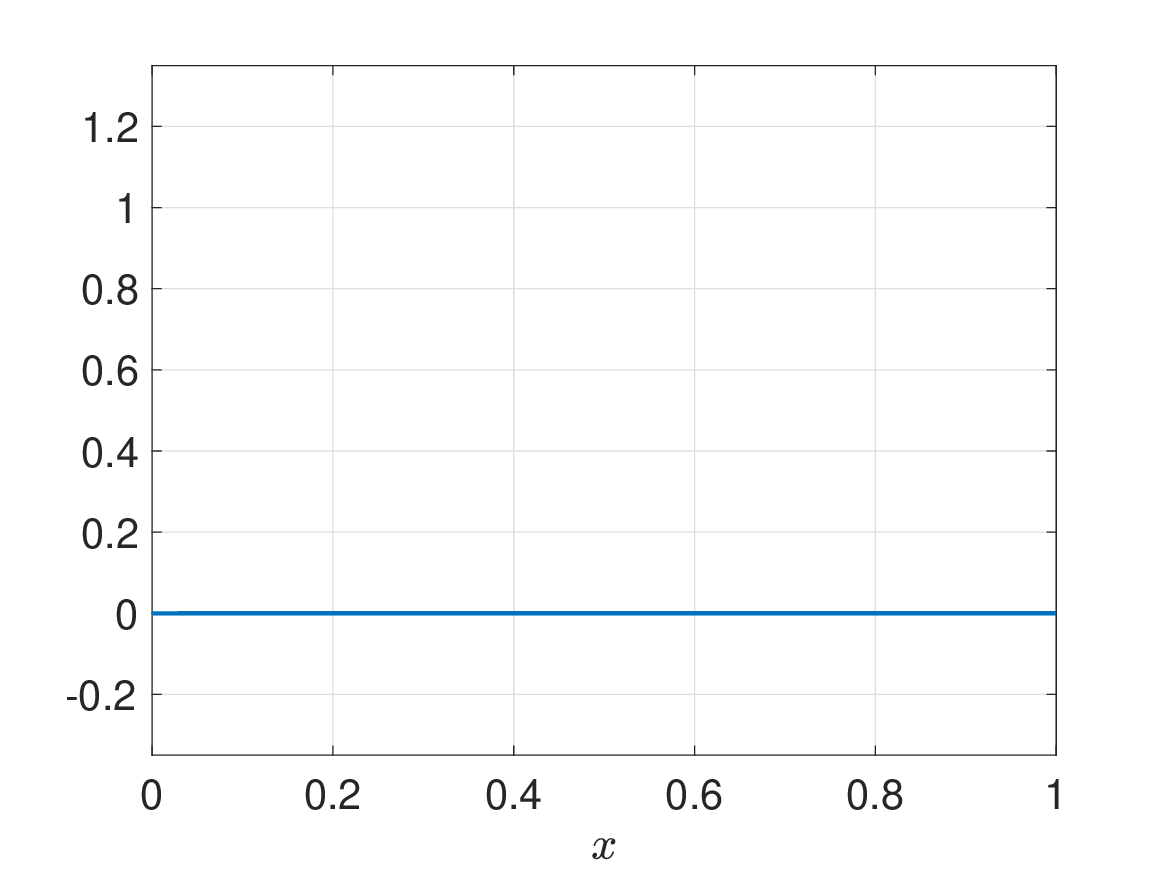}} & \resizebox{38mm}{!}{\includegraphics{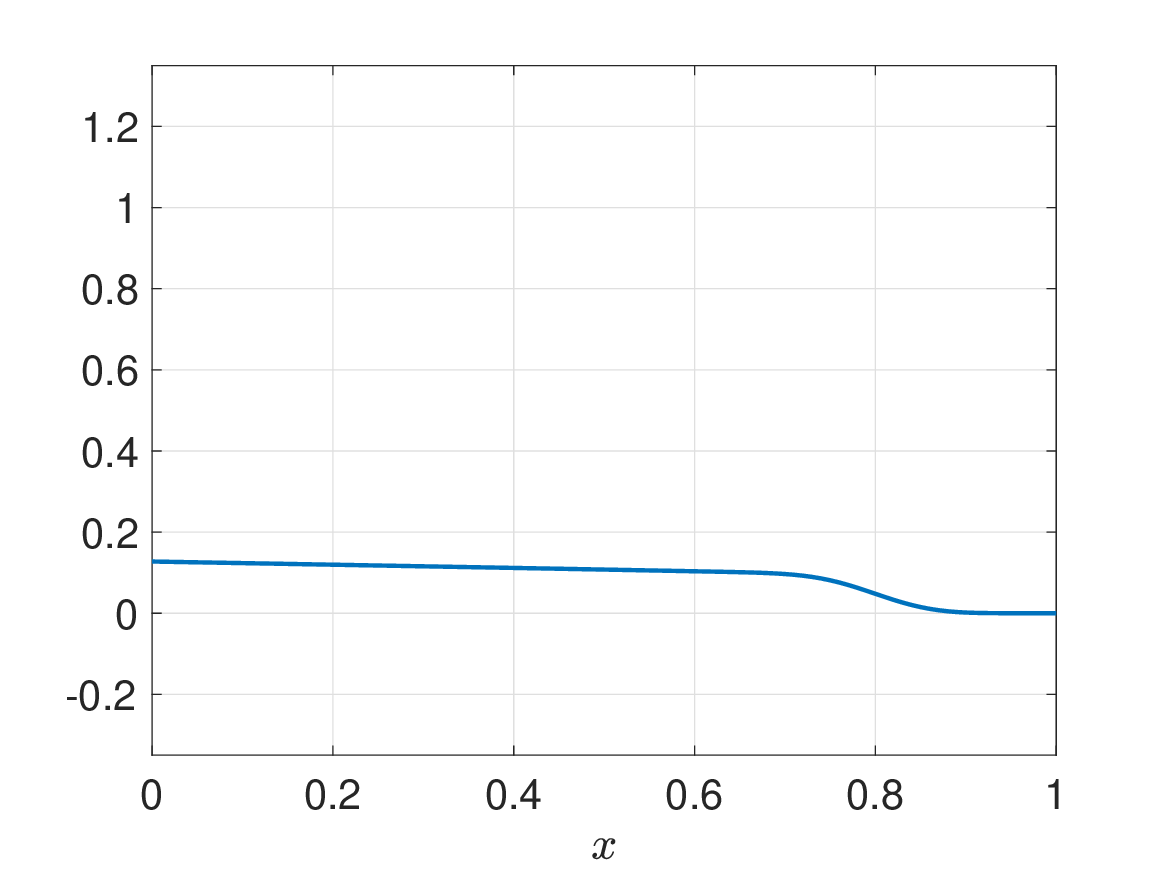}} & \resizebox{38mm}{!}{\includegraphics{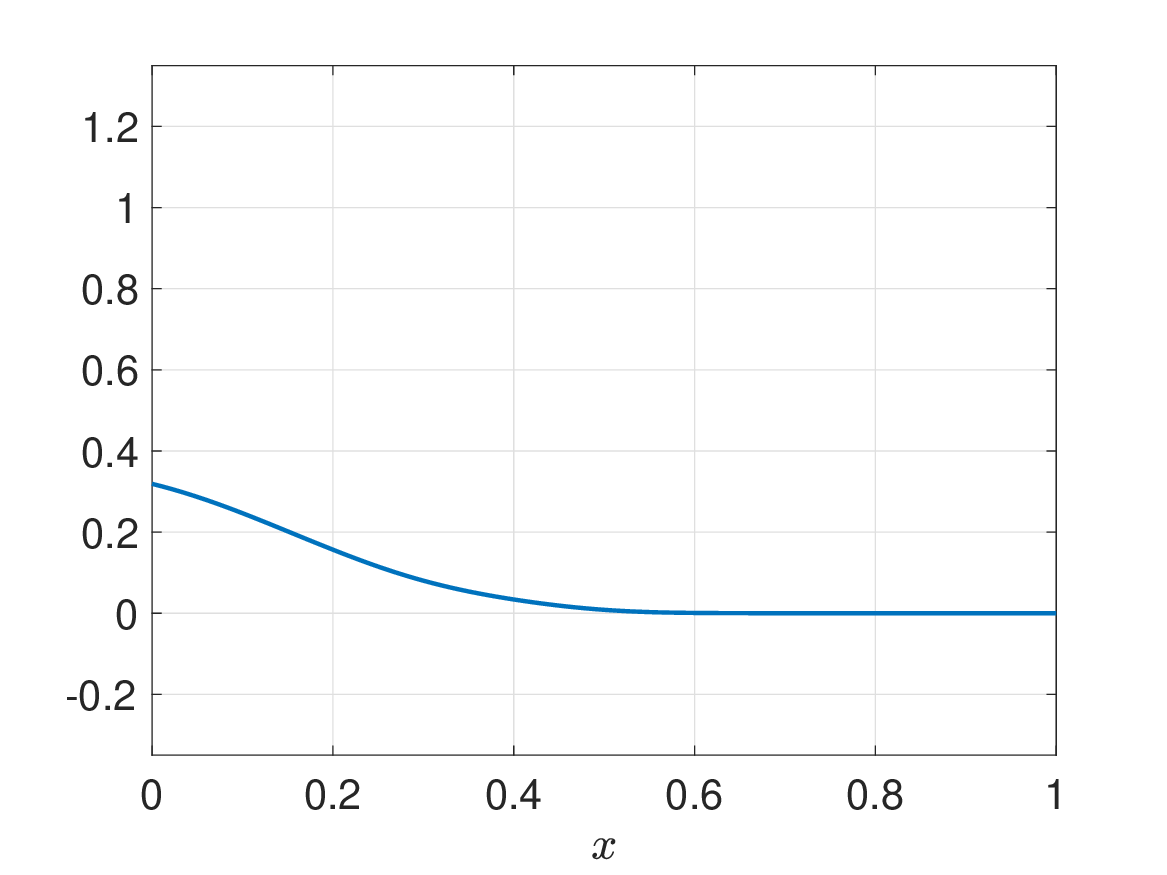}} & \resizebox{38mm}{!}{\includegraphics{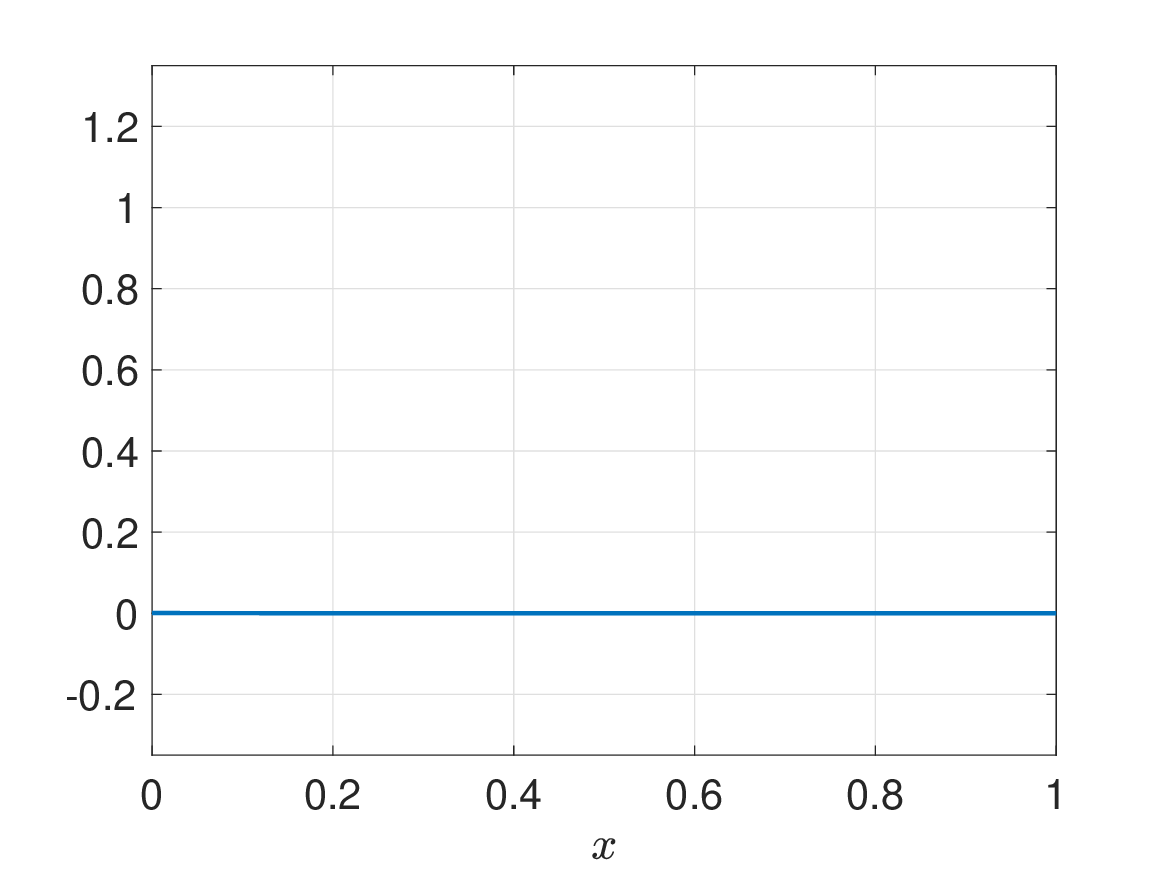}} \\
{\small $\Phi^0$} &     \resizebox{38mm}{!}{\includegraphics{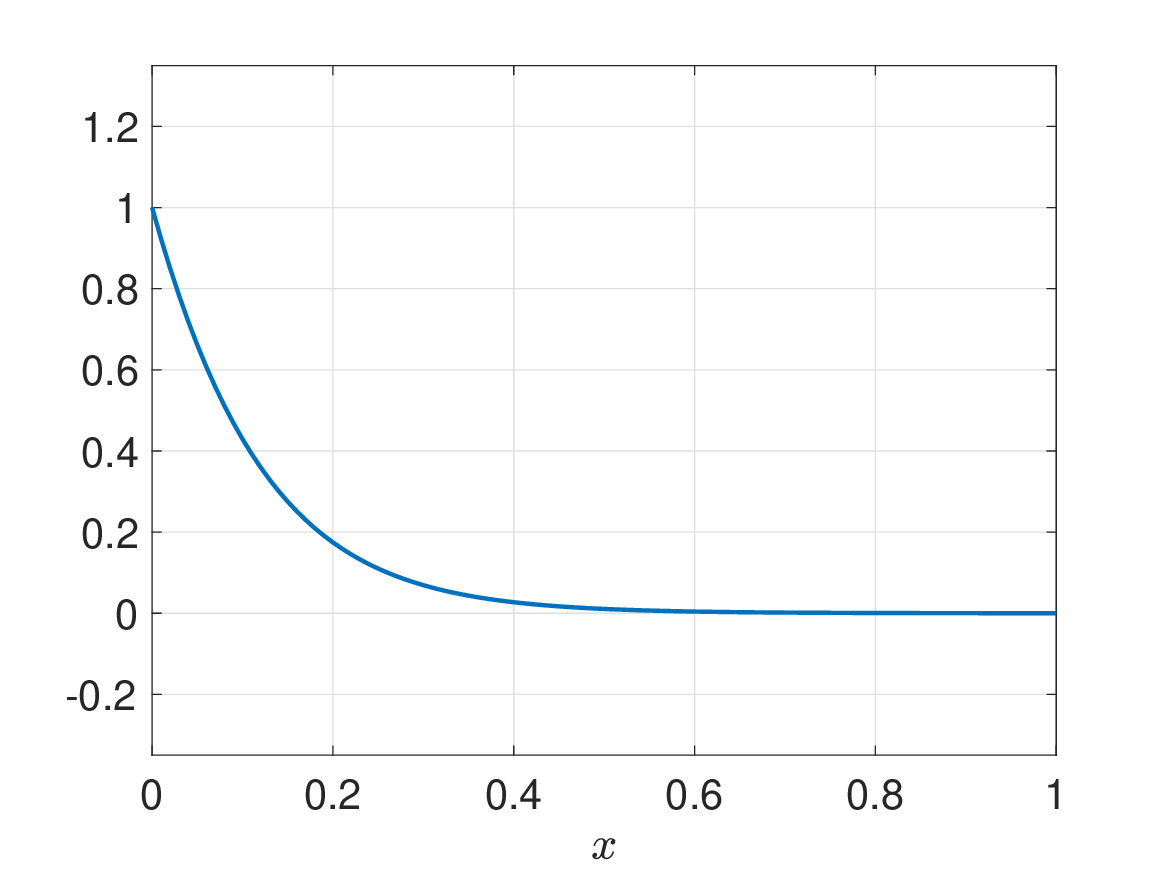}} & \resizebox{38mm}{!}{\includegraphics{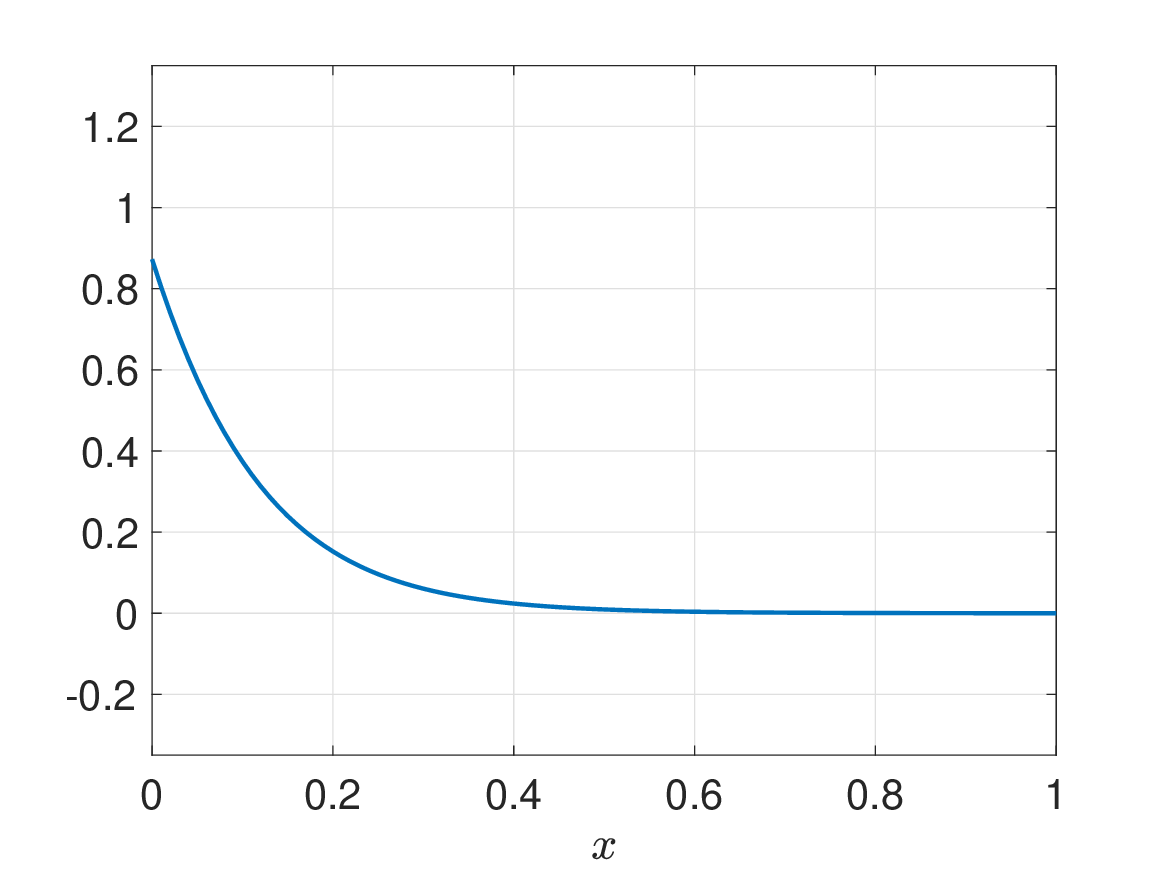}} & \resizebox{38mm}{!}{\includegraphics{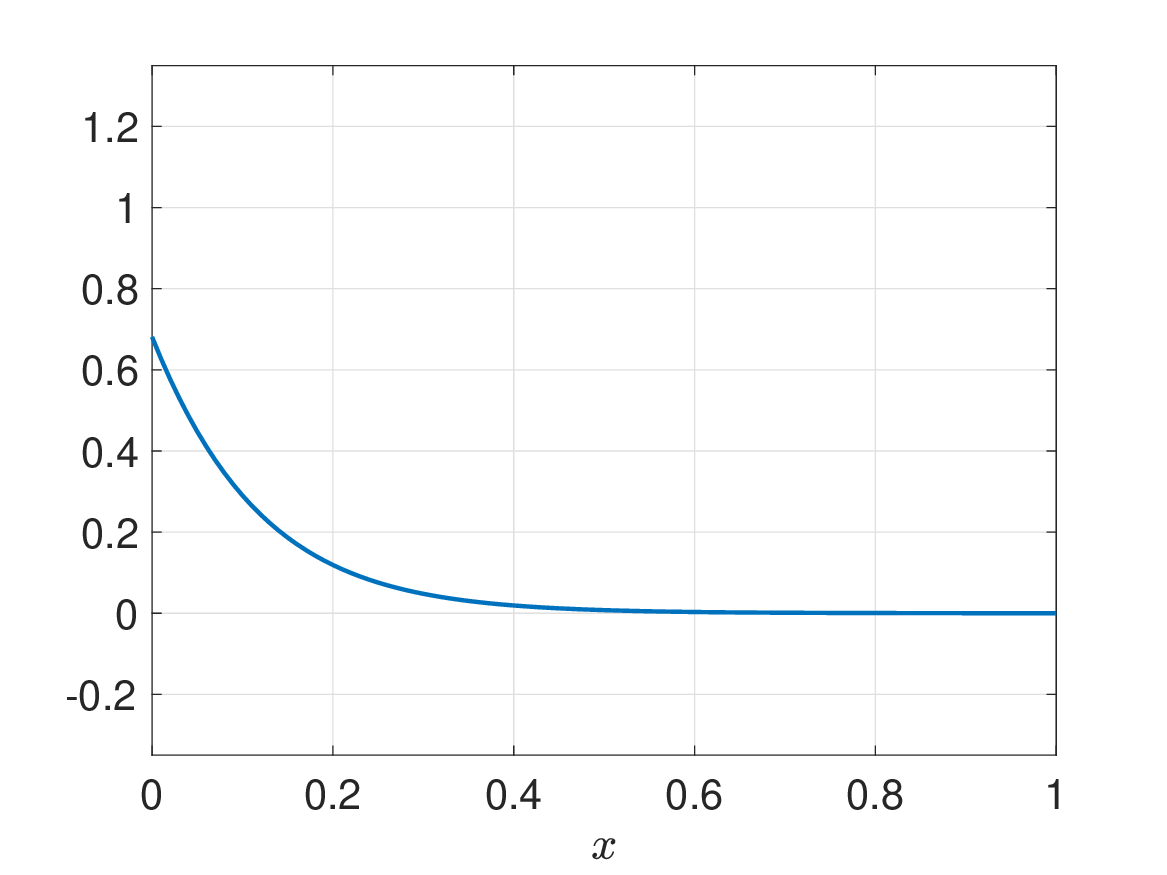}} & \resizebox{38mm}{!}{\includegraphics{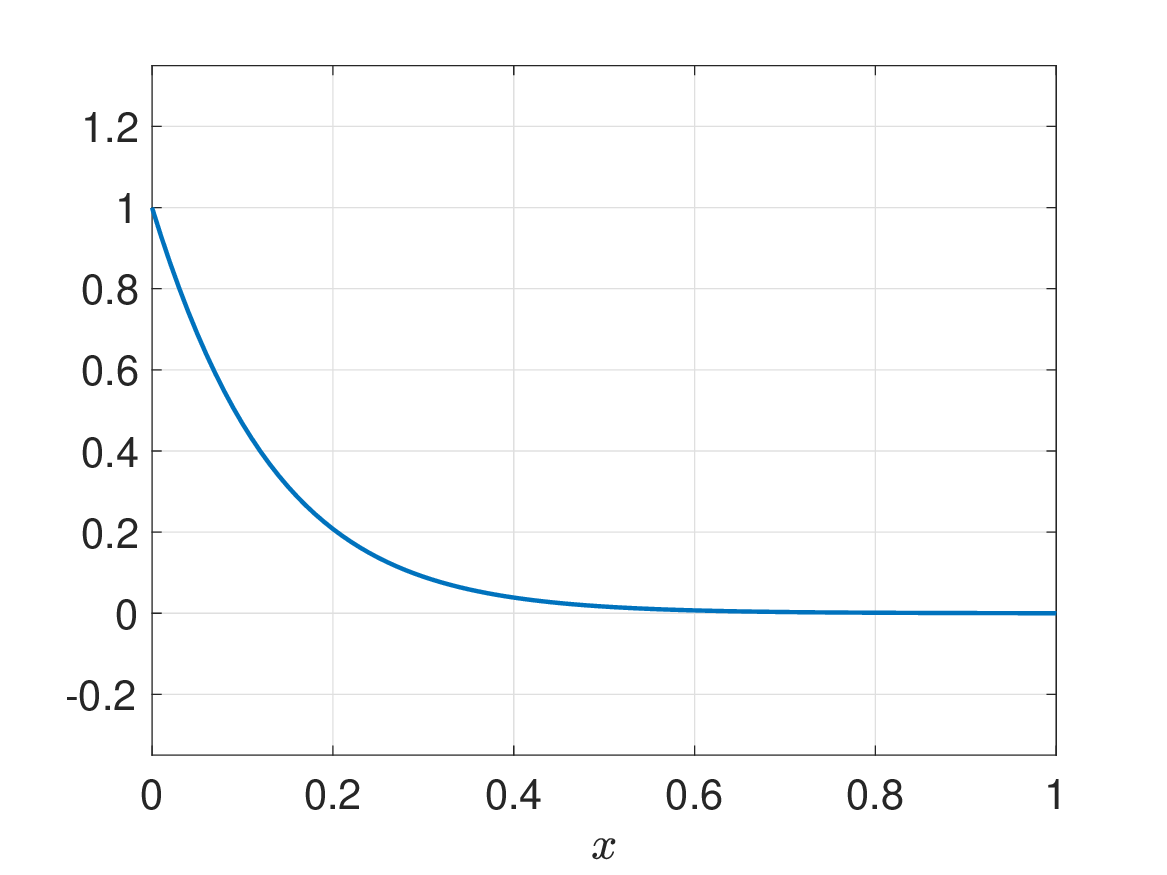}} 
  \end{tabular}  
  \caption{\small{The plots of $\phi$, $\phi^0$, and $\Phi^0$ for $\ep = 10^{-2}$ with the initial and boundary conditions in Case (I).}}\label{ex3_phi}
\end{figure}

\begin{figure}[h]
  \centering
  \begin{tabular}{m{1mm}m{34mm}m{34mm}m{34mm}m{34mm}}
\hspace{15mm} & \hspace{15mm} $t=0$ & \hspace{15mm} $t=0.1$ & \hspace{15mm} $t=0.2$ & \hspace{15mm} $t=1$\\ 
   {\small $f$} & \resizebox{38mm}{!}{\includegraphics{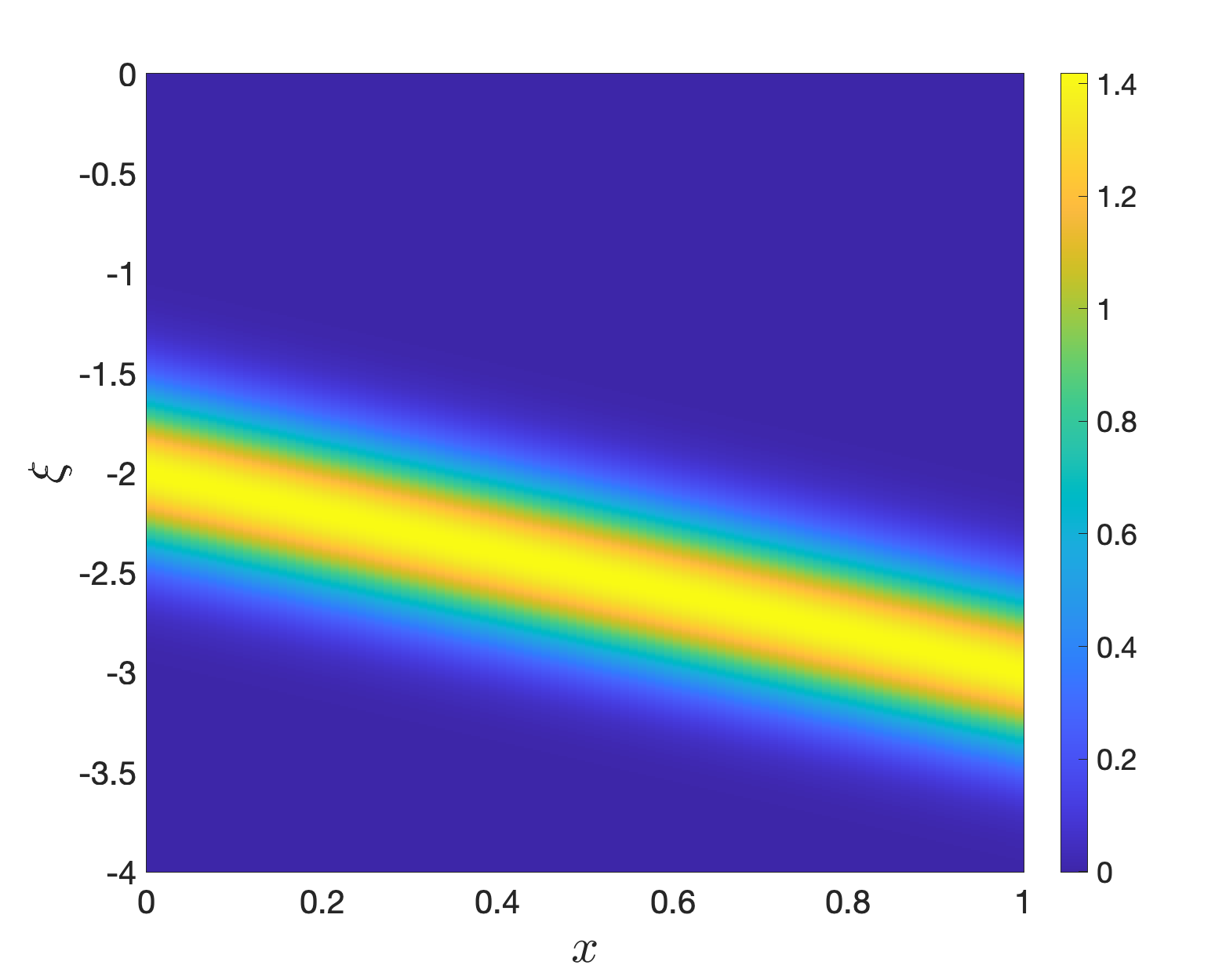}} & \resizebox{38mm}{!}{\includegraphics{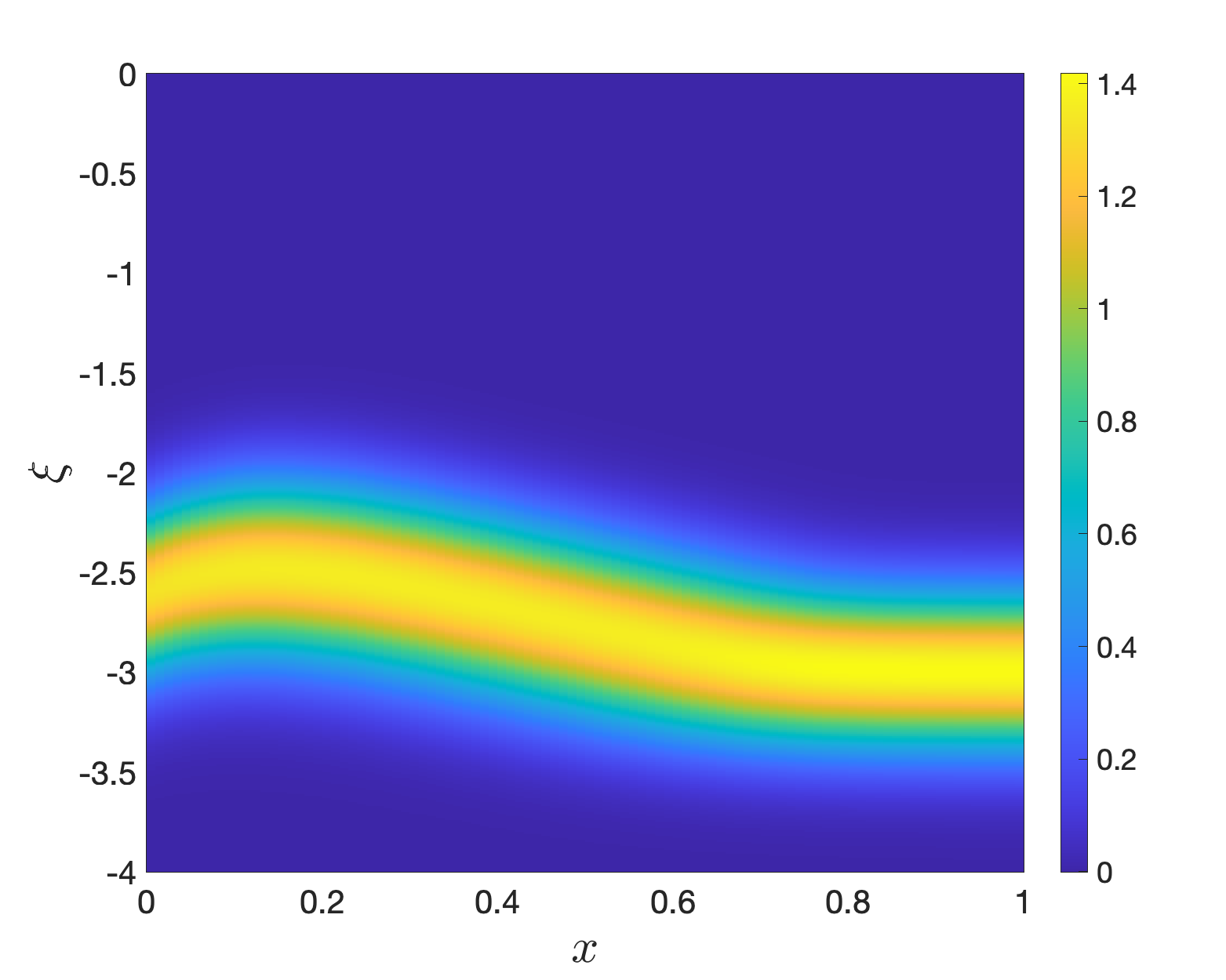}} & \resizebox{38mm}{!}{\includegraphics{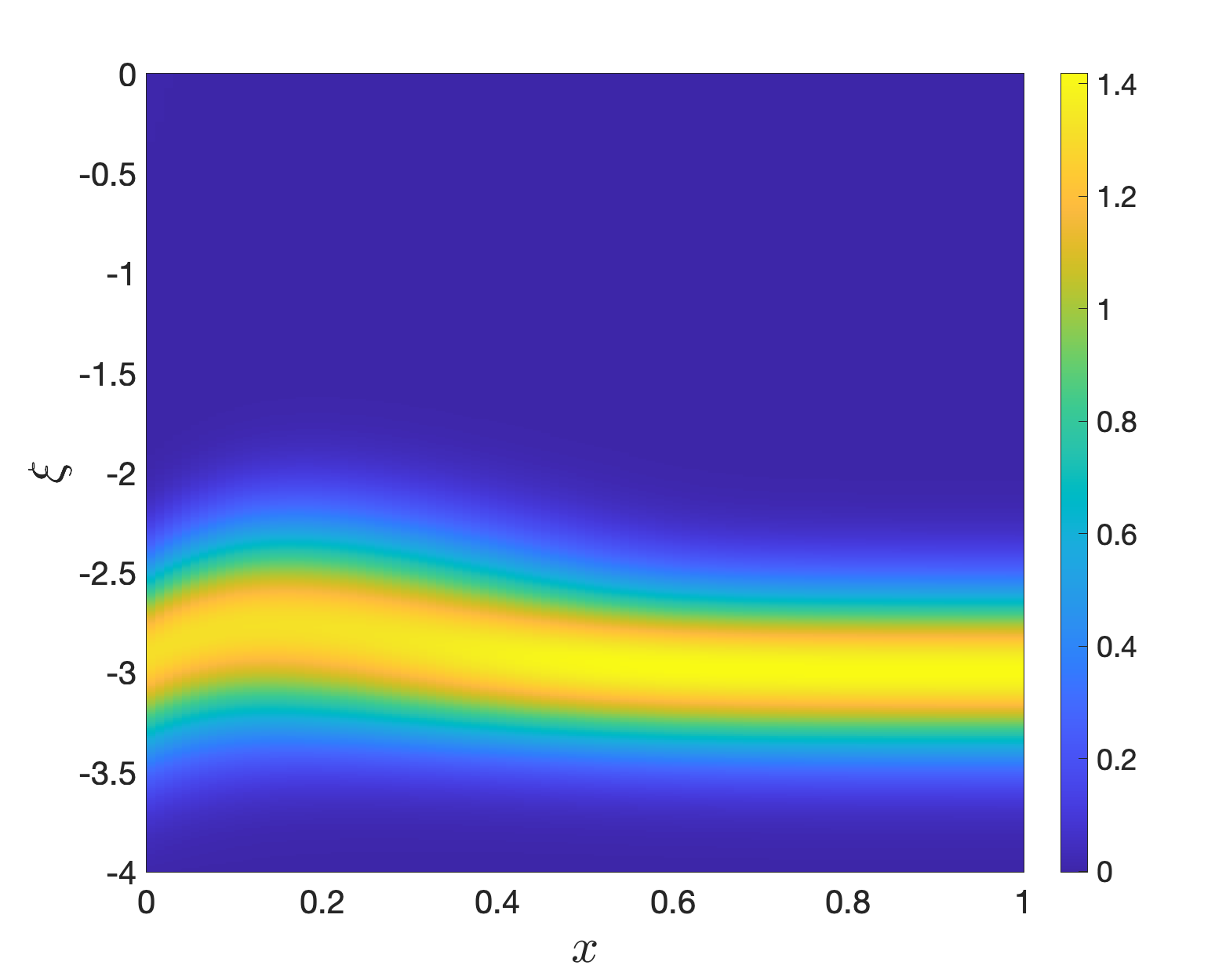}} & \resizebox{38mm}{!}{\includegraphics{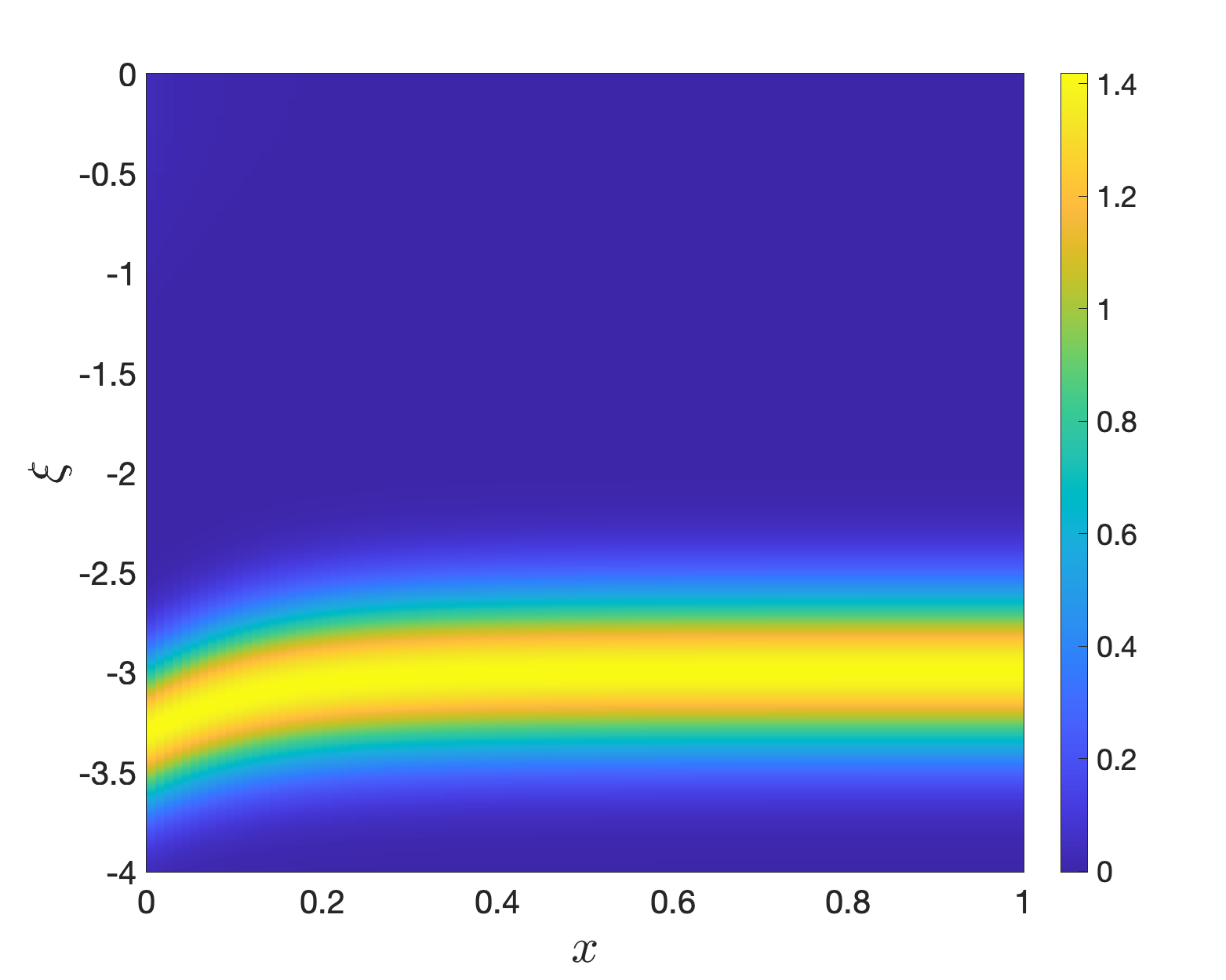}} \\
  {\small $f^0$} &   \resizebox{38mm}{!}{\includegraphics{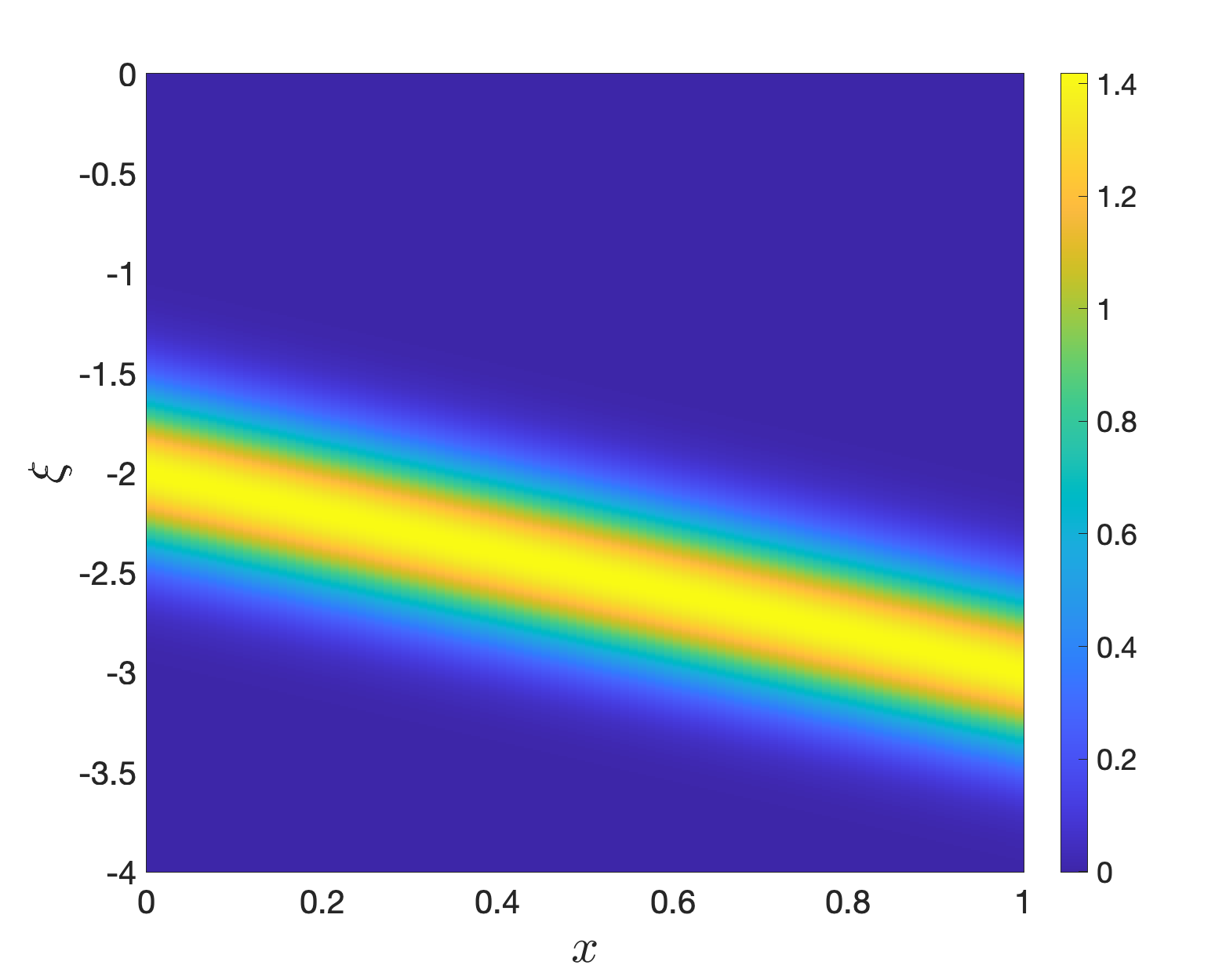}} & \resizebox{38mm}{!}{\includegraphics{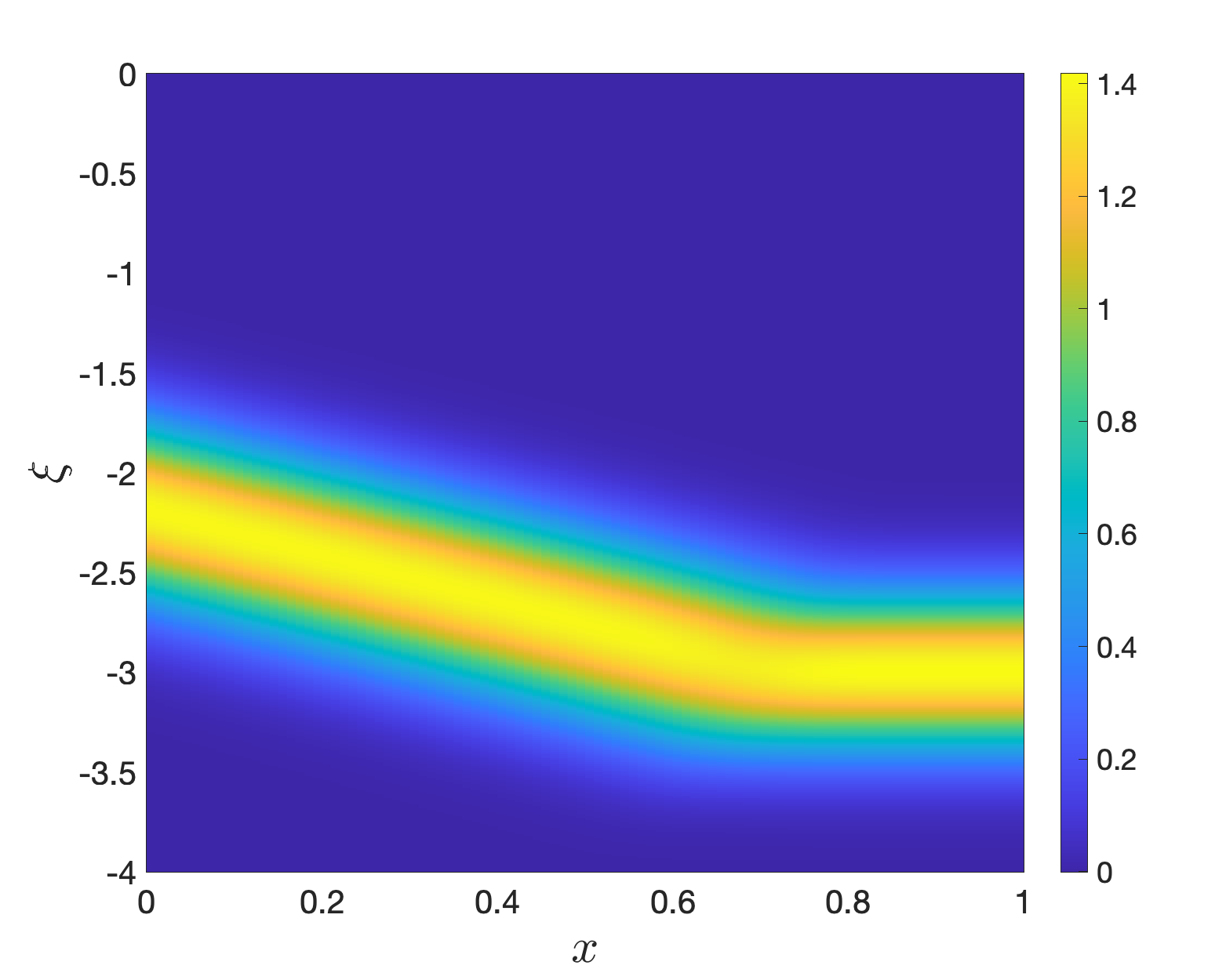}} & \resizebox{38mm}{!}{\includegraphics{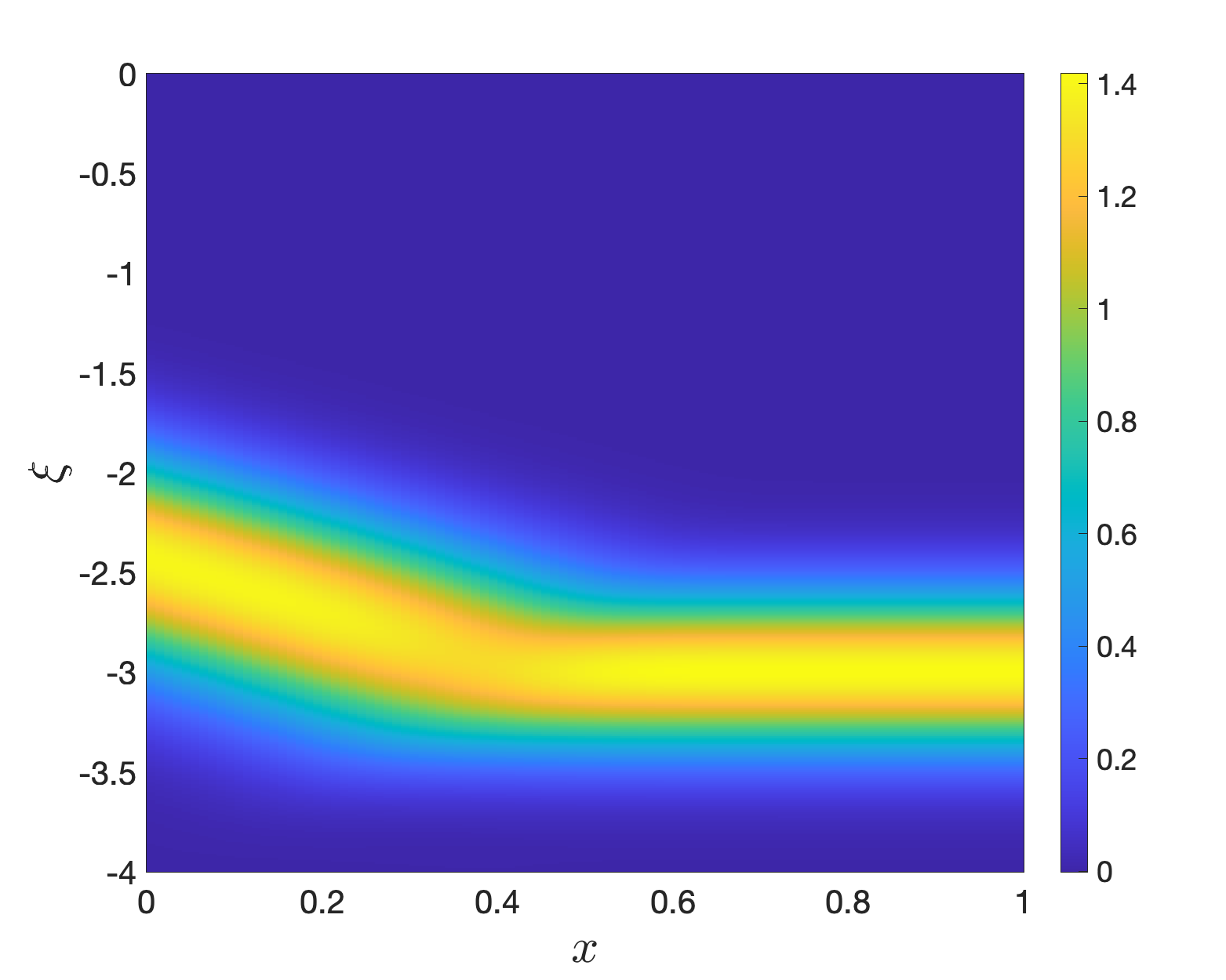}} & \resizebox{38mm}{!}{\includegraphics{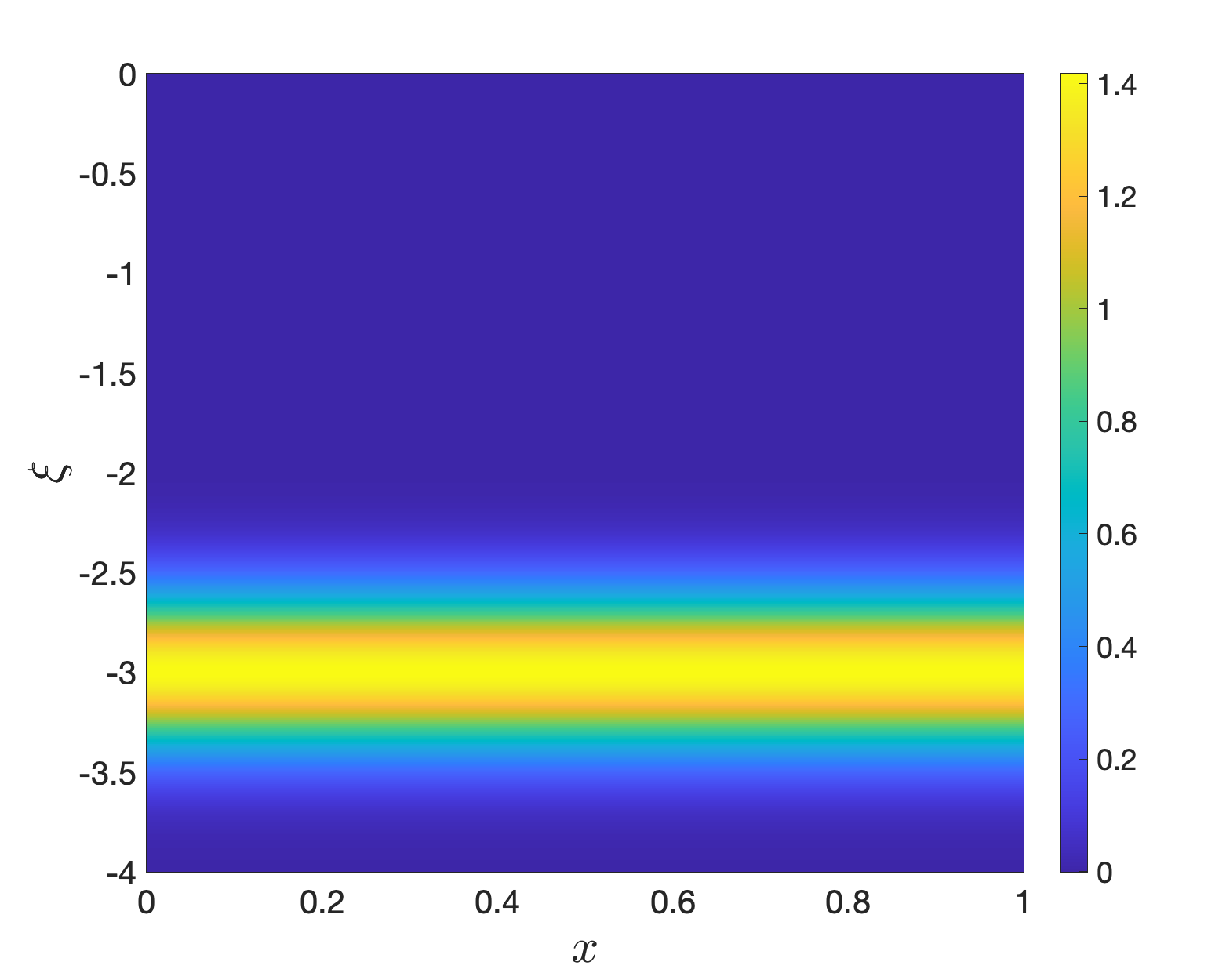}} \\
{\small $F^0$} &     \resizebox{38mm}{!}{\includegraphics{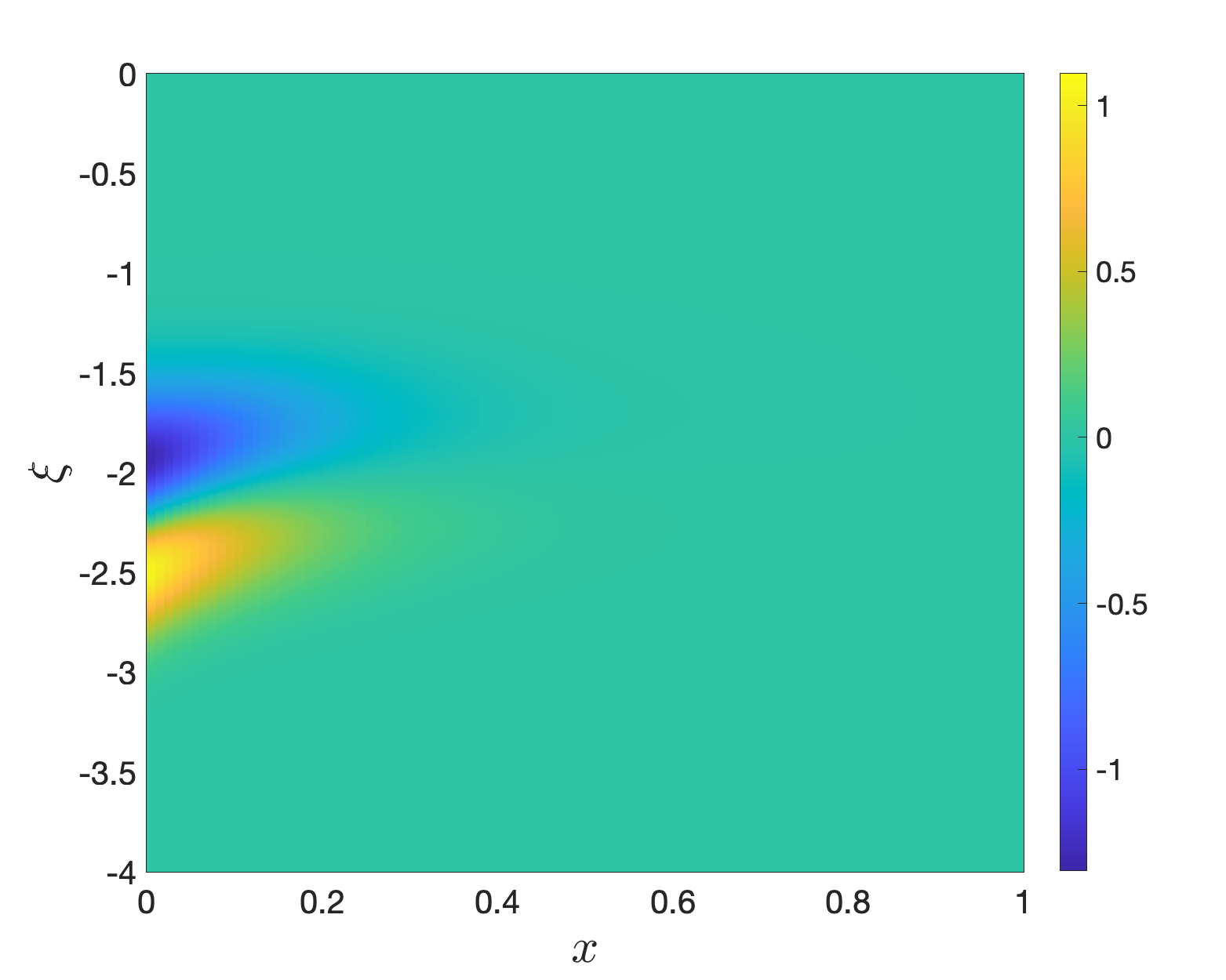}} & \resizebox{38mm}{!}{\includegraphics{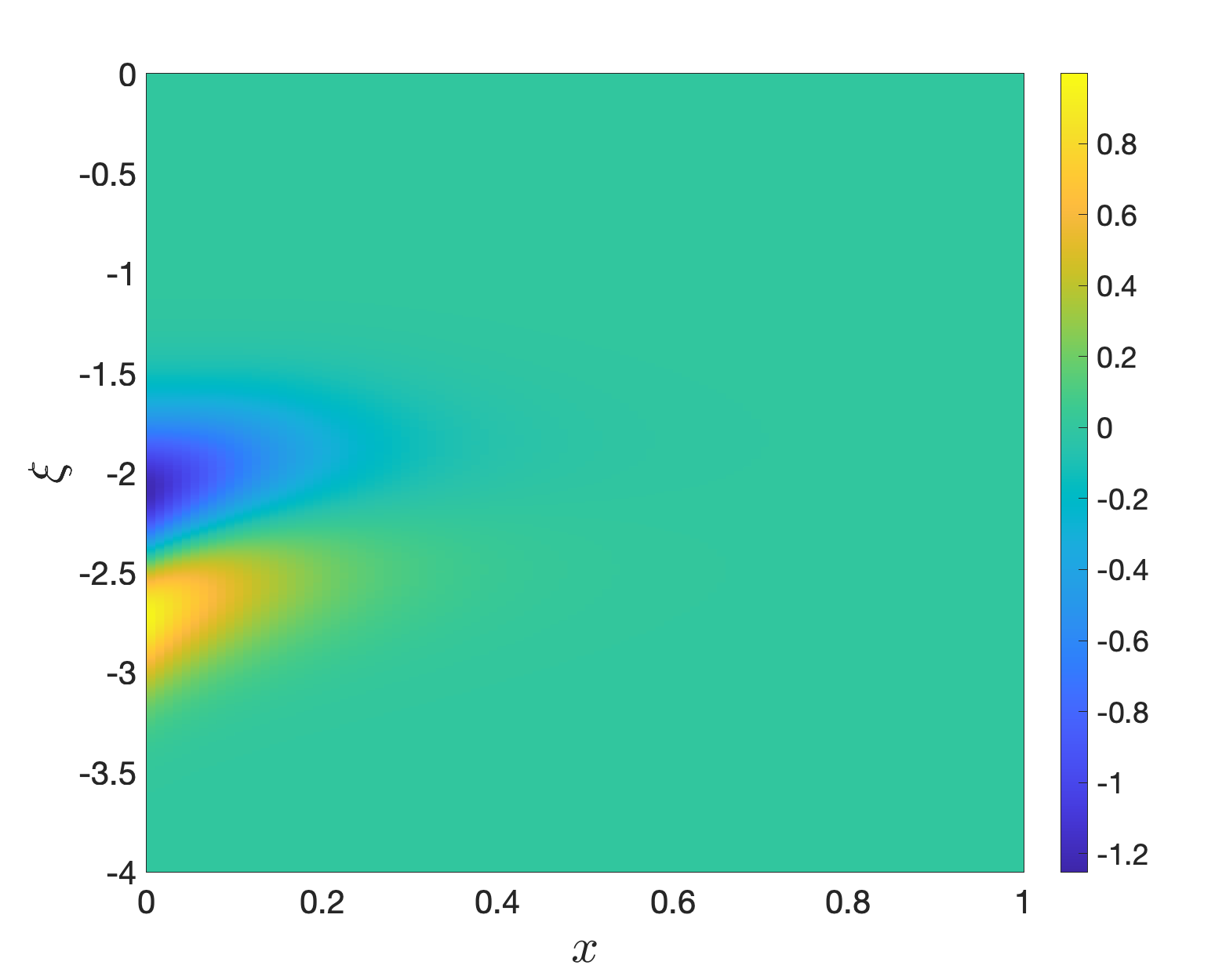}} & \resizebox{38mm}{!}{\includegraphics{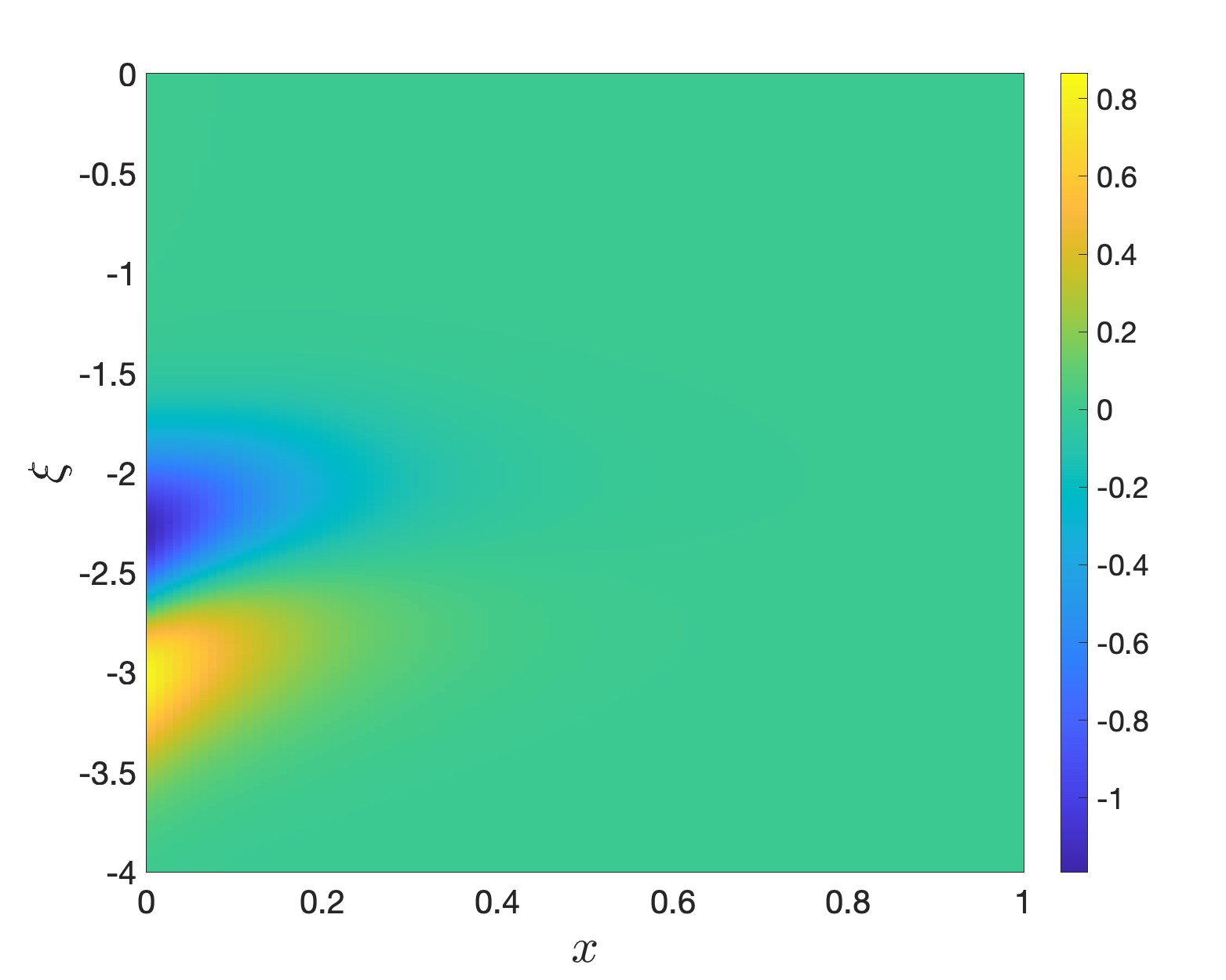}} & \resizebox{38mm}{!}{\includegraphics{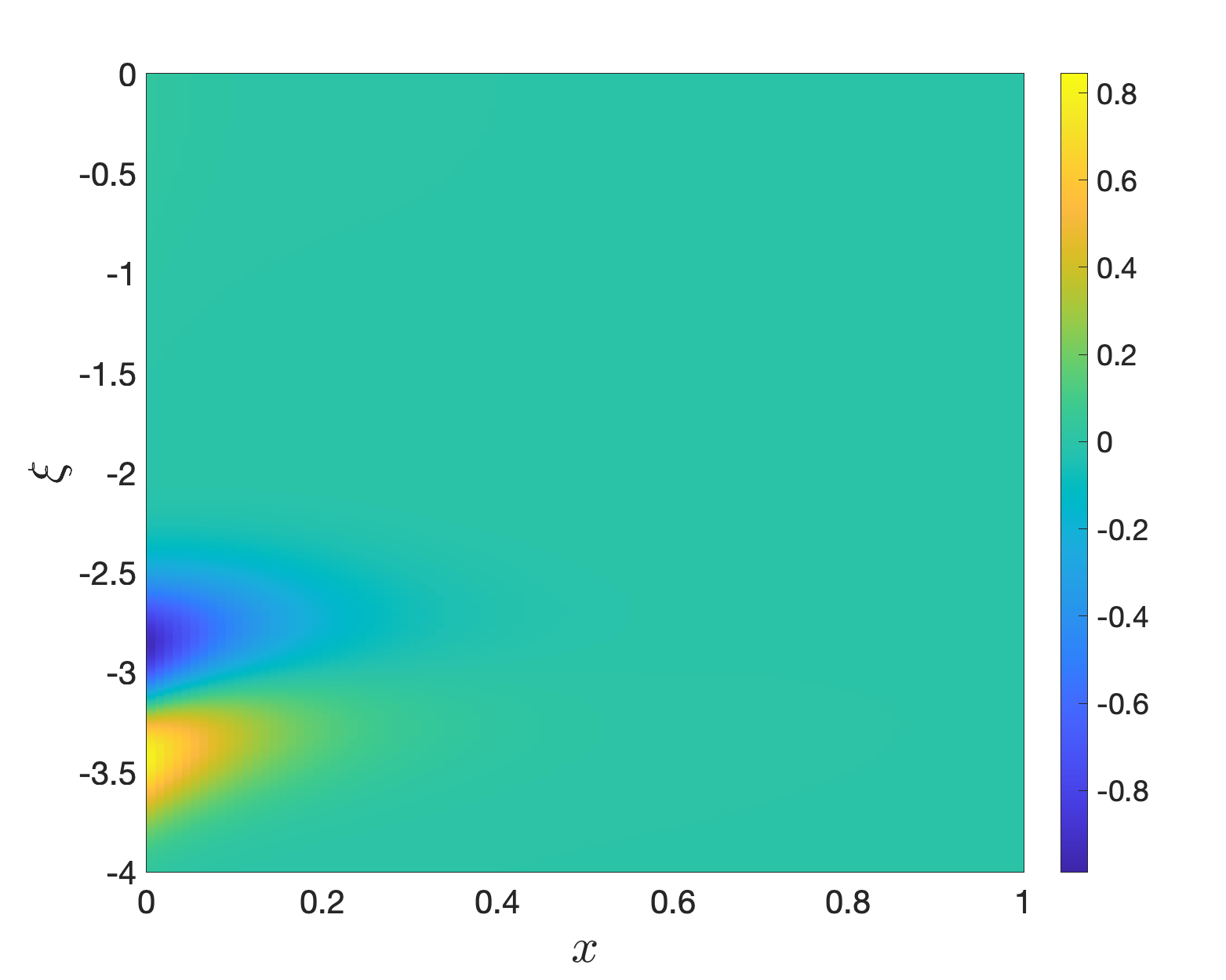}} 
  \end{tabular}  
  \caption{\small{The plots of $f$, $f^0$, and $F^0$ for $\ep = 10^{-2}$  with the initial and boundary conditions in Case (II).}}\label{ex4_f}
\vspace{5mm}
  \centering
  \begin{tabular}{m{1mm}m{34mm}m{34mm}m{34mm}m{34mm}}
\hspace{15mm} & \hspace{15mm} $t=0$ & \hspace{15mm} $t=0.1$ & \hspace{15mm} $t=0.2$ & \hspace{15mm} $t=1$\\ 
   {\small $\phi$} & \resizebox{38mm}{!}{\includegraphics{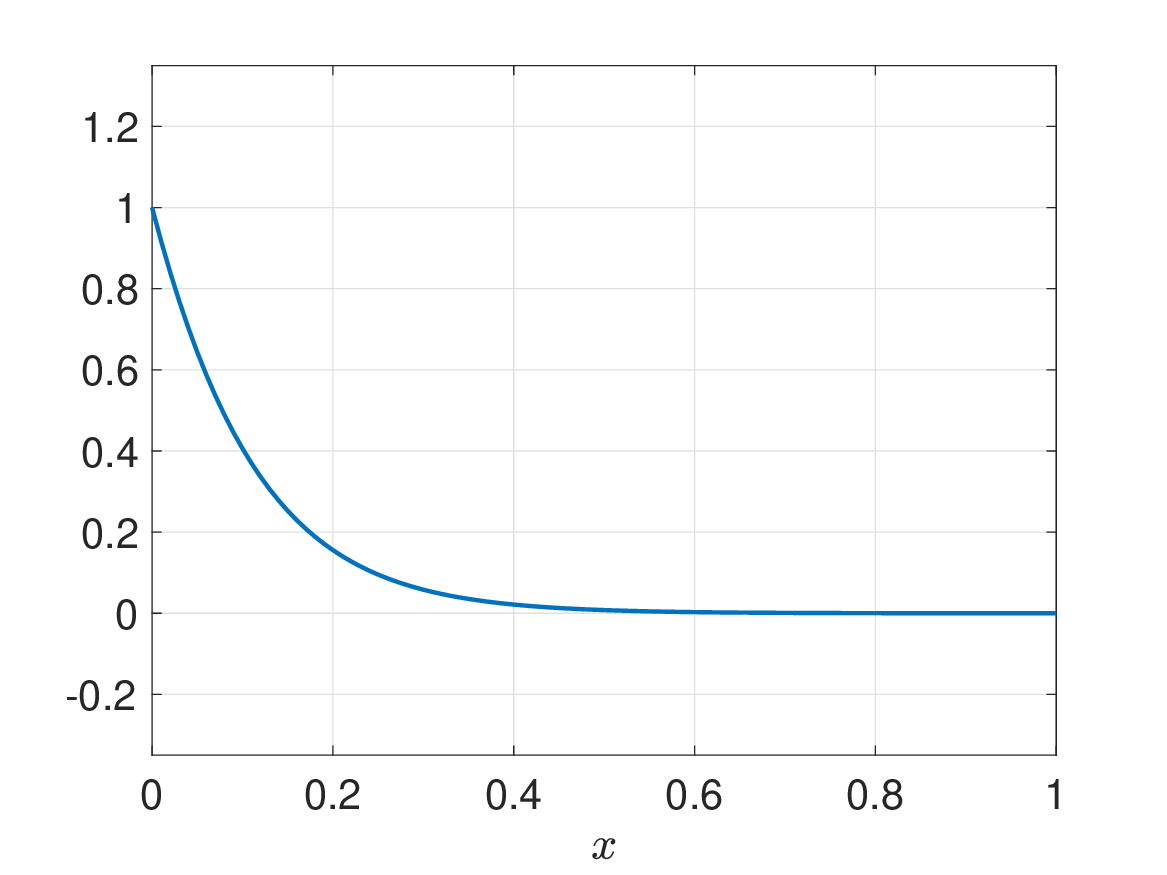}} & \resizebox{38mm}{!}{\includegraphics{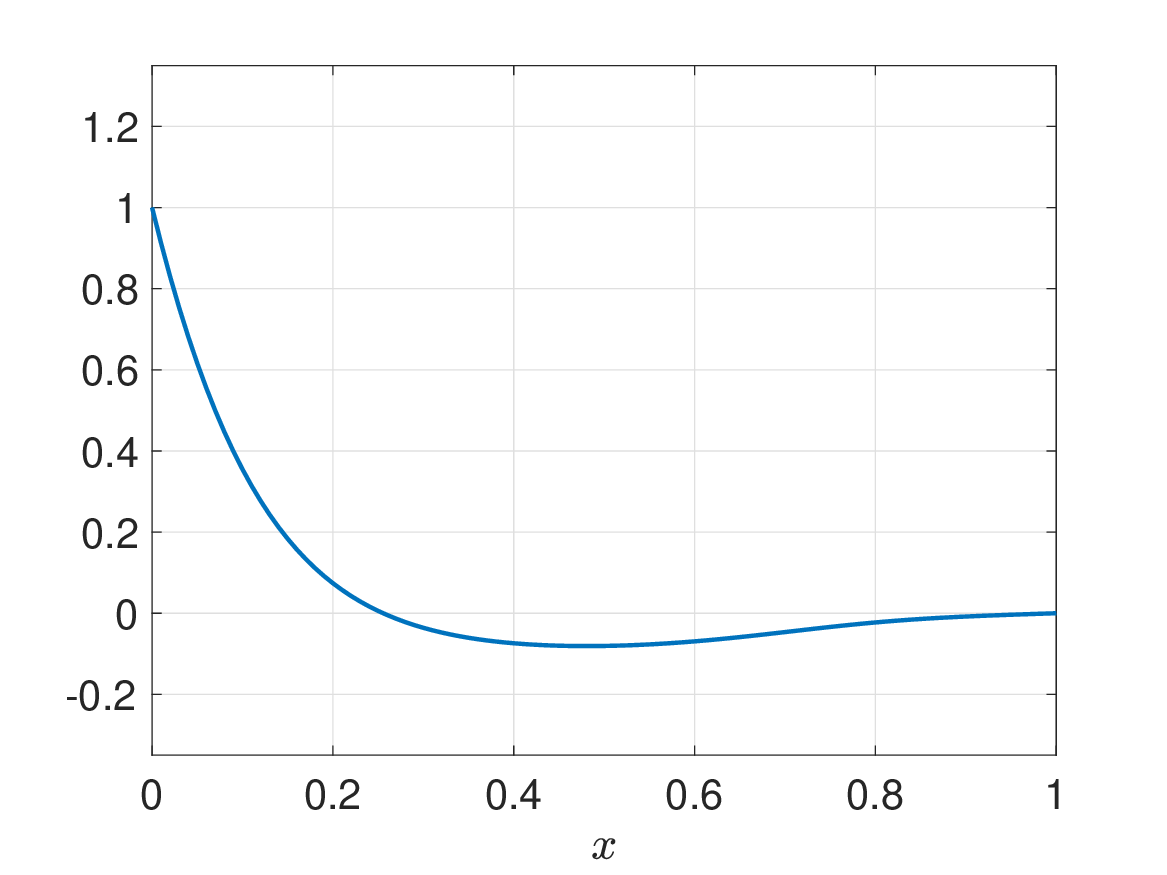}} & \resizebox{38mm}{!}{\includegraphics{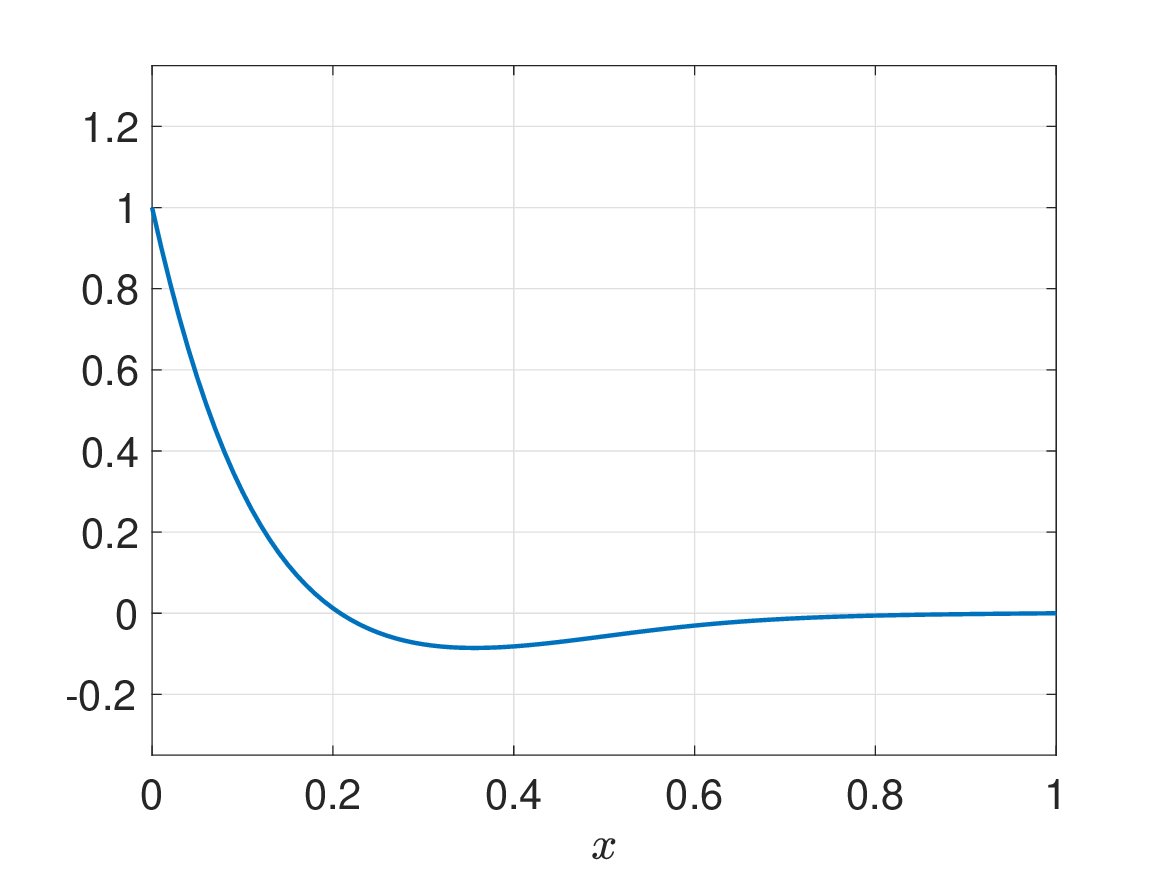}} & \resizebox{38mm}{!}{\includegraphics{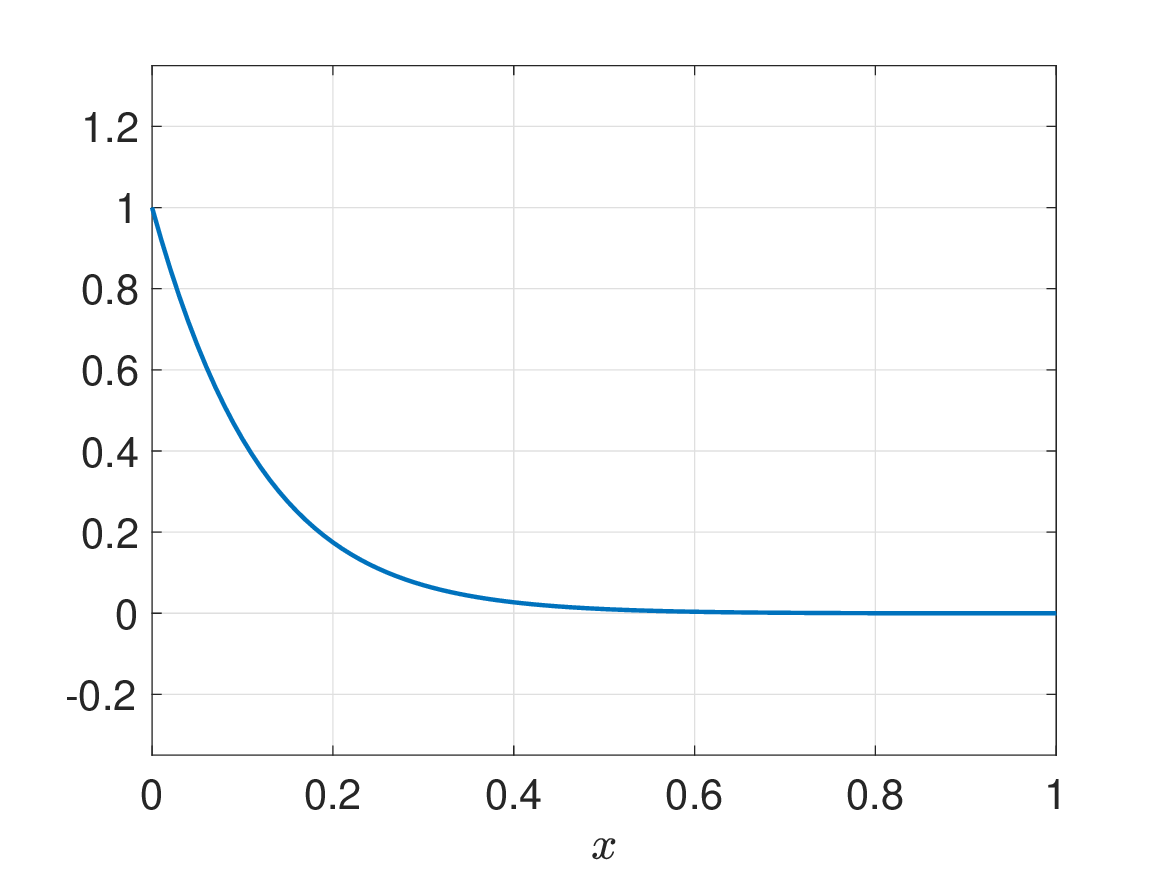}} \\
  {\small $\phi^0$} &   \resizebox{38mm}{!}{\includegraphics{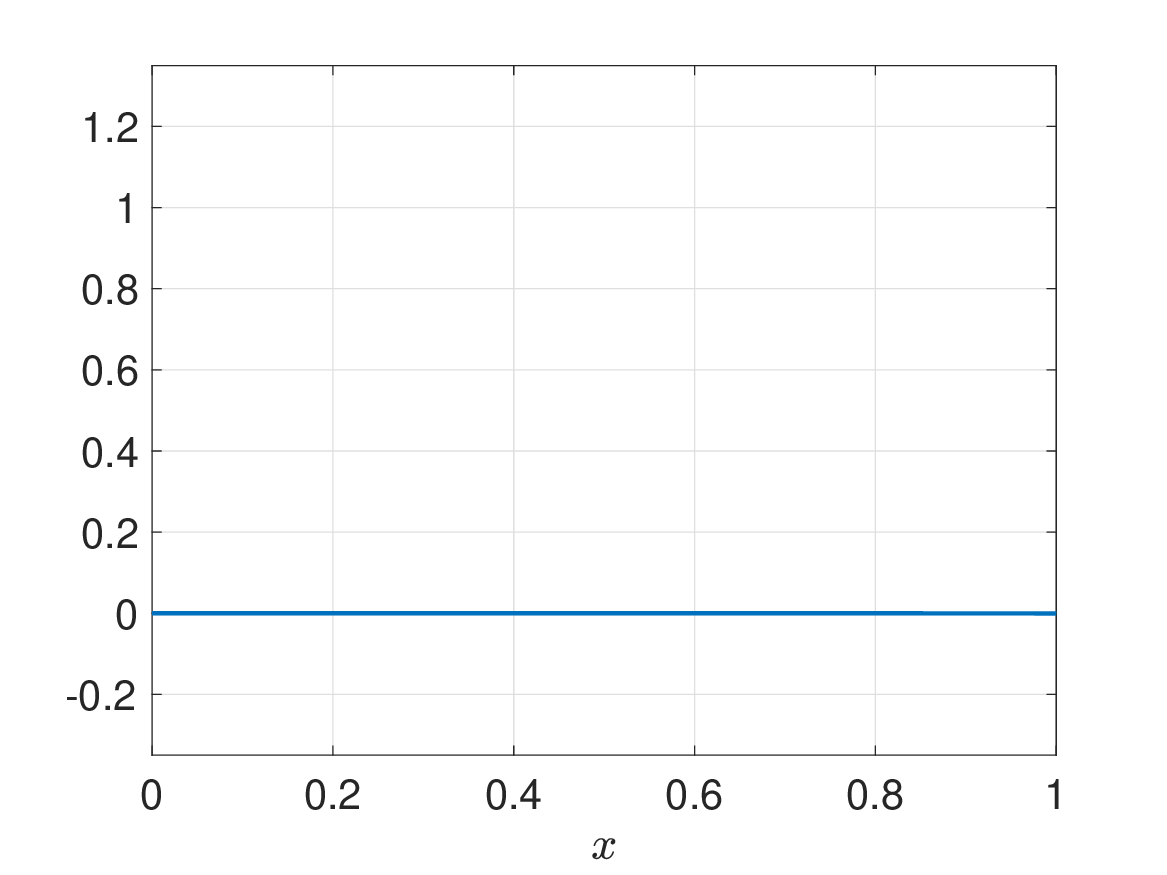}} & \resizebox{38mm}{!}{\includegraphics{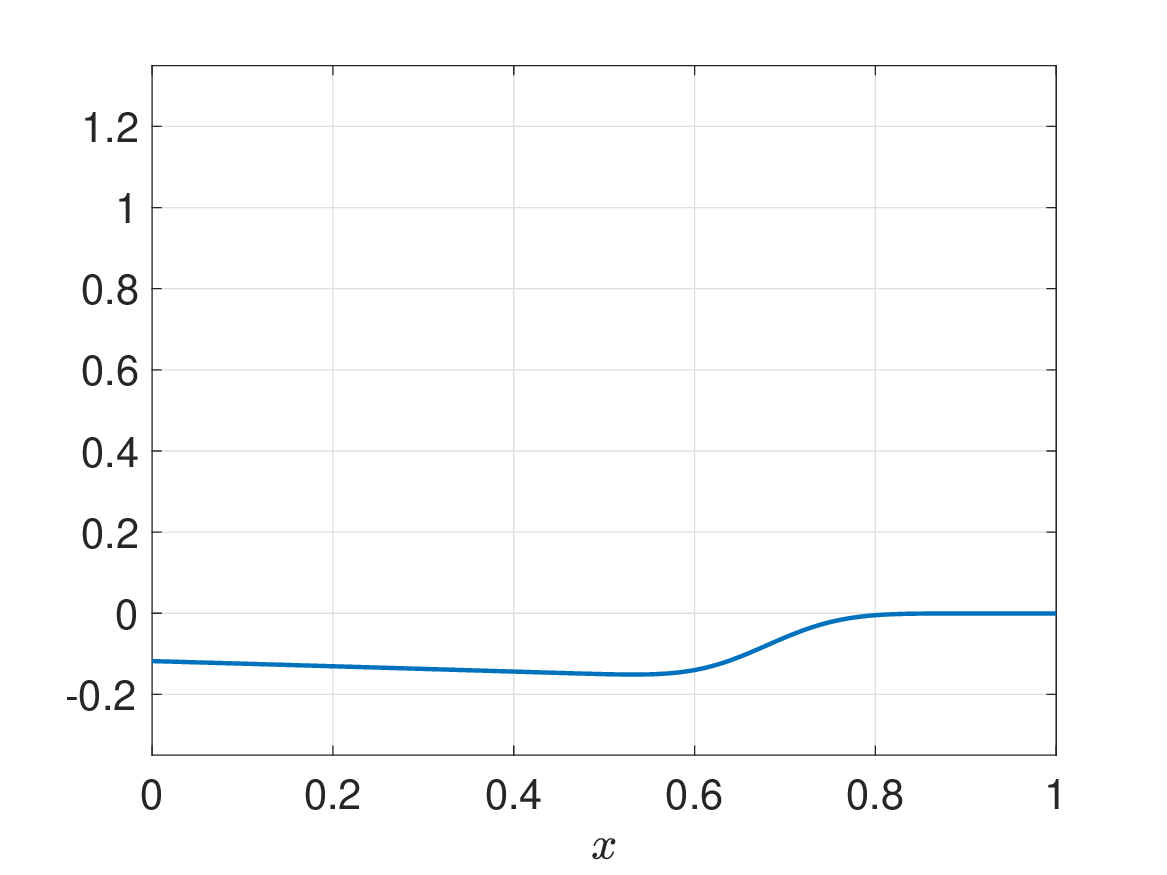}} & \resizebox{38mm}{!}{\includegraphics{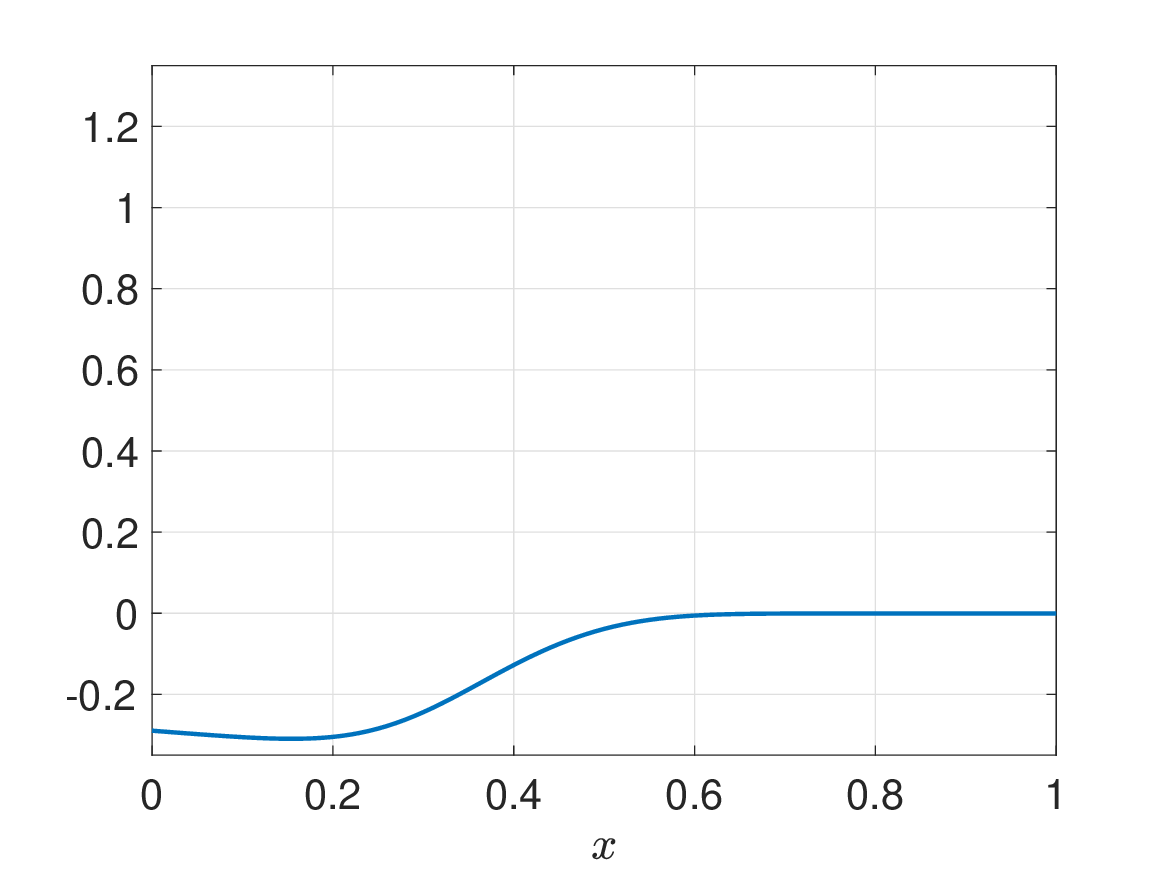}} & \resizebox{38mm}{!}{\includegraphics{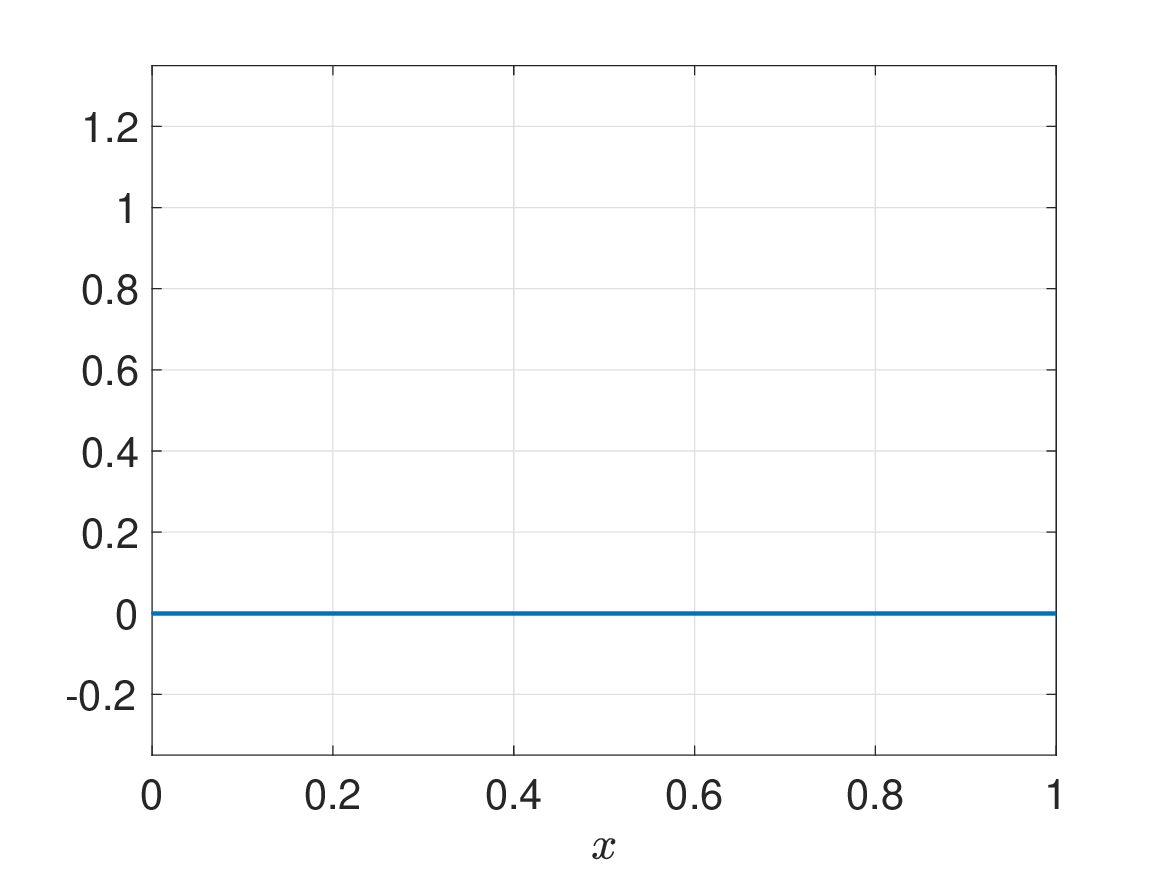}} \\
{\small $\Phi^0$} &     \resizebox{38mm}{!}{\includegraphics{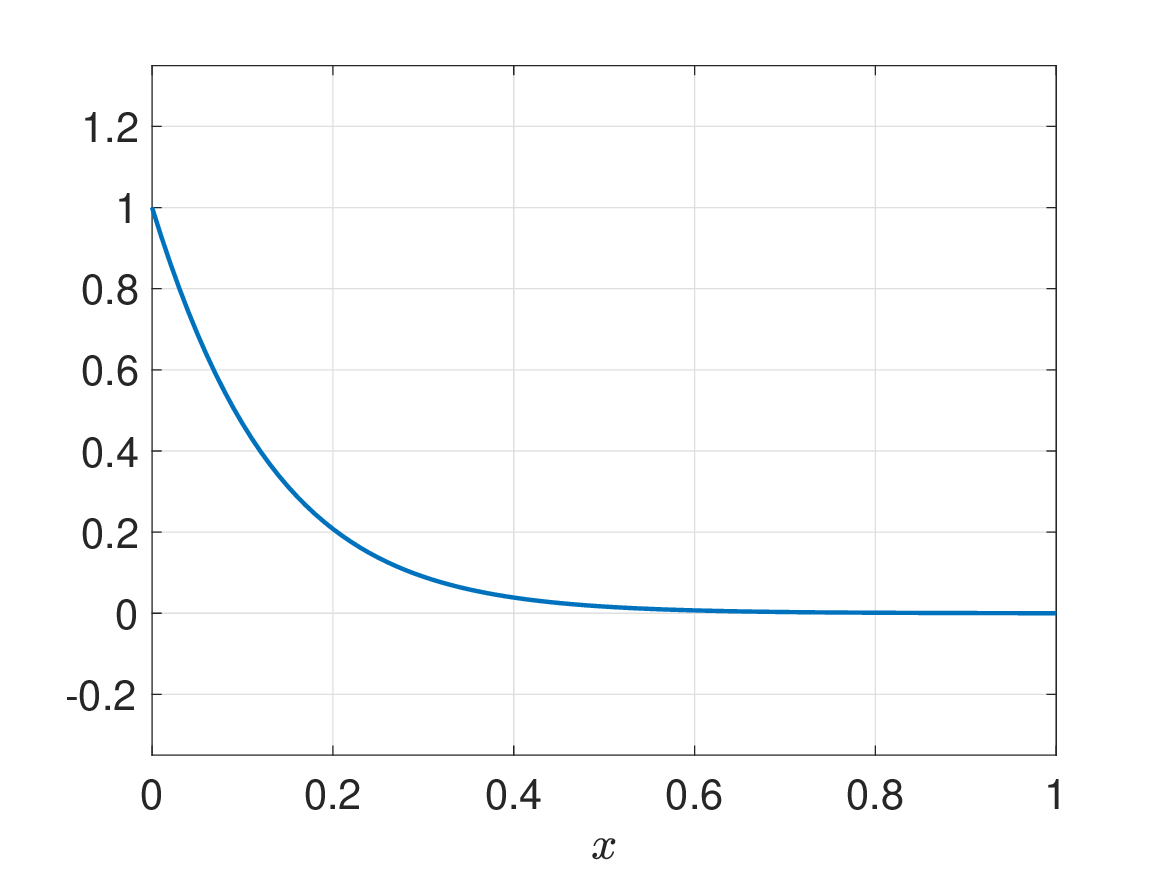}} & \resizebox{38mm}{!}{\includegraphics{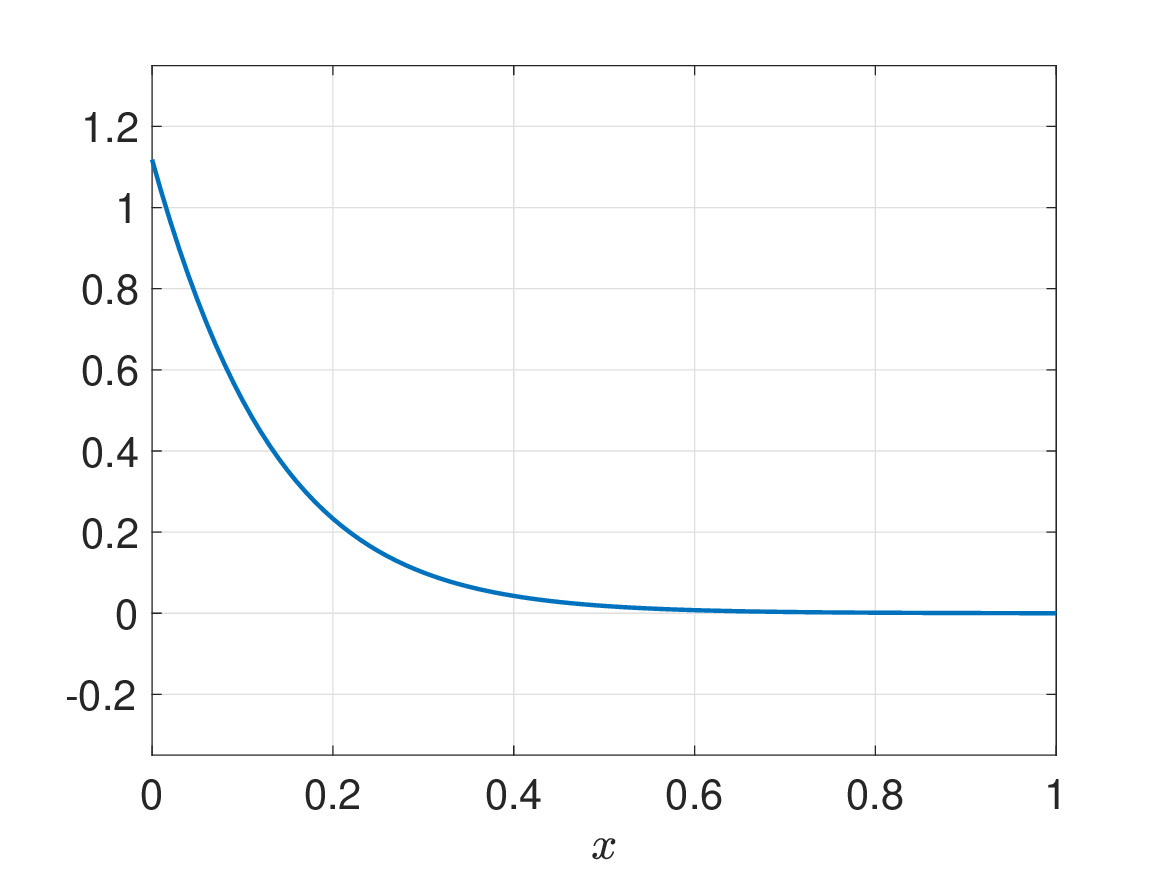}} & \resizebox{38mm}{!}{\includegraphics{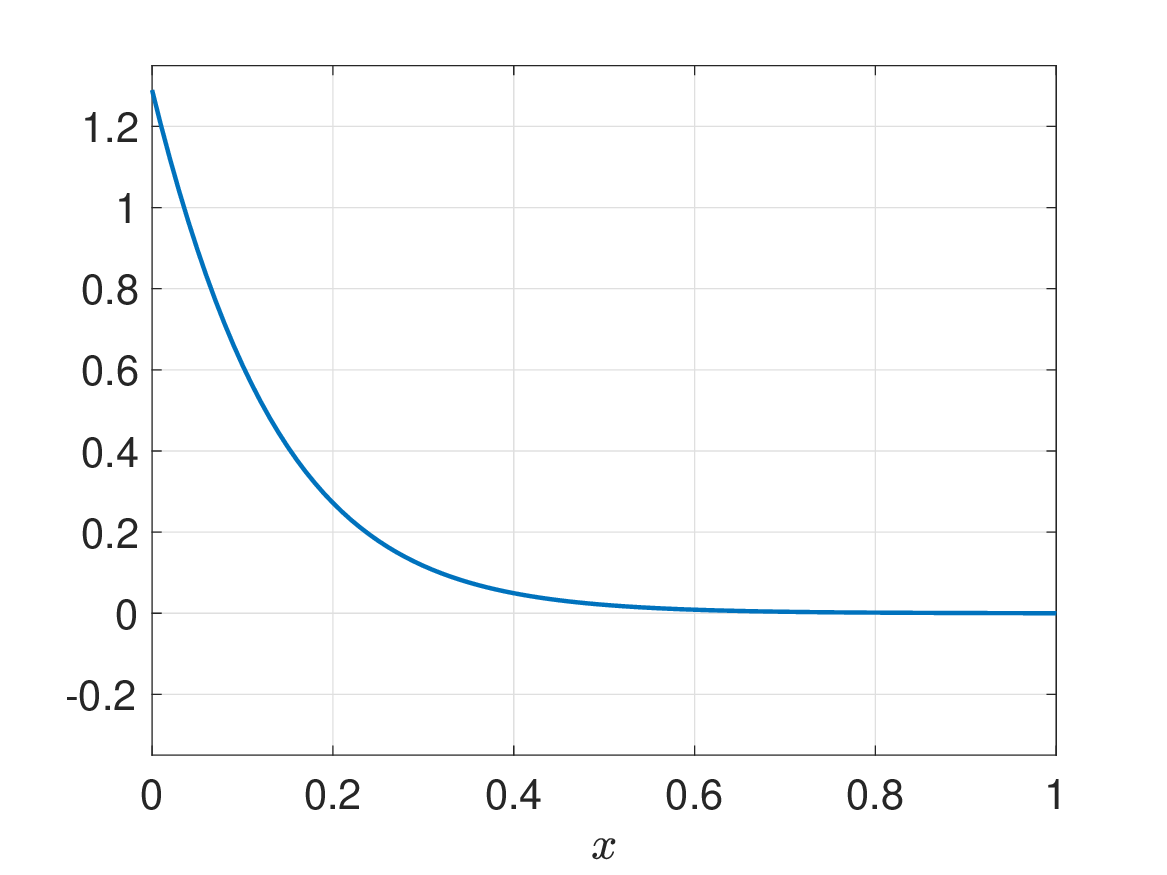}} & \resizebox{38mm}{!}{\includegraphics{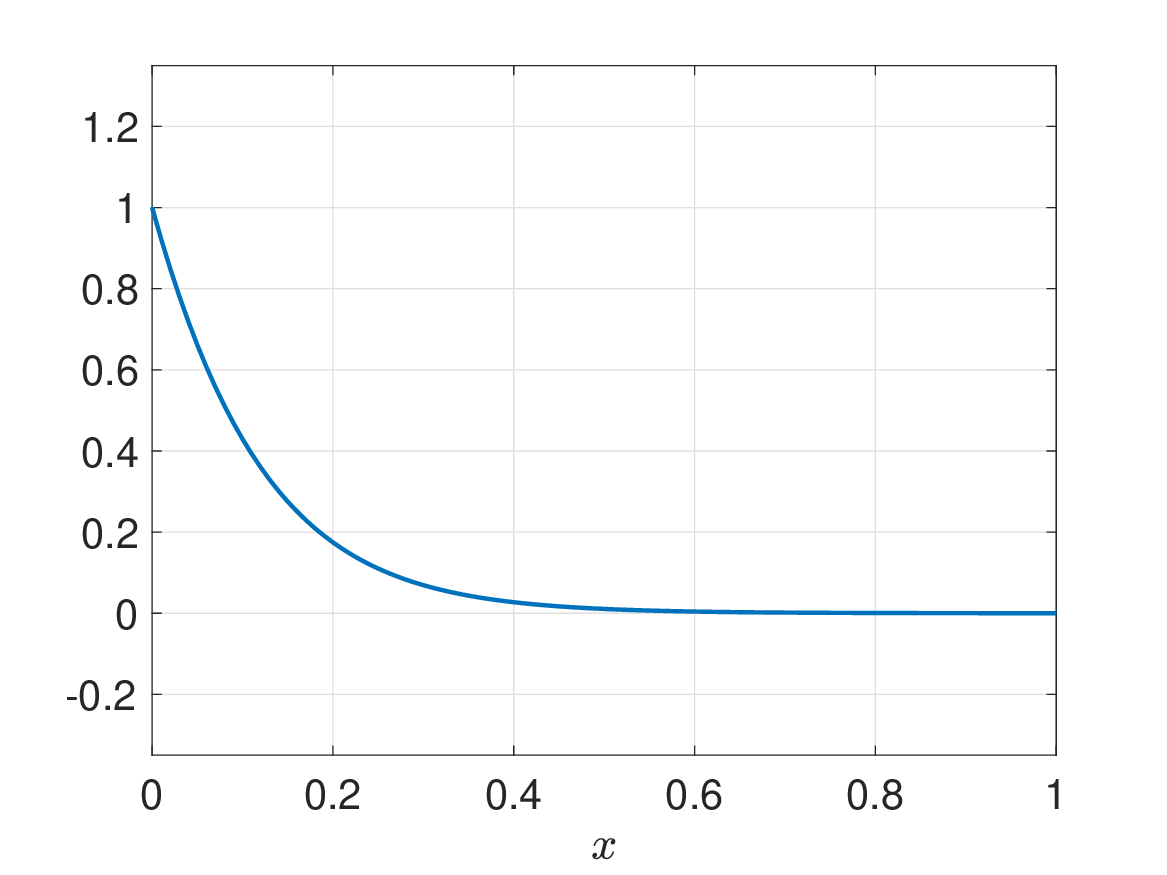}} 
  \end{tabular}  
  \caption{\small{The plots of $\phi$, $\phi^0$, and $\Phi^0$ for $\ep = 10^{-2}$  with the initial and boundary conditions in Case (II).}}\label{ex4_phi}
\end{figure}

\begin{figure}[h]
  \centering
  \begin{tabular}{m{2mm}m{43mm}m{43mm}m{43mm}}
   {\footnotesize (I)} & \resizebox{48mm}{!}{\includegraphics{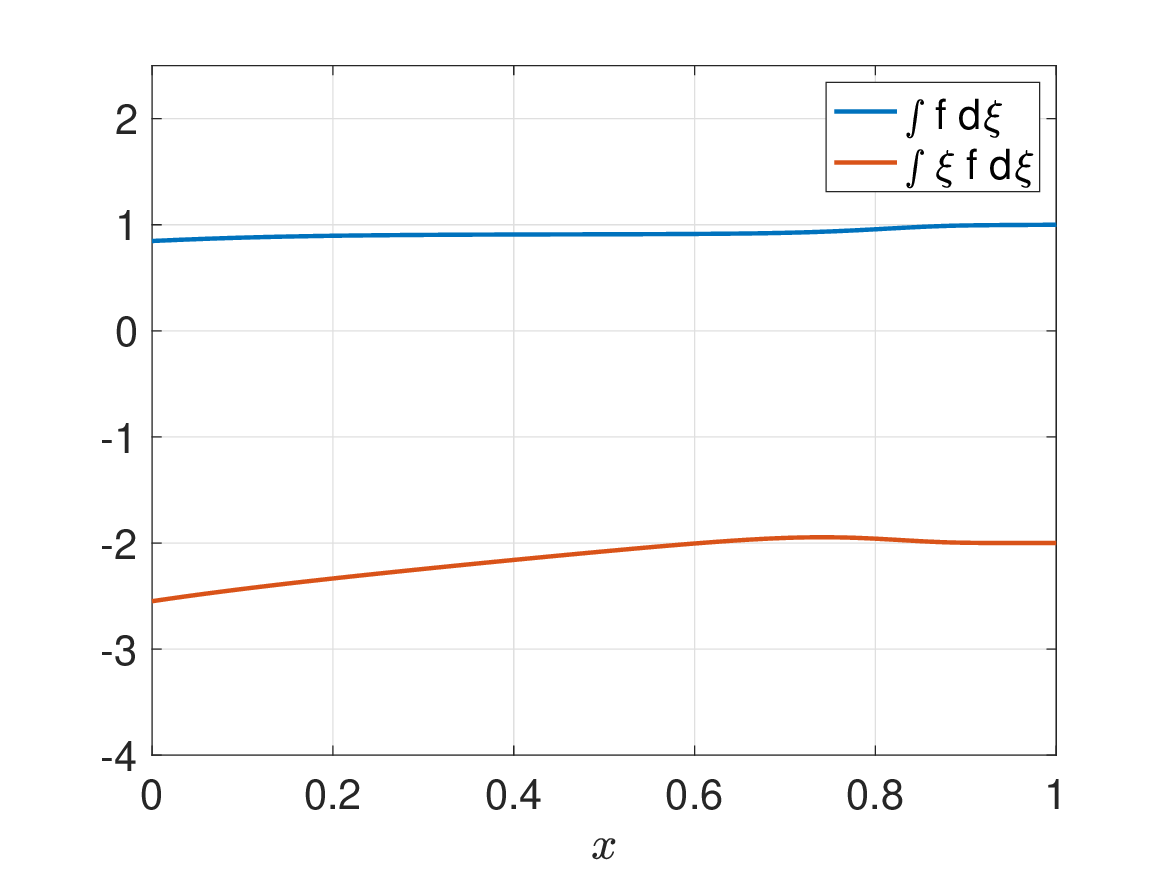}} & \resizebox{48mm}{!}{\includegraphics{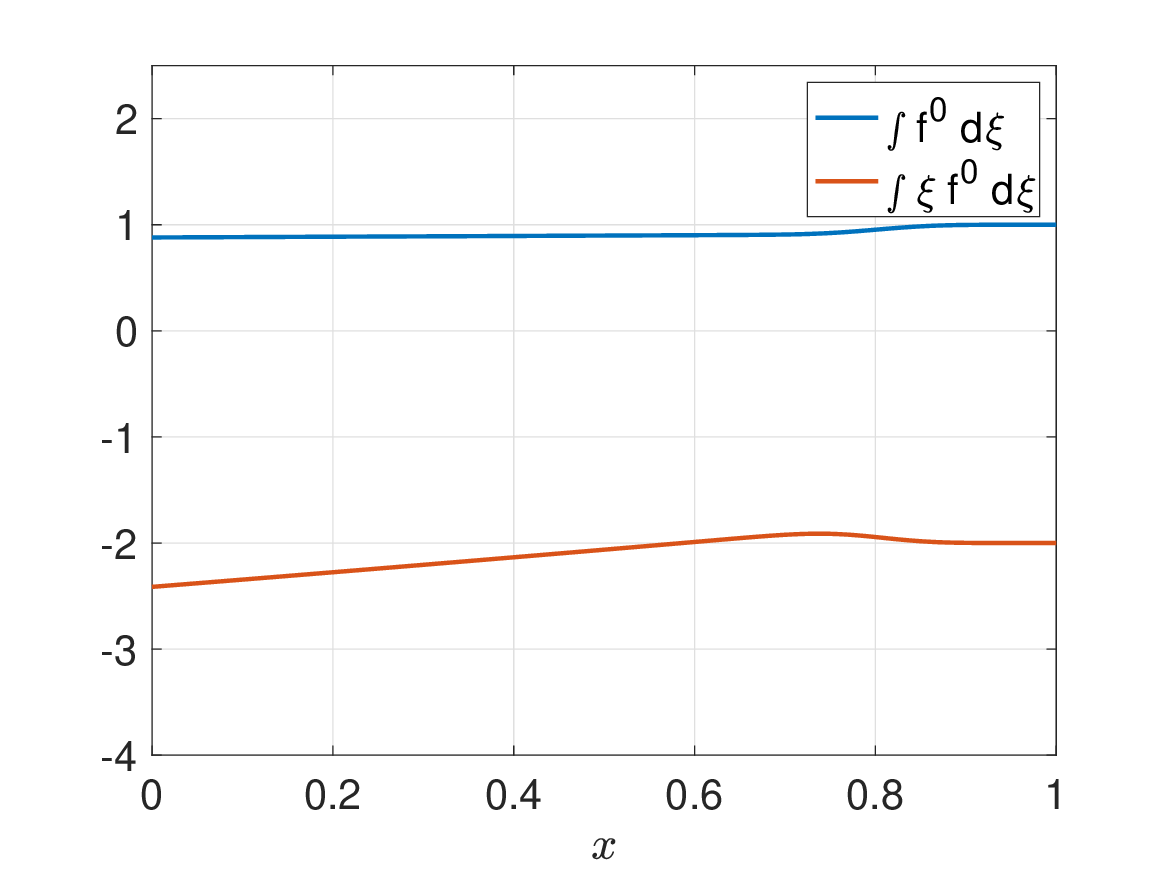}} & \resizebox{48mm}{!}{\includegraphics{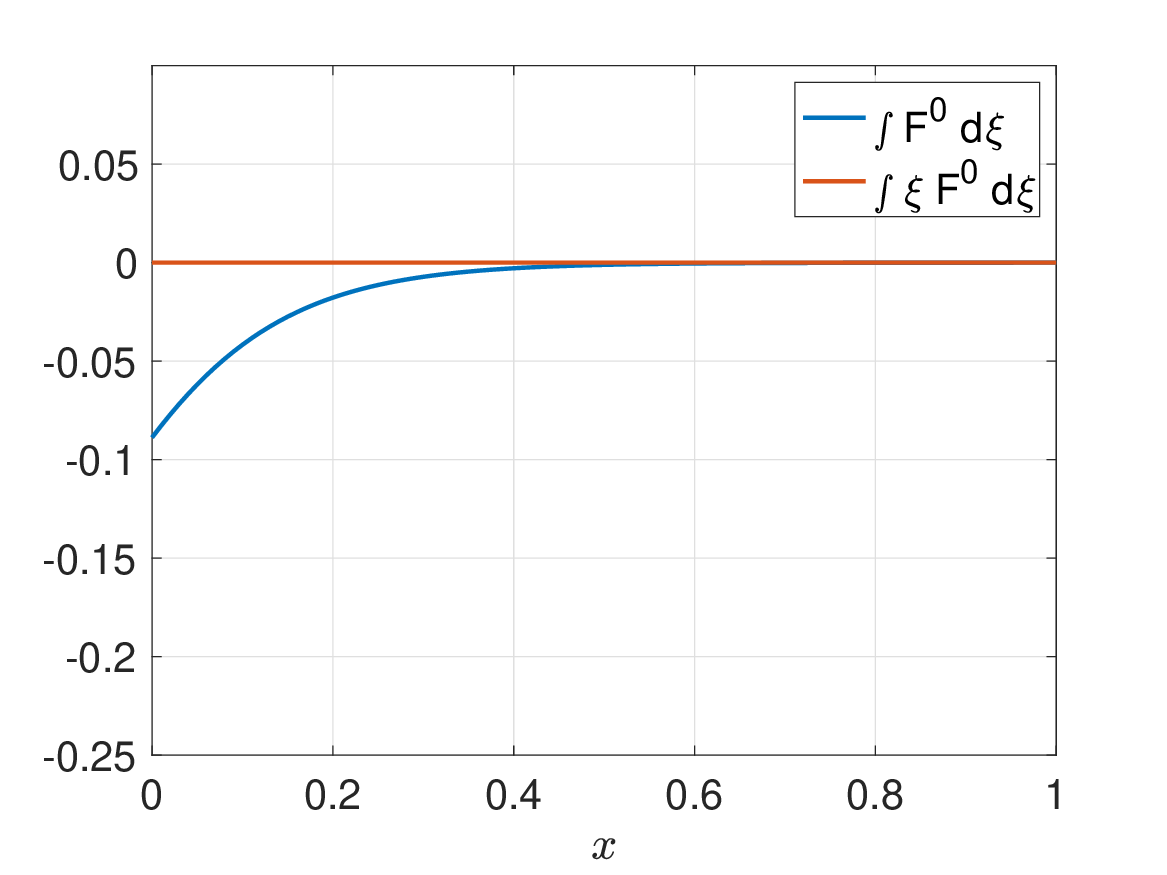}} \\
  {\footnotesize (II)} & \resizebox{48mm}{!}{\includegraphics{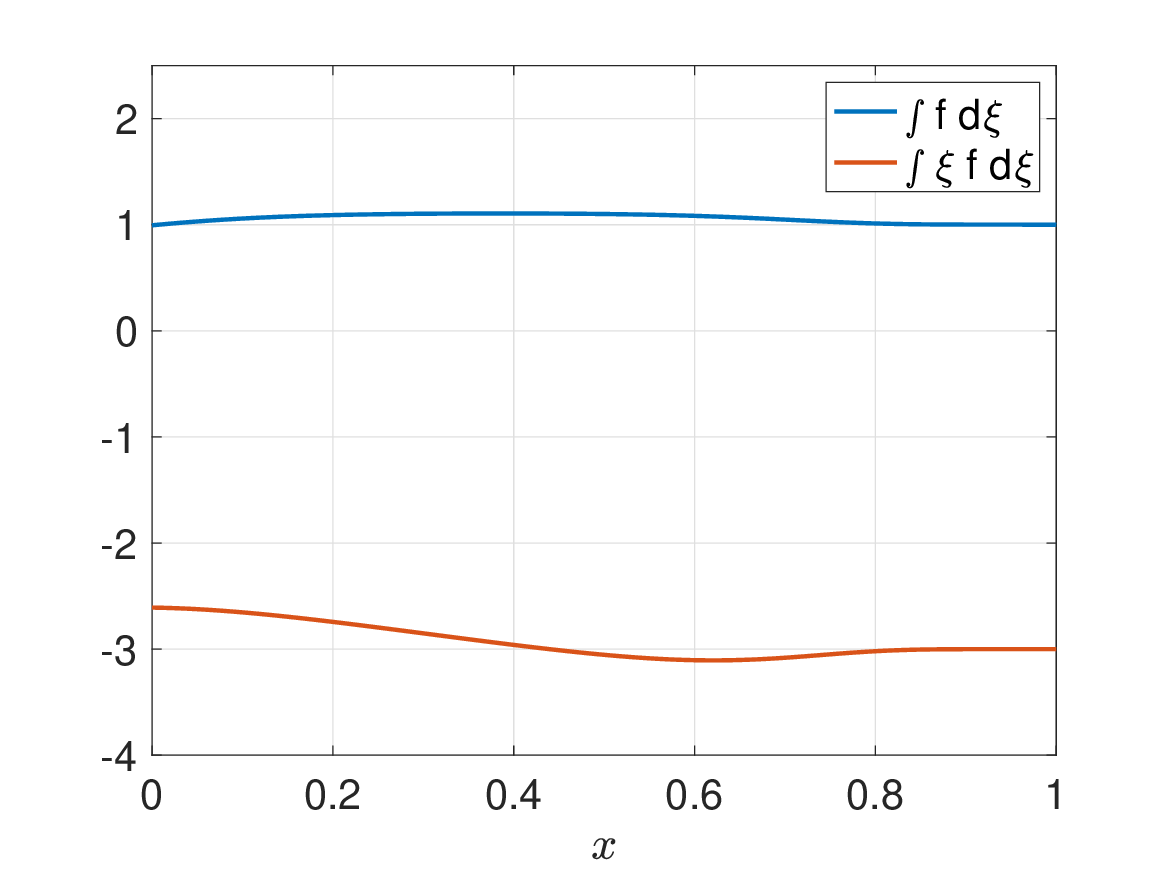}} & \resizebox{48mm}{!}{\includegraphics{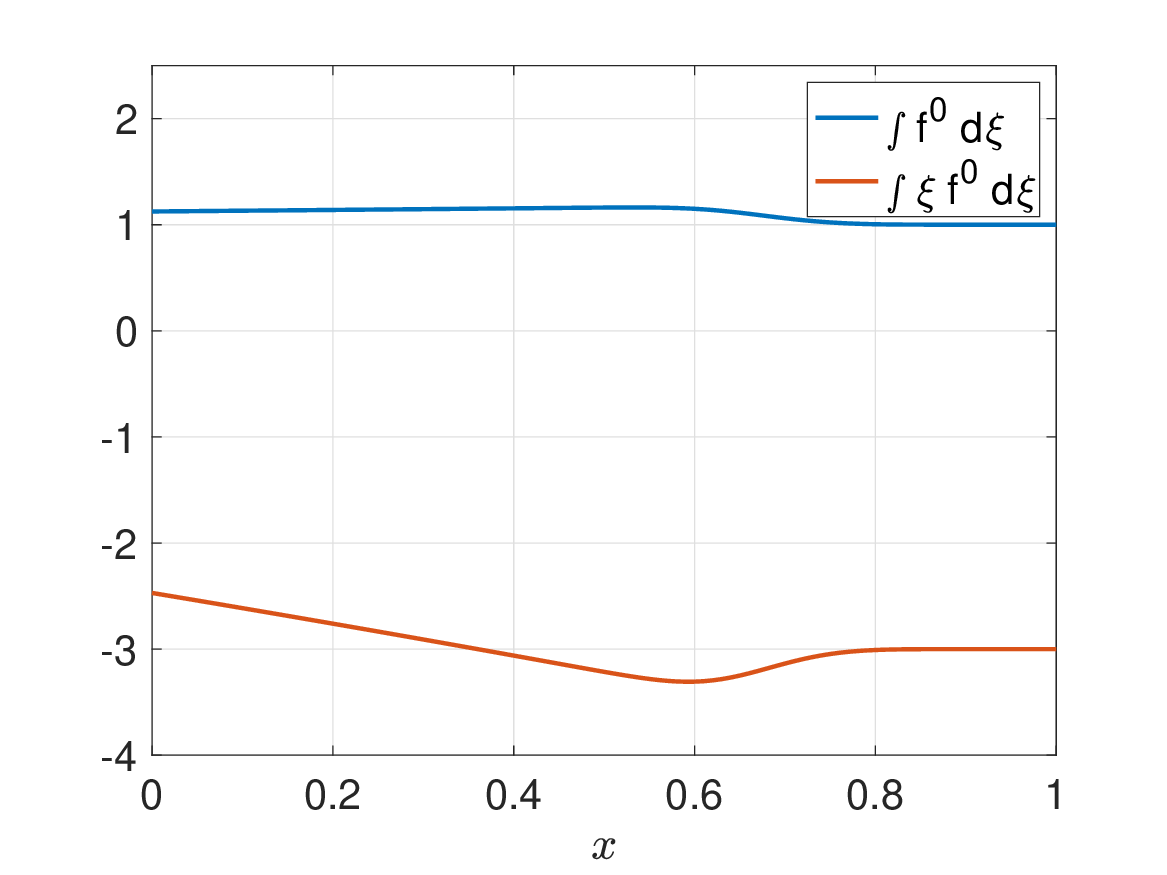}} & \resizebox{48mm}{!}{\includegraphics{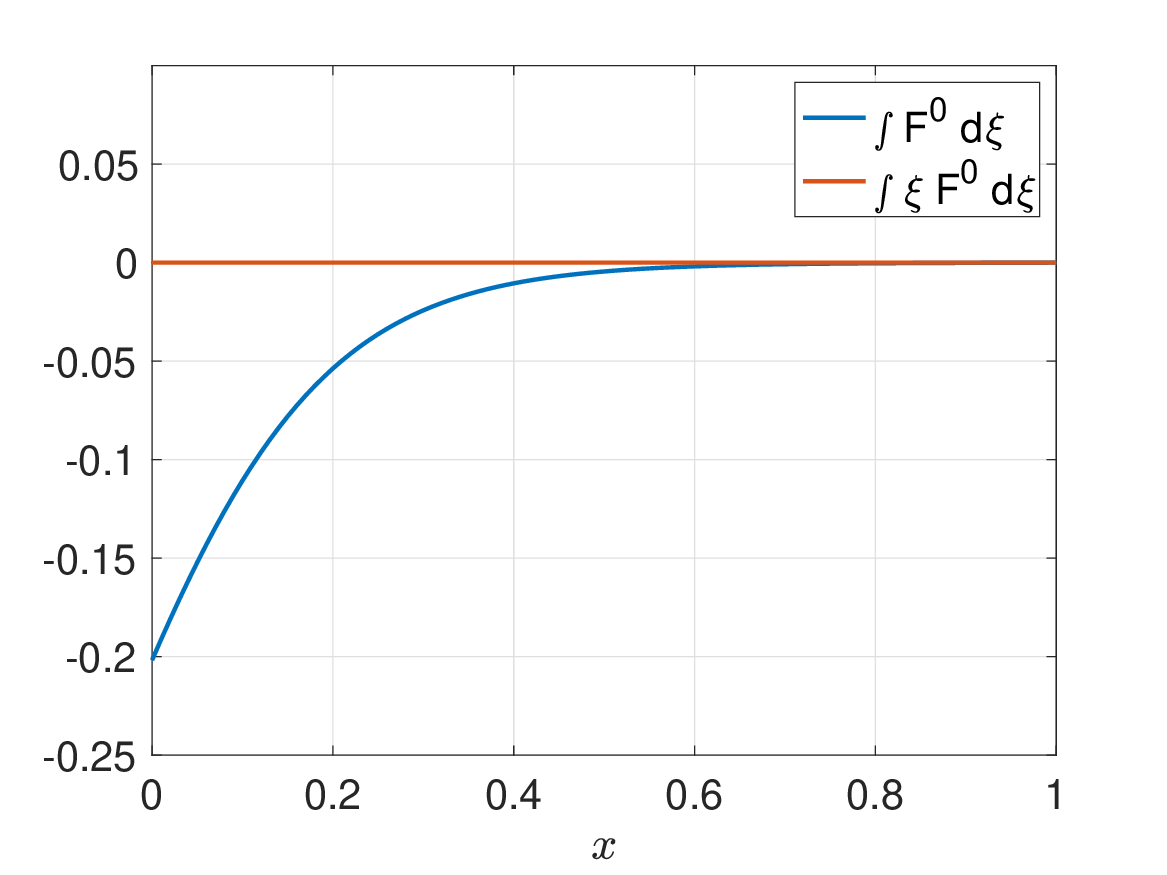}} 
  \end{tabular}  
  \caption{\small{The plots of $\int (f,f^0,F^0) d\xi$ and  $\int \xi (f,f^0,F^0) d\xi$ at $t=0.1$ for $\ep = 10^{-2}$ with the initial and boundary conditions in Cases (I) and (II).}}\label{ex3_ex4}
\end{figure}

\begin{figure}[h]
  \begin{center}
\begin{tabular}{ccc}
$f(0,x,\xi)$ \hspace{-9mm} & $f(0.1,x,\xi)$ \hspace{-9mm} & $f(0.2,x,\xi)$\\
    \resizebox{50mm}{!}{\includegraphics{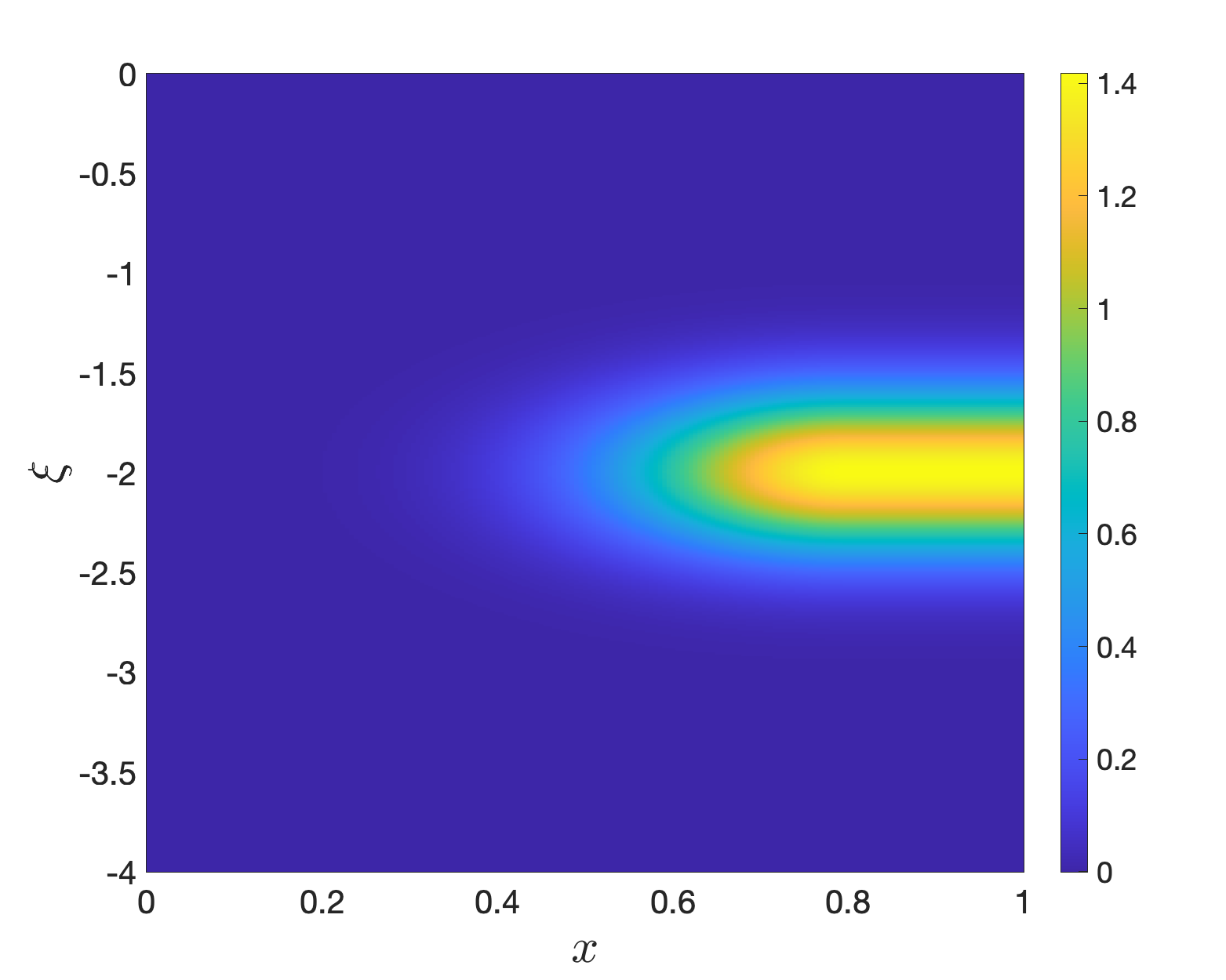}} \hspace{-9mm} & 
    \resizebox{50mm}{!}{\includegraphics{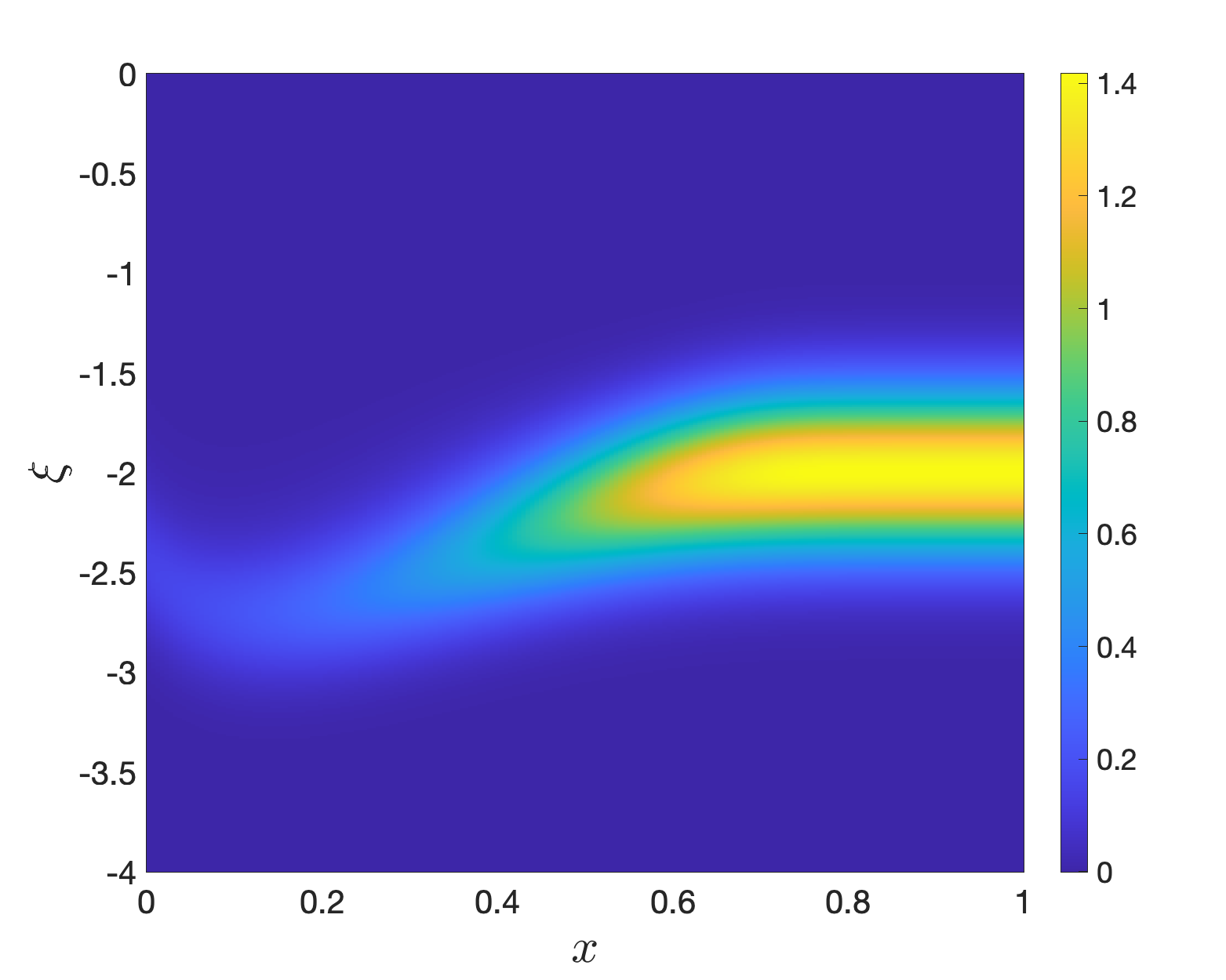}} \hspace{-9mm} &
    \resizebox{50mm}{!}{\includegraphics{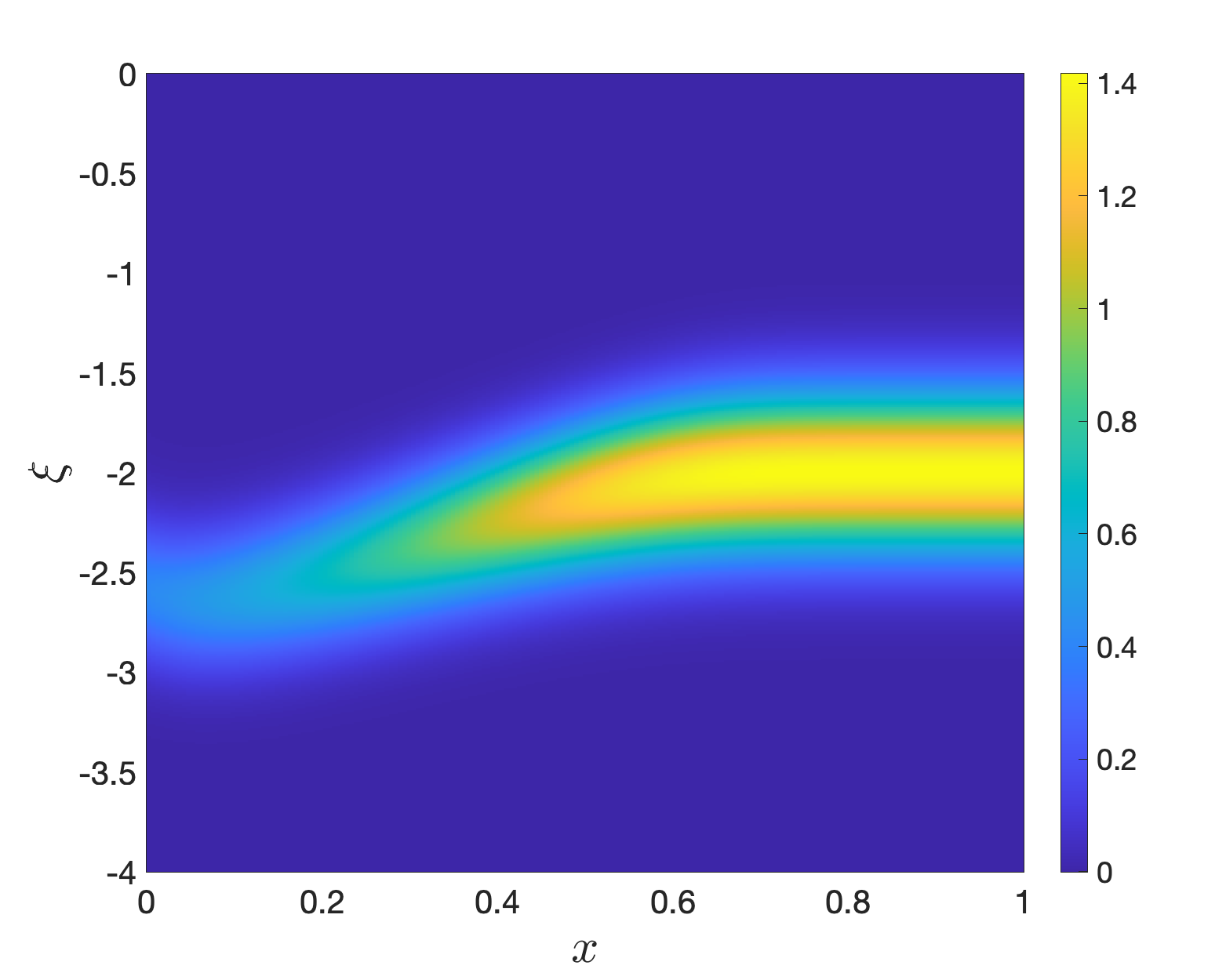}} \\
    $f(0.3,x,\xi)$ \hspace{-9mm} & $f(0.6,x,\xi)$ \hspace{-9mm} & $\tilde{f}(x,\xi)$ \\
    \resizebox{50mm}{!}{\includegraphics{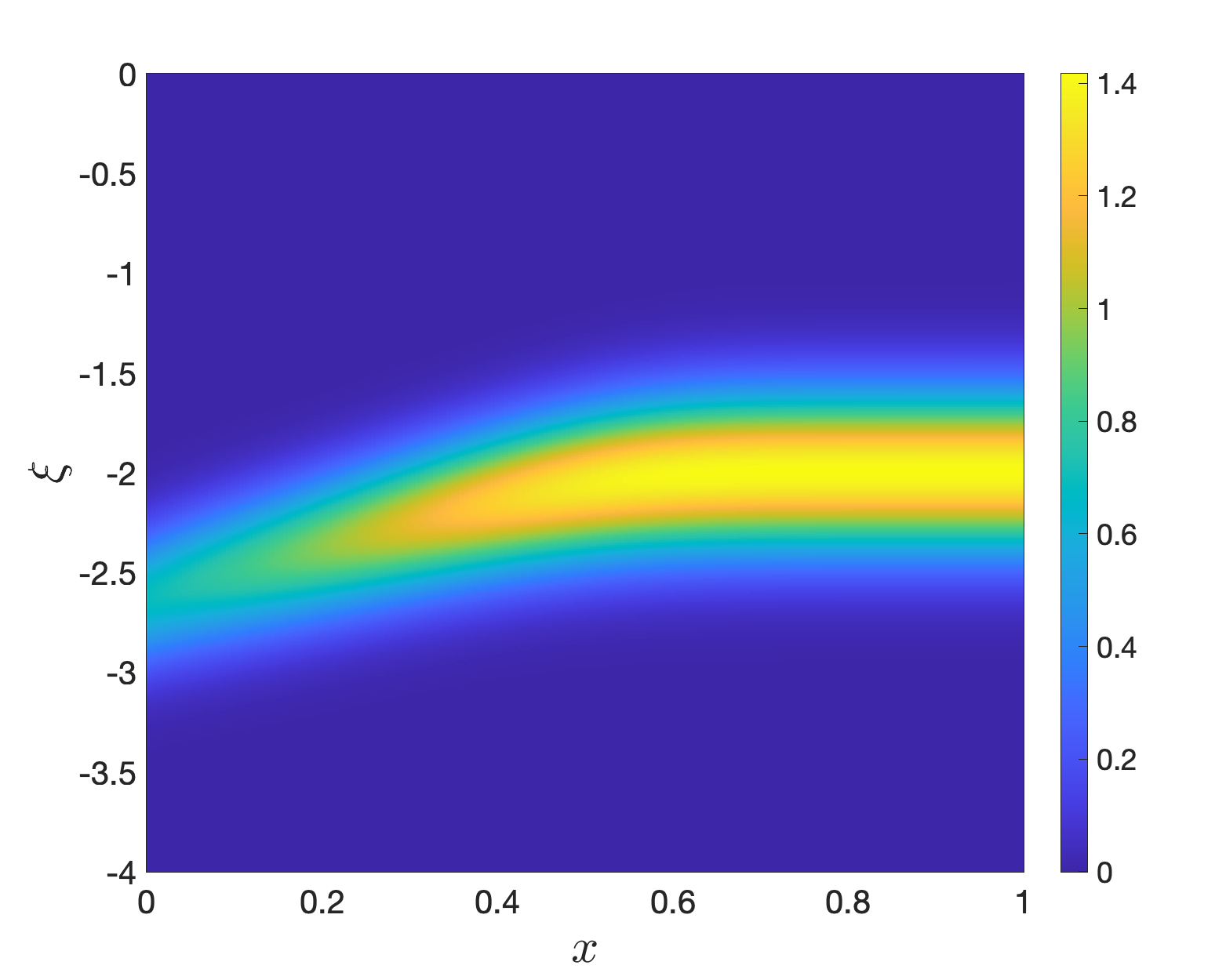}} \hspace{-9mm} &
    \resizebox{50mm}{!}{\includegraphics{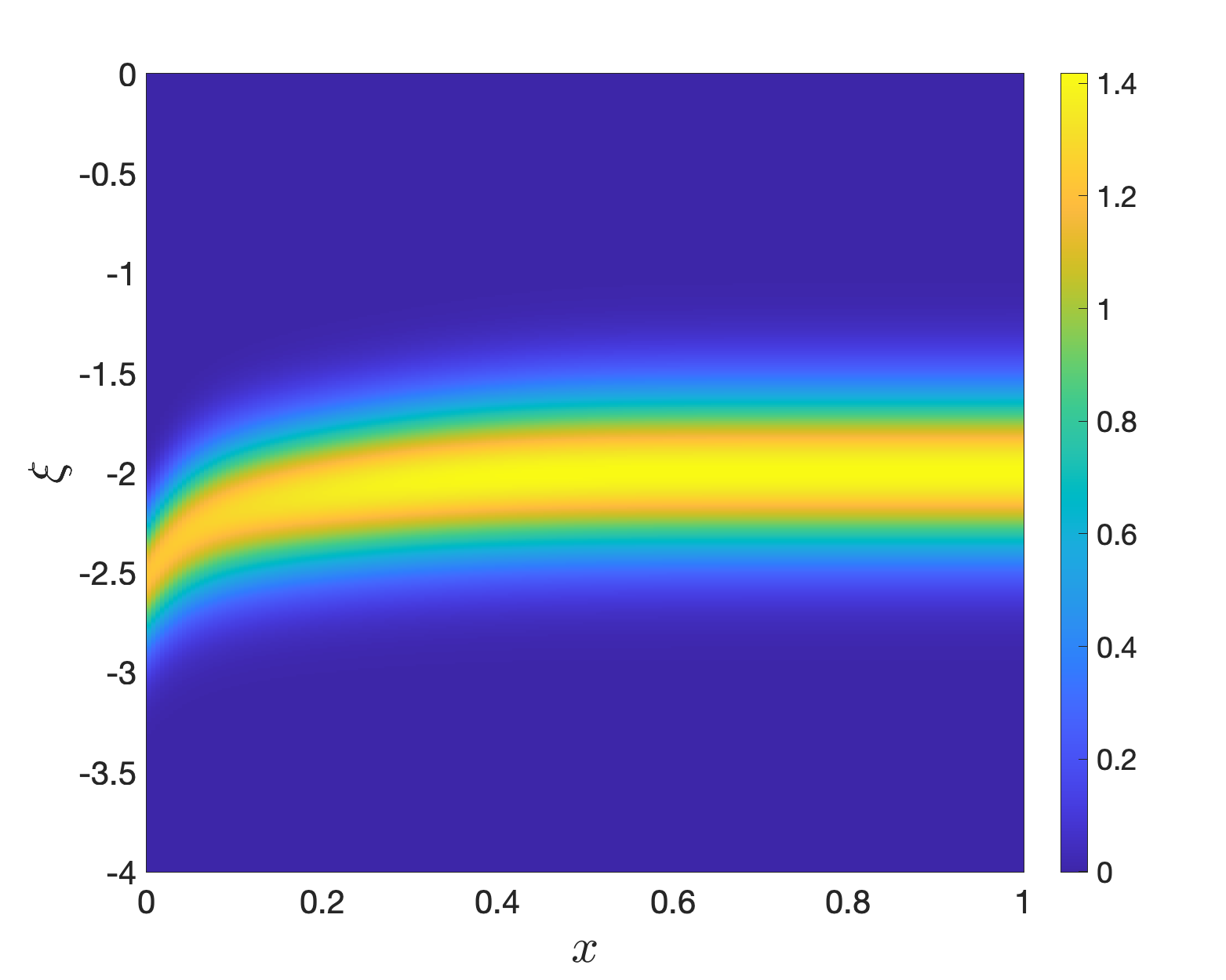}} \hspace{-9mm} & 
    \resizebox{50mm}{!}{\includegraphics{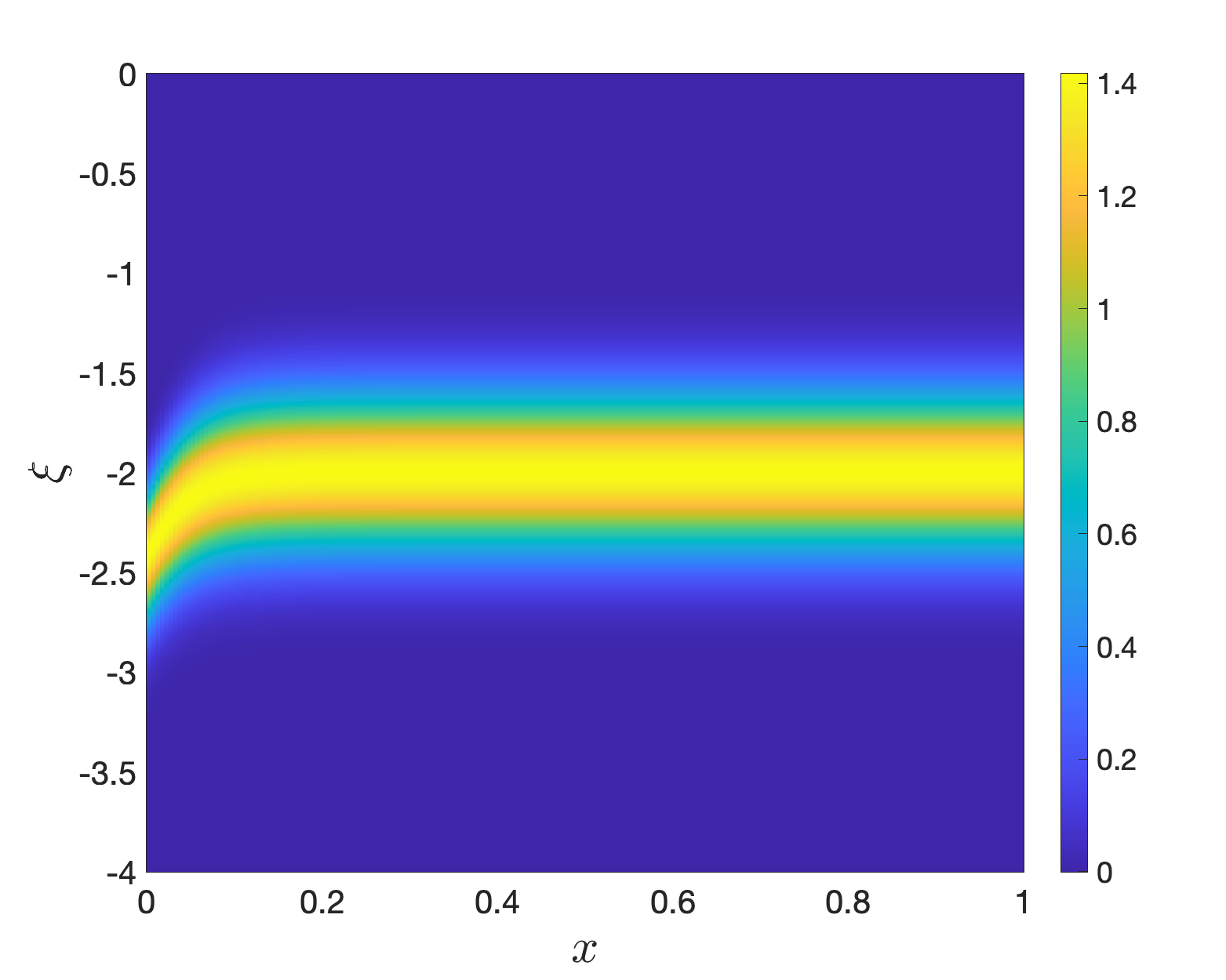}} \\
\end{tabular}  
    \begin{tabular}{ccc}
       $t=0$ \hspace{-9mm} & $t=0.1$ \hspace{-9mm} & $t=0.2$\\
       \resizebox{50mm}{!}{\includegraphics{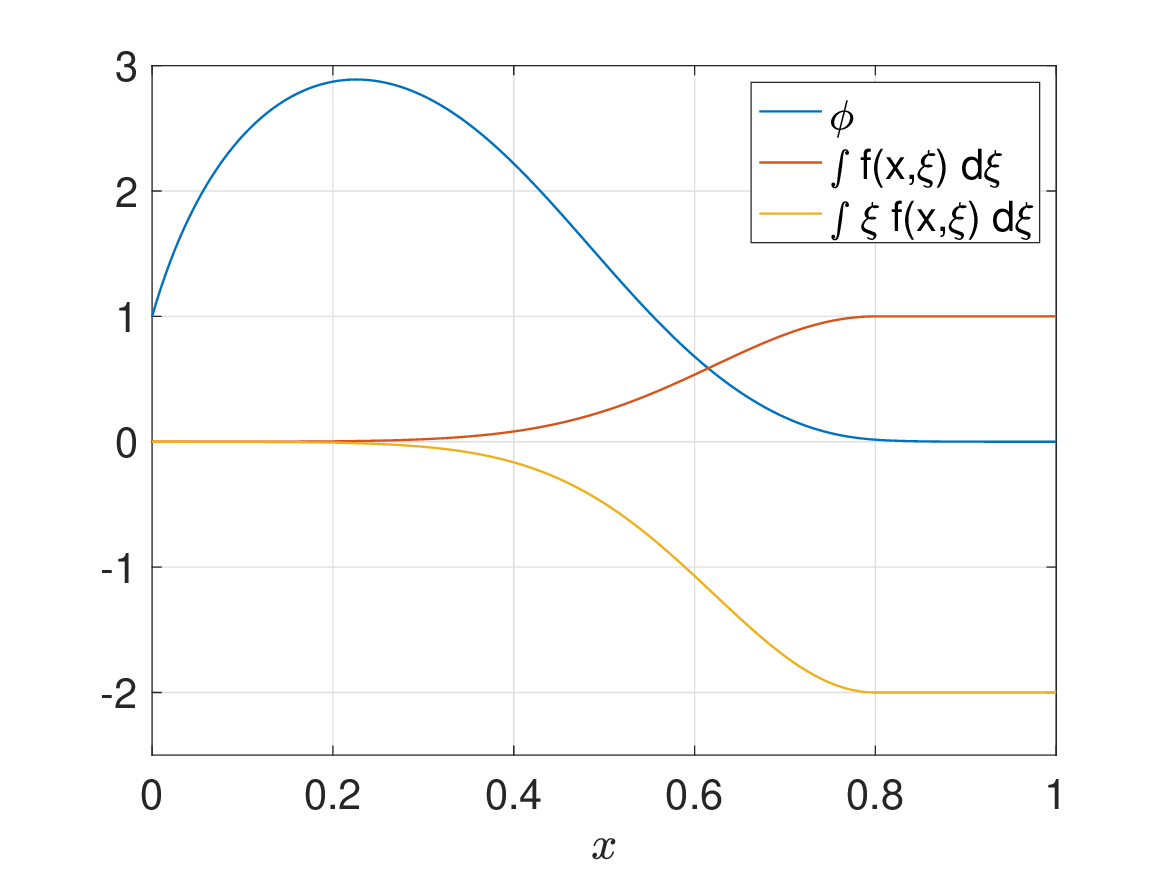}}  \hspace{-9mm} & 
       \resizebox{50mm}{!}{\includegraphics{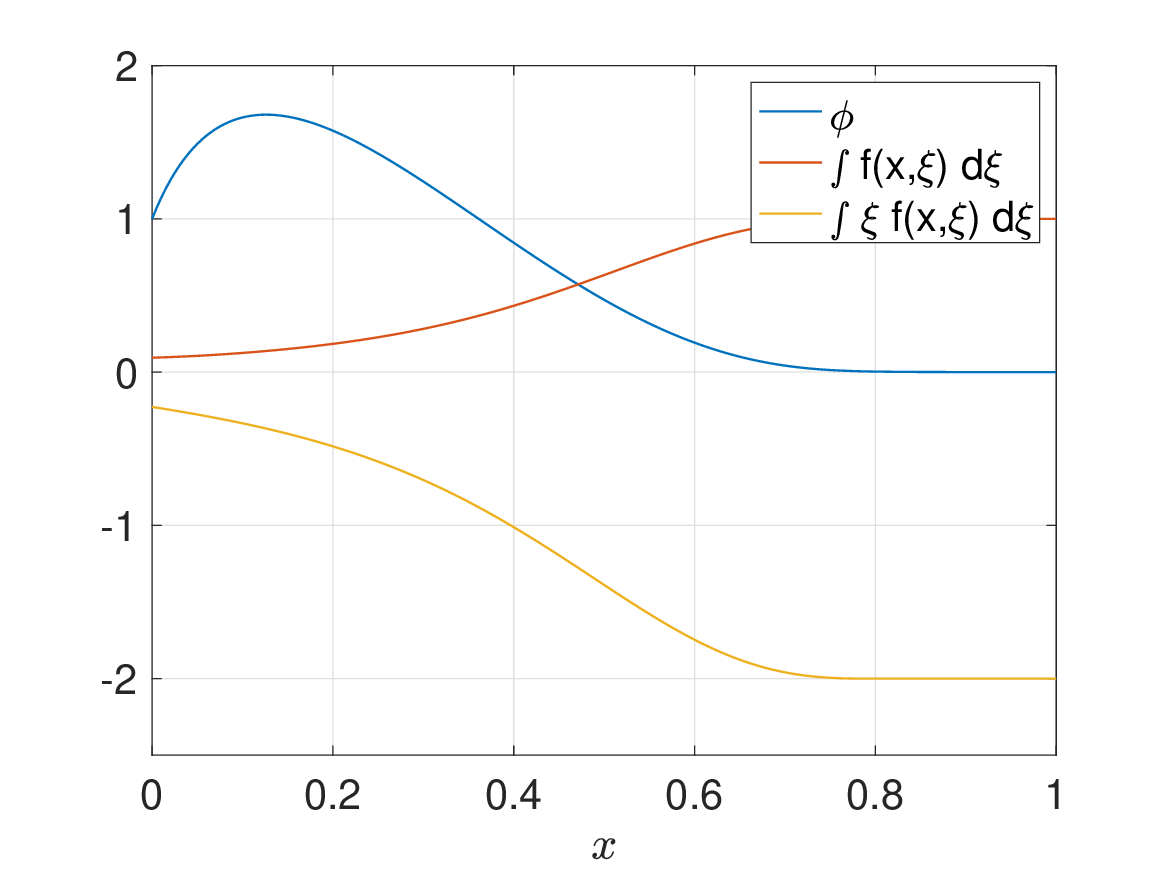}}  \hspace{-9mm} &
       \resizebox{50mm}{!}{\includegraphics{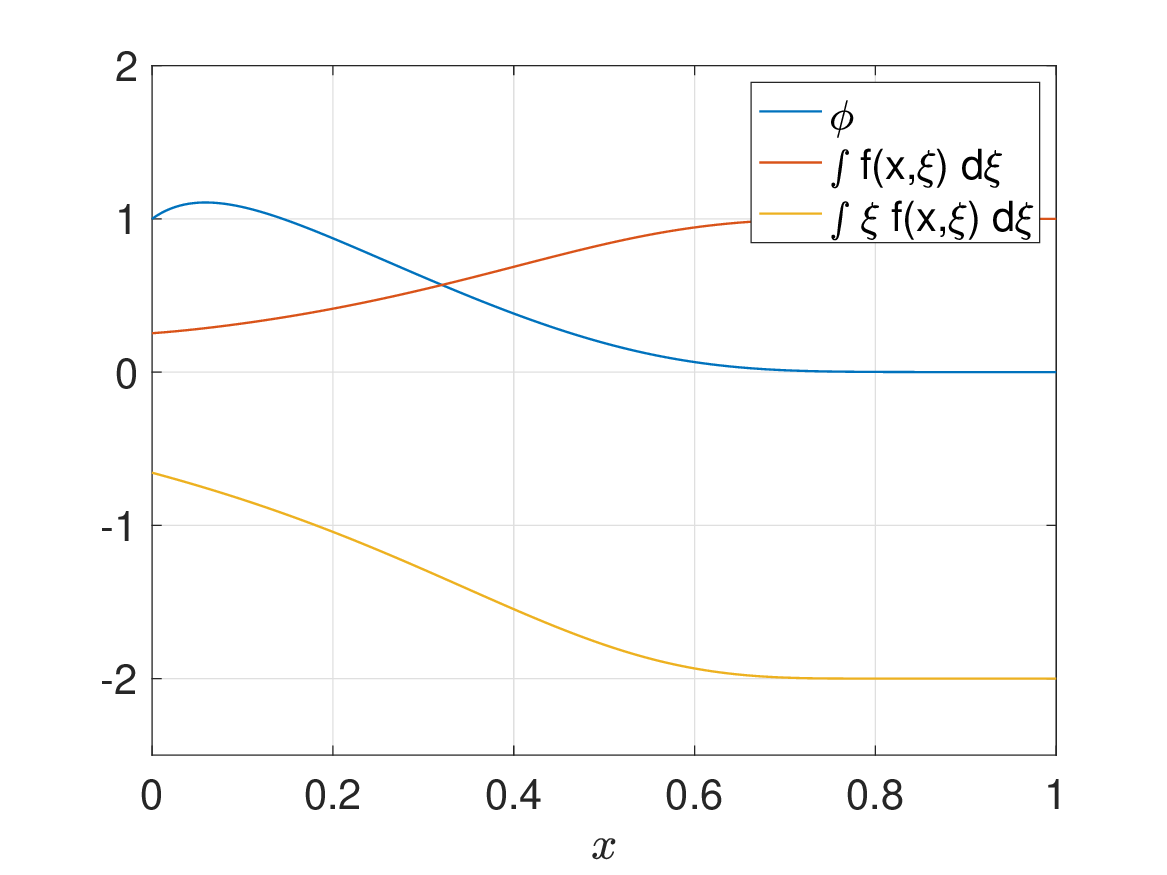}} \\ 
       $t=0.3$ \hspace{-9mm} & $t=0.6$ \hspace{-9mm} & $t=\infty$ (i.e., stationary)\\       
       \resizebox{50mm}{!}{\includegraphics{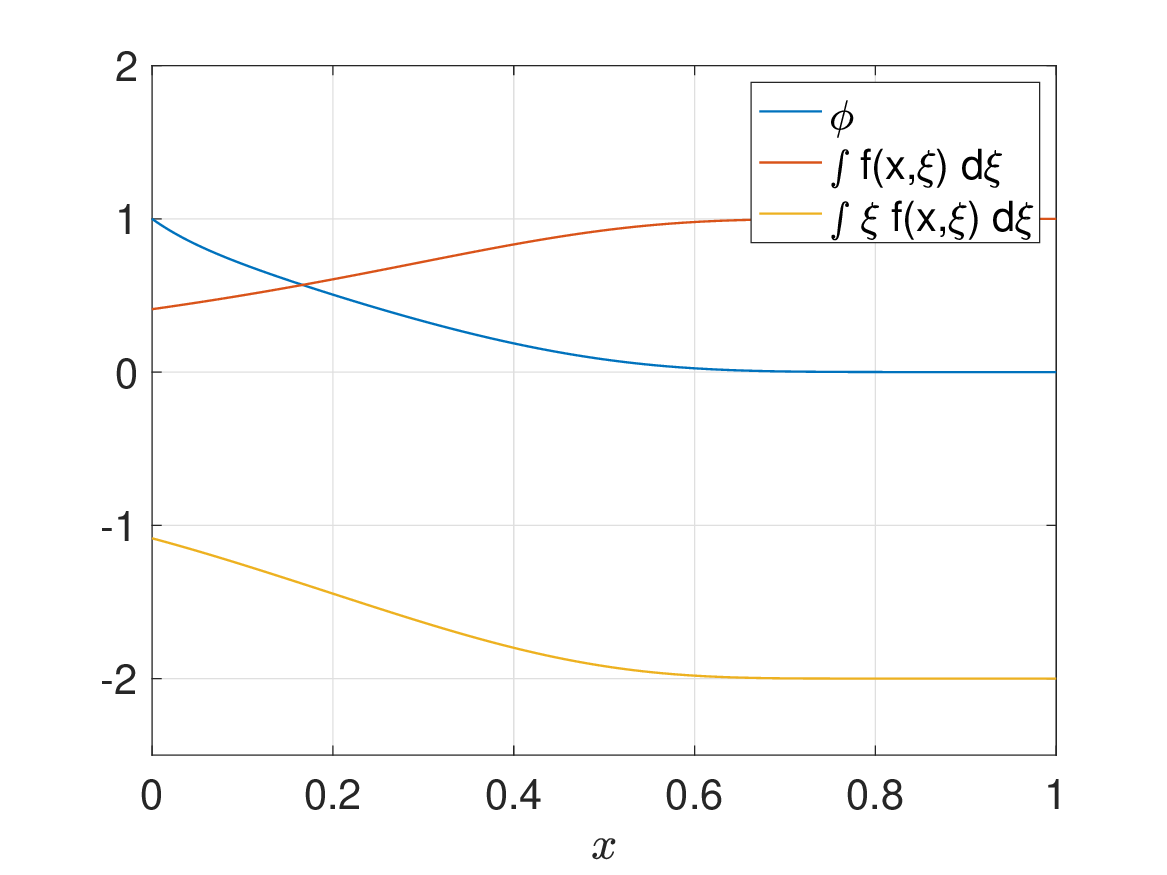}}  \hspace{-9mm} &
       \resizebox{50mm}{!}{\includegraphics{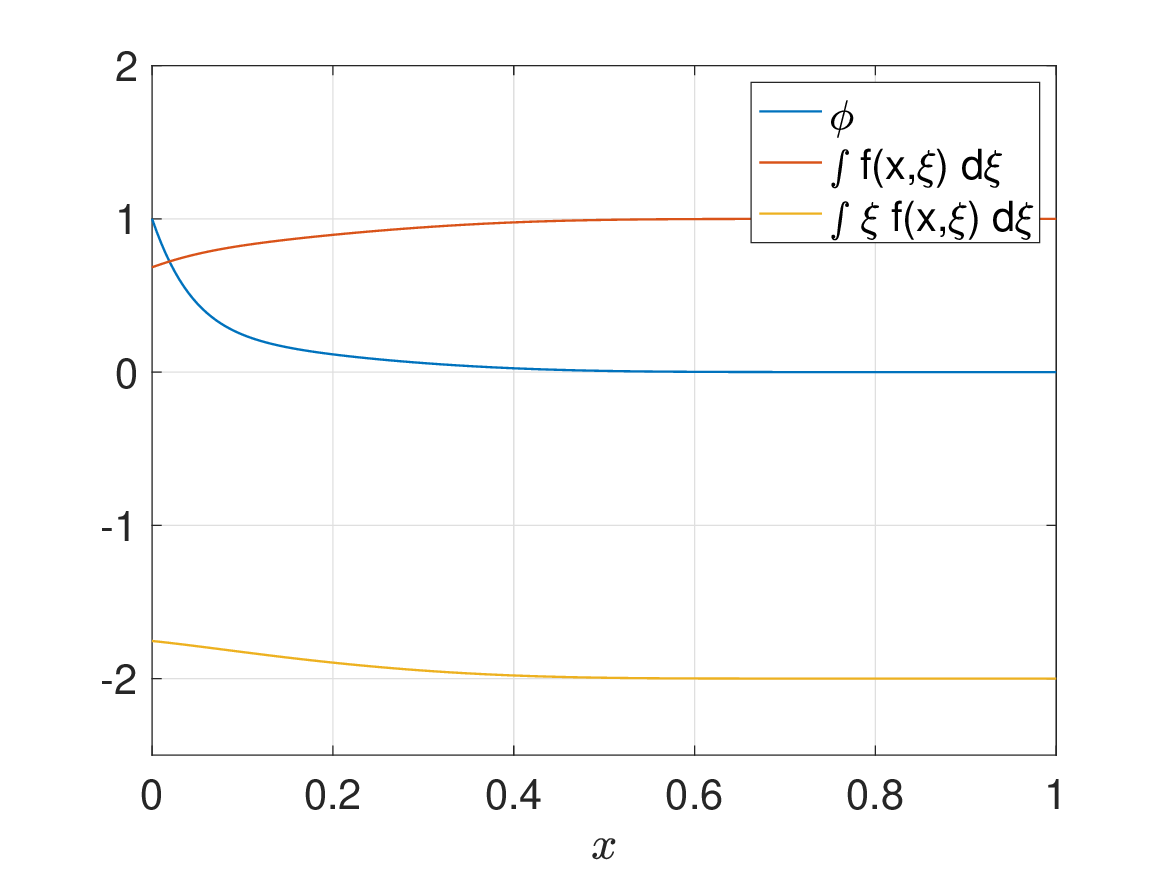}}  \hspace{-9mm} & 
       \resizebox{50mm}{!}{\includegraphics{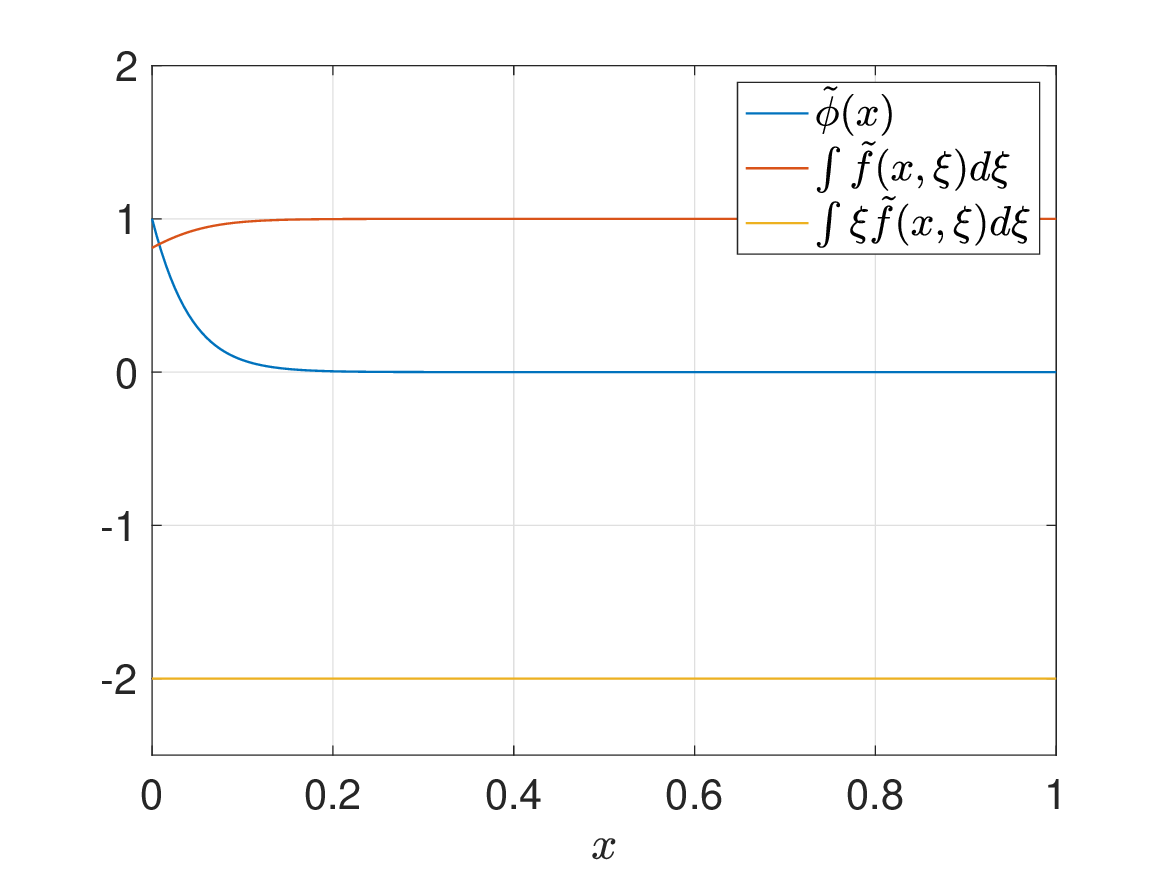}} \\
    \end{tabular}
    \caption{\small{The plots of  $f$, $\phi$, $\int f d\xi$, and $\int \xi f d\xi$ for $\ep = 10^{-3}$.}}\label{ex10}
  \end{center}
\end{figure}

Let
\begin{align*}
f^n &= \sum_{j=-M}^M a^n_j(x) \exp\left(2ij\pi\frac{\xi^2}{L^2}\right).
 \end{align*}
We substitute this into  (\ref{sta_eq_time}), multiply the resultant by $\exp(-2ik\pi\frac{\xi^2}{L^2})$, and integrate it over $(-L,0)$ in $\xi$,
and let $z^n_1 = \phi^n$, $z^n_2 = \phi^n_x$, $z^n_3=a^n_0$, $z^n_{2l+2} = a^n_{-l}$, and $z^n_{2l+3} = a^n_{l}$, where $l=1,2,\ldots M$ to obtain the following first-order system:
\begin{subequations}\label{first_order_sys_time}
\begin{align}
\begin{split}
\frac{dz^n_1}{dx} &= z^n_2,\\
\frac{dz^n_2}{dx} &= \ep^{-1}\left(  m_0z^n_3 + \sum_{j=1}^M m_{-j} z^n_{2j+2} + \sum_{j=1}^M m_{j} z^n_{2j+3}   - e^{-z^n_1} \right),\\
\frac{dz^n_3}{dx} &= \frac{2}{L^2} \left(  m_0\frac{z^n_3-z^{n-1}_{3} }{\Delta t} + \sum_{j=1}^M m_{-j} \frac{z^n_{2j+2}-z^{n-1}_{2j+2}}{\Delta t} + \sum_{j=1}^M m_{j} \frac{z^n_{2j+3}-z^{n-1}_{2j+3}}{\Delta t} \right),\\
\frac{dz^n_{2l+2}}{dx} &= \frac{4 i l \pi}{L^2}z^n_2z^n_{2l+2}  \\
& \quad + \frac{2}{L^2} \left(  m_l\frac{z^n_3-z^{n-1}_{3} }{\Delta t} + \sum_{j=1}^M m_{-j+l} \frac{z^n_{2j+2}-z^{n-1}_{2j+2}}{\Delta t} + \sum_{j=1}^M m_{j+l} \frac{z^n_{2j+3}-z^{n-1}_{2j+3}}{\Delta t} \right),\\
\frac{dz^n_{2l+3}}{dx} &= -\frac{4 i l \pi}{L^2}z^n_2z^n_{2l+3} \\
& \quad + \frac{2}{L^2} \left(  m_{-l}\frac{z^n_3-z^{n-1}_{3} }{\Delta t} + \sum_{j=1}^M m_{-j-l} \frac{z^n_{2j+2}-z^{n-1}_{2j+2}}{\Delta t} + \sum_{j=1}^M m_{j-l} \frac{z^n_{2j+3}-z^{n-1}_{2j+3}}{\Delta t} \right).
\end{split}
\end{align}
The boundary conditions are the same as in \eqref{first_order_sys_bdry}, i.e.,
\begin{align}
\begin{split}
&z_1(0) = 1,\ z_1(1) = 0,\\
&z_{2l+3}(1) =  -\frac{2}{L^2}\int_{-L}^0 \sigma(\xi)\exp\left(-2il\pi\frac{\xi^2}{L^2}\right)\xi d\xi,\ l=0,1,\ldots,M,\\
&z_{2l+2}(1) = -\frac{2}{L^2}\int_{-L}^0 \sigma(\xi)\exp\left(2il\pi\frac{\xi^2}{L^2}\right)\xi d\xi,\ l=1,\ldots,M.
\end{split}
\end{align}

  The initial conditions are 
\begin{align}\label{first_order_sys_bdry_time}
\begin{split}
&z^0_{2l+3}(x) = -\frac{2}{L^2}\int_{-L}^0 f_0(x,\xi)\exp\left(-2il\pi\frac{\xi^2}{L^2}\right)\xi d\xi,\ l=0,1,\ldots,M,\\
&z^0_{2l+2}(x) = -\frac{2}{L^2}\int_{-L}^0 f_0(x,\xi)\exp\left(2il\pi\frac{\xi^2}{L^2}\right)\xi d\xi,\ l=1,\ldots,M.
\end{split}
\end{align}
\end{subequations}
For the discrete problem (\ref{first_order_sys_time}), we use $\Delta t = 0.01$. 
In Figures \ref{ex2_vari_eps_f}--\ref{ex4_phi} and Figure \ref{ex10}, we choose the initial data $f_{0}$ as follows:
\begin{itemize}
\item In Figures \ref{ex2_vari_eps_f}--\ref{ex2_vari_eps_err}, $f_0(x,\xi) = \sigma(\xi)$;
\item In Figures \ref{ex3_f}--\ref{ex3_phi}, $f_0(x,\xi) = \sigma(\xi+1-x)$;
\item In Figures \ref{ex4_f}--\ref{ex4_phi}, $f_0(x,\xi) = \sigma(\xi-1+x)$;
\item In Figure \ref{ex10}, $f_0(x,\xi) = \left( \exp\big(-\frac{(0.8-x)^2}{0.1\!\! \times\!\! 0.8^2}\big)\!\! \times \! \mathcal{H}(0.8-x) \! + \! \mathcal{H}(x-0.8) \right)\!\! \times \! \sigma(\xi)$,
\end{itemize}
where $\mathcal{H}(x)$ is the Heaviside function.

Until now, we employed MATLAB's {\sf bvp4c} for solving the two-point boundary value problems (\ref{first_order_sys}), (\ref{first_order_sys_time}), and the aforementioned boundary layer equations.

As for the limit problem (\ref{vp2}), treated as one-point boundary value problems, various methods like leapfrog and Lax-Friedrichs are applicable. Among these, we use the straightforward upwind methods. Let $f^{0,n} = f^{0,n}(x,\xi) = f^0(t_n,x,\xi)$ and $\phi^{0,n}=\phi^{0,n}(x)=\phi^{0}(t_n,x)$, where $t_n = n\Delta t$ and $n=0,1,2,\ldots$. Then, the limit problem (\ref{vp2}) is discretized as
\begin{align}
\begin{split}
&f^{0,n+1} = f^{0,n} 
+ \frac{\Delta t}{\Delta x} \Big(\xi \big(f^{0,n}(x_{r-1},\xi)-f^{0,n}(x_r,\xi) \big)+ \big(\phi^{0,n}(x_{r-1})-\phi^{0,n}(x_r)\big) f^{0,n}_{\xi} \Big),
\label{limit_sys_upwind1}
\end{split}
\end{align}
where $x_{r}=r\Delta x$ and $r = 1,\ldots,N$ with $N=1/\Delta x$ as well as
\begin{align*}
\phi^{0,n} = -\log \left(\int_{\mathbb R}  f^{0,n}(x,\xi) d\xi\right).
\end{align*}
The boundary conditions are
\begin{align*}
f^{0,n}(1,\xi) = \sigma(\xi),\quad \phi^{0,n}(1) = 0.
\end{align*}
It is essential to highlight that upwind methods initiate the numerical solution updates from the inflow, at $x=x_N=1$.

Let
\begin{align*}
f^{0,n} &= \sum_{j=-M}^M a^n_j(x) \exp\left(2ij\pi\frac{\xi^2}{L^2}\right).
 \end{align*}
We substitute this into (\ref{limit_sys_upwind1}), multiply the result by $\xi\exp(-2ik\pi\frac{\xi^2}{L^2})$, and integrate over $(-L,0)$ in $\xi$, and let $z^n_{1,r}=a^n_0(x_r)$, $z^n_{2l,r} = a^n_{-l}(x_r)$, and $z^n_{2l+1,r} = a^n_{l}(x_r)$, where $l=1,2,\ldots M$,
$x_r=r\Delta x$,  and $r=0,1,\ldots,N$  with $N=1/\Delta x$ to obtain the following discrete system: 
\begin{subequations}\label{limit_sys_discr}
\begin{align}
\begin{split}
z^{n+1}_{1,r} &= z^{n}_{1,r} - \frac{2}{L^2} \frac{\Delta t}{\Delta x} \Big(  (F_1(z^n_{r-1},0) - F_1(z^n_{r},0)) + ( \phi^{0,n}(x_{r-1})-\phi^{0,n}(x_r) ) F_2(z^n_r,0) \Big),\\
z^{n+1}_{2l,r} &= z^{n}_{2l,r} - \frac{2}{L^2} \frac{\Delta t}{\Delta x} \Big( (F_1(z^n_{r-1},l) - F_1(z^n_{r},l)) + ( \phi^{0,n}(x_{r-1})-\phi^{0,n}(x_r) ) F_2(z^n_r,l) \Big),\\
z^{n+1}_{2l+1,r} &= z^{n}_{2l+1,r} - \frac{2}{L^2} \frac{\Delta t}{\Delta x} \Big(  (F_1(z^n_{r-1},-l) - F_1(z^n_{r},-l)) + ( \phi^{0,n}(x_{r-1})-\phi^{0,n}(x_r) ) F_2(z^n_r,-l) \Big),
\end{split}
\end{align}
where 
\begin{align*}
k_j &= \int_{-L}^0 \xi^2\exp\left(2ij\pi\frac{\xi^2}{L^2}\right) d\xi = 
\left\{
\begin{array}{ll}
\displaystyle \frac{L^3}{3} & \text{for $j=0$,}\\[7pt]
 \displaystyle \frac{L^2 i}{4j\pi}(-L+m_{j}) & \text{for $j\neq 0$,} 
\end{array}
\right.\\
F_1(z^n_r,l) &= k_{l} z^n_{1,r} + \sum_{j=1}^M k_{-j+l} z^n_{2j,r} + \sum_{j=1}^M k_{j+l} z^n_{2j+1,r},\\
F_2(z^n_r,l) &= \sum_{j=1}^M k_{-j+l} \frac{-4 i j \pi}{L^2} z^n_{2j,r} + \sum_{j=1}^M k_{j+l} \frac{4 i j \pi}{L^2} z^n_{2j+1,r}.
\end{align*}
The boundary conditions are  
\begin{align}
\begin{split}
&z^n_{2l+1,N} =  -\frac{2}{L^2}\int_{-L}^0 \sigma(\xi)\exp\left(-2il\pi\frac{\xi^2}{L^2}\right)\xi d\xi,\ l=0,1,\ldots,M,\\
&z^n_{2l,N} = -\frac{2}{L^2}\int_{-L}^0 \sigma(\xi)\exp\left(2il\pi\frac{\xi^2}{L^2}\right)\xi d\xi,\ l=1,\ldots,M,
\end{split}
\end{align}
and 
the initial conditions are
\begin{align}
\begin{split}
&z^0_{2l+1,r} = -\frac{2}{L^2}\int_{-L}^0 f_0(x_r,\xi)\exp\left(-2il\pi\frac{\xi^2}{L^2}\right)\xi d\xi,\ l=0,1,\ldots,M,\\
&z^0_{2l,r} = -\frac{2}{L^2}\int_{-L}^0 f_0(x_r,\xi)\exp\left(2il\pi\frac{\xi^2}{L^2}\right)\xi d\xi,\ l=1,\ldots,M,
\end{split}
\end{align}
\end{subequations}
where $f_0(x,\xi)$'s are given just after (\ref{first_order_sys_time}). 
For the discrete problem (\ref{limit_sys_discr}), we use $\Delta t = 0.001$ and $\Delta x = 0.01$.
 
\section*{Acknowledgments}
C.-Y. J. was supported by the National Research Foundation of Korea(NRF) grant
funded by the Korea government(MSIT) (No. 2023R1A2C1003120).
B. K.  was supported by the National Research Foundation of Korea(NRF) grant funded by the Korea government(MSIT)(grant No. 2022R1A4A1032094). 
M. S. was supported by JSPS KAKENHI Grant Numbers 21K03308.

\end{document}